\documentstyle[12pt]{article}
\def\fnote#1#2{\begingroup\def\thefootnote{#1}\footnote{#2}\addtocounter
{footnote}{-1}\endgroup}
\def\inbar{\vrule height1.5ex width.4pt depth0pt}
\def\IB{\relax{\rm I\kern-.18em B}}
\def\IC{\relax\,\hbox{$\inbar\kern-.3em{\rm C}$}}
\def\ID{\relax{\rm I\kern-.18em D}}
\def\IE{\relax{\rm I\kern-.18em E}}
\def\IF{\relax{\rm I\kern-.18em F}}
\def\IG{\relax\,\hbox{$\inbar\kern-.3em{\rm G}$}}
\def\IH{\relax{\rm I\kern-.18em H}}
\def\II{\relax{\rm I\kern-.18em I}}
\def\IK{\relax{\rm I\kern-.18em K}}
\def\IL{\relax{\rm I\kern-.18em L}}
\def\IM{\relax{\rm I\kern-.18em M}}
\def\IN{\relax{\rm I\kern-.18em N}}
\def\IO{\relax\,\hbox{$\inbar\kern-.3em{\rm O}$}}
\def\IP{\relax{\rm I\kern-.18em P}}
\def\IQ{\relax\,\hbox{$\inbar\kern-.3em{\rm Q}$}}
\def\IR{\relax{\rm I\kern-.18em R}}
\def\ZZ{\relax{\sf Z\kern-.4em Z}}
\def\bz{\bar z}
\def\bPhi{\bar \Phi} \def\bth{\bar \theta}
\def\cA{{\cal A}}   
  \def\cH{{\cal H}} \def\cI{{\cal I}}
   
  \def\cP{{\cal P}} 
  
\def\a{\alpha}     
 \def\l{\lambda} \def\bPhi{\bar \Phi}
   \def\th{\theta}

\def\beq{\begin{equation}}  \def\eeq{\end{equation}}
\def\tabroom{\hbox to0pt{\phantom{\Huge A}\hss}}
\def\notin{\ \hbox{{$\in$}\kern-.51em\hbox{/}}}
\def\nn{\nonumber}
\def\lleq#1{\label{#1}\eeq}

\def\del{\partial}
\def\fnote#1#2{\begingroup\def\thefootnote{#1}\footnote{#2}\addtocounter
{footnote}{-1}\endgroup}
\def\tabroom{\hbox to0pt{\phantom{\Huge A}\hss}}
\def\plot#1#2{\vskip\parskip
                  \vbox{\hrule width\hsize
                        \hbox{\kern-0.2pt\vrule height#1
                              \vbox{\hfill}\kern-0.6pt
                              \vrule}\hrule width\hsize}
    \setbox0=\hbox{#2} \dimen0=\wd0 \divide\dimen0 by 2
    \setbox0=\hbox{\kern-\dimen0 #2}
    \dimen3=#1}

\def\hmark{\kern-0.2pt\lower10pt\hbox{\vrule height 5pt}}
\def\leftscalemark{\vbox{\hrule width5pt}}
\def\rightscalemark{\kern-5pt\vbox{\hrule width5pt}}

\def\Place#1#2#3{
  \count10=#1 \advance\count10 by 960
  \dimen1=\hsize \divide\dimen1 by 1920 \multiply\dimen1 by \count10
  \dimen2=\dimen3 \divide\dimen2 by 400 \multiply\dimen2 by #2
  \vbox to
0pt{\kern-\parskip\kern-12truept\kern-\dimen2
    \hbox{\kern\dimen1#3}\vss}\nointerlineskip}

\def\datum#1#2{\Place{#1}{#2}{\copy0}}
\def\bea{\begin{eqnarray}}
\def\eea{\end{eqnarray}}

\begin{document}
\hfill {HD--THEP--92--13}
\vskip .01truein
\hfill{TUM--TP--142/92}
\vskip .01truein
\hfill {CERN--TH--6459/92}
\vskip .7truein
\centerline{\LARGE LANDAU--GINZBURG STRING VACUA}
\vskip .4truein
\centerline{\sc Albrecht Klemm
                  \fnote{\dag}{Supported by the
                   Deutsche Forschungsgemeinschaft}
                  \fnote{\star}{E-mail: aklemm@physik.tu-muenchen.de}}
\vskip .2truein
\centerline{\it Fakult\"at f\"ur Theoretische Physik, Technische
                Univerist\"at M\"unchen}
\centerline{\it James Franck Str., D--8046 Garching, FRG}
\vskip .4truein
\centerline{\sc Rolf Schimmrigk
                    \fnote{\diamondsuit}{E-mail: netah@cernvm.bitnet}}
\vskip .2truein
\centerline{\it Theory Division, CERN}
\centerline{\it CH--1211 Geneva 23, Switzerland}
\vskip .1truein
\centerline{and}
\vskip .1truein
\centerline{\it Institut f\"ur Theoretische Physik, Universit\"at
                Heidelberg}
\centerline{\it Philosophenweg 16, D--6900 Heidelberg, FRG}

\vskip .8truein
\centerline{\bf ABSTRACT}
\vskip .1truein

We investigate a class of (2,2) supersymmetric string vacua which may
be represented as Landau--Ginzburg theories with a quasihomogeneous
potential which has an isolated singularity at the origin. There are
at least three thousand distinct models in this class. All vacua of
this type lead to Euler numbers which lie in the range
$-960 \leq \chi \leq 960$.
The Euler characteristics do not pair up completely hence the space
of Landau--Ginzburg ground states is not mirror symmetric even though
it exhibits a high degree of symmetry. We discuss
in some detail the relation between Landau--Ginzburg models and
Calabi--Yau
manifolds and describe a subtlety regarding Landau--Ginzburg potentials
with an arbitrary number of fields. We also show that the use
of topological identities makes it possible to relate Landau--Ginzburg
theories to types of Calabi--Yau manifolds for which the usual
Landau--Ginzburg framework does not apply.

\vskip .1truein
\noindent
CERN--TH--6459/92

\noindent
4/92
\renewcommand\thepage{}
\vfill
\eject

\parskip .1truein
\parindent=20pt
\pagenumbering{arabic}
\baselineskip 20pt
\noindent
\section{Introduction}

\noindent
Most of the insight into the vacuum structure of string
theory that has been gained over the last few years
derives from different types of explicit constructions. A recent
example is the discovery of mirror symmetry via the construction of
a large class of Landau--Ginzburg (LG) vacua \cite{cls} on the one hand
and via orbifolding of tensor models of exactly solvable minimal
theories \cite{gp} on the other.

Both of these constructions describe only part of the moduli space
of the string and it has become evident that mirror symmetry holds for
a wider class of vacua than provided  by both of these
constructions. It is at present unclear what the most general
framework is for mirror symmetry since  no derivation from first
principles is known, instead its analysis is tied to the specific
constructions mentioned above. Even though the space of phase
orbifolds of tensor models of $N=2$ superconformal minimal theories
implemented with the diagonal affine invariant is closed under  mirror
symmetry  it represents a relatively small subclass of all known vacua.
The  class of exactly solvable ground states discovered sofar  is
certainly not the most general framework in which mirror symmetry can be
formulated.

Neither is the class of Landau--Ginzburg vacua described in
ref. \cite{cls}. The construction presented there was constrained by the
fact  that it focused attention on potentials which
contained only five scaling variables and were of a special form such
as to guarantee the existence of an isolated
singularity, i.e. that they define so--called catastrophes or
singularities. The motivation for this condition comes from the fact
that such theories have a finite number of physical states and are
therefore particularly simple.
It turns out that this class of polynomials is also of considerable
interest to mathematicians who have made an effort to classify the
explicit form such potentials can have \cite{agzv}.
Unfortunately the classification turns out to be involved and
has not been completed yet
\fnote{1}{Recent progress in this direction has made in \cite{maha}.}.

Even though the explicit form of the Landau--Ginzburg potential is
essential for a detailed analysis of a given model it does not have
to be specified in order to gain partial information about the
theory based on it. If, for example, we are interested only in the
spectrum of the model we only need to know the anomalous weights of
the scaling variables and to check that the particular combination
of weights
allows for the existence of a polynomial with an isolated singularity.
We use this fact to complete the analysis begun in \cite{cls}
by constructing all vacuum configurations that are based on LG
potentials.

The paper is organized as follows: in Section 2 we briefly review the
constraints a Landau--Ginzburg potential has to satisfy in order to
define a string vacuum. We also discuss the relation between LG theories
and Calabi--Yau manifolds since this will turn out to be useful
for certain aspects of the construction.
In Section 3 we explain how it is
possible to complete the construction of all LG vacua without an
explicit classification via
a somewhat indirect construction, based on a Bertini--type theorem.
In Section 4 we discuss some of the main features of this class of vacua
and in Section 5 we conclude with some general remarks.

\vskip .4truein
\noindent
\section{ Landau--Ginzburg Vacua}

\noindent
In this section we briefly review the properties of the
 Landau--Ginzburg vacua we wish to investigate \cite{mvw}\cite{lvwg}.
Consider a string ground state based on a Landau--Ginzburg theory
which we
assume to be$ N=2$ supersymmetric  since we demand $N=1$ spacetime
supersymmetry.
Using a superspace formulation in terms of the coordinates
$(z,\bz,\th^+,\bth^+,\th^-,\bth^-)$ the action takes the form
\beq
\cA = \int d^2zd^4\th~K(\Phi_i,\bPhi_i) + \int d^2zd^2\th^- ~W(\Phi_i)
                                        + \int d^2zd^2\th^+ ~W(\bPhi_i)
\eeq
where $K$ is the K\"ahler potential and the superpotential $W$ is a
holomorphic function of the chiral superfields $\Phi_i$.
Since the groundstates of the bosonic potential are the critical points
of the superpotential of the LG--theory
 we demand the existence of such. The type of critical points we
need is determined by the fact that we wish to keep the fermions in
the theory massless, hence we assume that the critical points are
completely degenerate. Furthermore we require that all critical points
are isolated since we wish to relate the finite dimensional ring of
monomials associated to such a singularity with the chiral ring of
physical states in the Landau--Ginzburg theory  in order to construct
the spectrum of the corresponding string vacuum.
Finally we demand the theory to be conformally  invariant from which
follows, relying on some  assumptions regarding the renormalization
properties of the theory,  that the Landau--Ginzburg potential is
quasihomogeneous, i.e. we require  that we can assign to each field
$\Phi_i$ a weight $q_i$ such that for any nonzero complex number
$\l \in \IC^{\star}$
\beq
W(\l^{q_1}\Phi_1,\dots,\l^{q_n}\Phi_n) =\l W(\Phi_1,\dots,\Phi_n).
\eeq
Thus we have formulated the class of potentials we need to consider:
quasihomogeneous polynomials that have an isolated, completely
degenerate singularity (which we can always shift to the origin).

Associated to each of the superpotentials $W(\Phi_i)$ is a
so--called catastrophe which is obtained by first truncating the
superfield $\Phi_i$ to its lowest bosonic component
$\phi_i(z,\bz)$ and then going to the
field theoretic limit of the string by assuming $\phi_i$ to be constant
$\phi_i=z_i$. Writing the weights as $q_i = k_i/d$ we will denote by
\beq
\IC_{(k_1,k_2,\dots,k_n)}[d]
\eeq
the set of all catastrophes described by the zero locus of polnomials
of degree $d$ in variables $z_i$ of weight $k_i$.

The affine varieties described by these polynomials are not compact
and hence it is necessary to implement a projection in order to
compactify these spaces. In Landau--Ginzburg speak this amounts to an
orbifolding of the theory with respect to a discrete group $\ZZ_d$ the
order
of which is the degree of the LG potential \cite{v}. The spectrum of the
orbifold theory will have contain twisted states which together with
the monomial ring of the potential describe the complete spectrum of the
corresponding Calabi--Yau manifold. We will denote the orbifold
of a Landau--Ginzburg theory by
\beq
\IC^{\star}_{(k_1,k_2,\dots,k_n)}[d]
\eeq
and call it a configuration.

In manifold speak the projection should lead to a three--dimensional
K\"ahler manifold with vanishing first Chern class. For a general
Landau--Ginzburg theory  no unambiguous universal prescription for doing
so has been found and as we will describe in Section 4 none can exist.
One way to compactify amounts   to simply impose projective
equivalence \beq
(z_1,....,z_n) \equiv (\l^{k_1} z_1,.....,\l^{k_n} z_n)
\lleq{proj}
which embeds the hypersurface described by the zero locus of the
polynomial into a weighted projective space $\IP_{(k_1,k_2,\dots,k_n)}$
with  weights  $k_i$. The set of hypersurfaces of degree $d$ embedded
in weighted projective space will be denoted by
\beq
\IP_{(k_1,k_2,\dots,k_n)}[d].
\eeq
For a potential with five scaling variables this
construction is completely sufficient in order to pass from the
Landau--Ginzburg theory to a string vacuum \cite{gvw} provided
$d=\sum_{i=1}^5 k_i$ which  is the condition that these hypersurfaces
have vanishing first Chern class.  For more than five variables however
this type of compactification does not lead to a string vacuum.

Even though the precise relation between LG theories and CY manifolds
is not known for the most general case certain facts are known.
Since LG theories with five variables describe a CY manifold embedded in
a four complex dimensional weighted projective space one might expect
e.g. that LG potentials with 6,7 etc variables describe manifolds
embedded in 5,6 etc dimensional weighted projective spaces. This is
not correct.

In  fact none of the models with more than five variables is related
to manifolds embedded in one weighted  projective  space. Instead they
describe Calabi--Yau manifolds embedded in products of weighted
projective space. A simple example is
furnished by the LG potential in six variables
\beq
W=\Phi_1 \Psi_1^2+\Phi_2 \Psi_2^2+\sum_{i=1}^3 \Phi_i^{12}+
\Phi_4^3
\eeq
which corresponds to the exactly solvable model described by the
tensor product of $N=2$ minimal theories at the levels
\beq
(22^2 \cdot 10\cdot 1)_{D^2\cdot A^2}
\eeq
where the subscripts indicate the affine invariants chosen for the
individual factors. This theory belongs to the LG configuration
\beq
\IC^{\star}_{(2,11,2,11,2,8)}[24]^{(3,243)}_{-480}
\lleq{lgform}
and is equivalent to the weighted CICY manifold in the configuration
\beq
\matrix{\IP_{(1,1,1,4,6)}\cr \IP_{(1,1)}\hfill\cr}
        \left [\matrix{1&12\cr 2&0\cr}\right]^{(3,243)}_{-480}
\lleq{cyform}
described by the intersection of the zero locus of the two potentials
\bea
p_1 &=& x_1^2y_1+x_2^2y_2 \nonumber \\
p_2 &=& y_1^{12}+y_2^{12}+y_3^{12}+y_4^3+y_5^2.
\eea
Here we have added a trivial factor $\Phi_5^2$ to the potential and
taken the field theory limit via $\phi_i(z,\bz)=y_i$ where $\phi_i$
 is the lowest component of the chiral superfield $\Phi_i$. The first
column in the degree matrix (\ref{cyform}) indicates that the first
polynomial is of bidgree (2,1) in the coordinates $(x_i,y_j)$
of the product of the projective line $\IP_1$ and the weighted
projective space $\IP_{(1,1,1,4,6)}$ respectively whereas the second
colums shows that the second polynomial is independent of the
projective line and of degree twelve in the coordinates of the
weighted $\IP_4$. The superscripts in (\ref{lgform}),(\ref{cyform})
describe the dimensions $(h^{(1,1)},h^{(2,1)})$ of the fields
corresponding to the cohomology groups $(H^{(1,1)},H^{(2,1)})$
whereas the subscript is the Euler number of the configuration.

It should be noted however that Landau--Ginzburg potentials in six
variables do not describe the most general complete intersection in
products of weighted spaces simply because not all of these manifolds
involve trivial factors, or put differently, quadratic monomials.
A simple example is the manifold that corresponds to the
Landau--Ginzburg theory
$(1\cdot 16^3)_{A\cdot E_7^3}$ with LG potential
\beq
W= \sum_{i=1}^3 \left(\Phi_i^3 + \Phi_i \Psi_i^3\right) + \Phi_4^3.
\eeq
This theory describes a
codimension 2 Calabi--Yau manifold embedded in
\beq
\matrix{\IP_2\cr \IP_3\cr} \left[\matrix{3 &0\cr
                                         1 &3\cr}\right].
\eeq
This space has 8 (1,1)--forms and 35 (2,1)--forms which correspond to
the possible complex deformations in the two polynomials
$p_1,p_2$ \cite{rs1}.

Associated to this Calabi--Yau manifold in a product of ordinary
projective
spaces is an auxiliary algebraic manifold in a weighted six--dimensional
projective space
\beq
\IP_{(2,2,2,3,3,3,3)}[9].
\eeq
obtained via the naive compactification (\ref{proj}) where the first
three
coordinates come from the fields $\Psi_i$ and the last four come from
the
$\Phi_i$. We want to compute the number of complex deformations of this
manifold, i.e. we want to compute the number of monomials of charge 1.

The most general monomial is of the form $\prod_i \Phi_i \prod \Psi_j$.
It is easy to show explicitly that there are precisely 35 monomials by
writing them down but the following remarks may suffice. There are
four different types of possible monomials, depending on whether they
contain the fields $\Phi_i$ not at all (I), linearly (II),
quadratically (III) or cubic (IV).  First note
that monomials of type (I) do not contribute to the marginal operators.
Monomials of type (II) have to contain cubic monomials in the $\Psi$.
Since there are three fields $\Psi_i$ available we obtain a total of
$ 40$ marginal operators of this type.
Because of the equation of motion nine of these are in fact in the ideal
and we are left with 31 complex deformation of this type. Monomials
quadratic in the $\Phi_i's$ again do not contribute wheras those cubic
in the $\Phi_i$ fields contribute the remaining 4. Indeed, there are
20 cubic monomials in terms of the four $\Phi_i$ but using the equations
of motion one finds that 16 of those are in the ideal.

Put  differently: for the total of 60=40+20 monomials of degree
nine (or charge 1) it is possible to fix the coefficients of nine of the
40 by allowing linear field redefinitions of the first three coordinates
and also to fix the coefficients of 16 of the 20 via linear field
redefinitions of the last four coordinates. Hence even though the ambient
space is singular and the manifold hits these singularities in a $\IP_2$
and in a cubic surface $\IP_3[3]$ the resolution does not contribute any
complex deformations because these surfaces are simply connected.

It should be emphasized that this manifold is not the physical internal
part
of a string ground state but plays an auxiliary role which allows to
discuss just  one particular sector of the string vacuum, namely the
complex  deformations.

 Before turning  to  the problem of constructing LG configurations
satisfying the  constraints described above we wish to make some remarks
regarding the validity of the
requirements formulated in the previous paragraph.

Even though the assumptions formulated in refs. \cite{mvw}\cite{lvwg}
and
reviewed above  seem rather reasonable and previous work shows that the
set of such  Landau--Ginzburg theories certainly is an interesting and
quite extensive class of models it is clear that it is not the most
general class of (2,2)--vacua. Although it provides a rather large set
of different models
\fnote{2}{The rather extravagant values that have been
          mentioned in the literature as the number of possible (2,2)
          vacua are based on extrapolations that do not take into
account
          the problem of overcounting that is generic to all of these
          different constructions.}
 which contains many classes of previously constructed vacua
\fnote{3}{Such as vacua constructed tensor models based on
    the ADE minimal models \cite{mvw}\cite{gvw}\cite{ls123}\cite{fkss}
       and level 1 Kazama--Suzuki models \cite{lvwg}\cite{fiq2}\cite{s}
       as well as G/H LG theories \cite{ls5} related to Kazama--Suzuki
       models of higher levels.}
there are known vacua which cannot be described in this framework.

The perhaps simplest example that does not fit into the classification
 above are the Calabi--Yau manifolds in
\beq
\IP_5[4~~2]
\eeq
described by the intersection of two hypersurfaces defined by a quartic
and a quadratic
polynomial in a five dimensional
projective space $\IP_5$ because of the purely quadratic polynomial
that appears as one  of the constraints defining the hypersurface.
 The requirement that the singularity
is completely degenerate in fact seems to exclude a
great many of the complete intersection Calabi--Yau manifolds (CICYs),
the complete class of which was constructed in ref. \cite{cdls}.
An important set of manifolds in that class that cannot be described
either by a superpotential of the type described above is defined by
polynomials of bidegree (1,4) and (1,1)
\beq
\matrix{\IP_1\cr \IP_4\cr}\left[\matrix{1&1\cr 1&4\cr}\right].
\eeq
The superpotential $W=p_1+p_2$ is neither quasihomogeneous nor does it
have  an isolated singularity
\fnote{4}{These manifolds are important in the context of possible
          phase transitions between Calabi--Yau string vacua \cite{cgh}
         via the process of splitting and contraction introduced in
         \cite{cdls}.}.
Thus it appears that there ought to be a generalization of the framework
described above which allows a modified LG description of these and other
string vacua. This however we leave for future work.

\vskip .4truein
\noindent
\section{Transversality of Catastrophes.}

\noindent
The most explicit way of constructing a Landau--Ginzburg vacuum is, of
course,
to exhibit a specific potential that satisfies all the conditions
imposed
by the requirement  that it ought to describe a consistent ground state
 of the string.
 Even though much effort has
gone into the classification of singularities of the type described in
the
previous section
such polynomials have not been yet been classified. The mathematicians
have
 classified
polynomials with at most three variables \cite{agzv} which is two short
of
the lowest  number of variables
that is needed in order to construct a vacuum that allows to formulate
a four--dimensional low energy effective theory
\fnote{5}{The complete list of K3 representations embedded in
$dim_{\IC} =3$ weighted projective space $\IP_{(k_1,k_2,k_3,k_4)}$ has
been obtained in \cite{reid}.}.

In ref.\cite{cls} a set of potentials in five variables was constructed
that  represents the obvious
generalization of the polynomials that appear in two--dimensional
catastrophes.
After imposing the conditions for these theories to describe
string vacua only a finite number of the infinite number of LG theories
survive and all these solutions were constructed. It was clear to the
authors of ref. \cite{cls} that their classification  of singularities
is
not complete,  even after restricting  to five variables, and that their
result should be
considered as a first step toward a proper classification.
It is indeed easy to construct polynomials not contained in the
classification of \cite{cls}, a simple one being furnished by the
example \cite{orlik}
 \beq
\IP_{(15,10,6,1)}[45]\ni
\left\{z_1(z_1^2+z_2^3+z_3^5)+z_4^{45}+z_2^2z_3^4z_4=0\right\}.
\eeq
This polynomial is not of
any of the types analyzed in \cite{cls} but it is transverse
nevertheless.

Knowledge of the explicit form of the potential of a LG  theory is very
useful information when it comes to the detailed analysis of such a
model.
It is however not necessary if only limited knowledge, such as the
computation of the spectrum of the theory, is required. In fact
the only ingredients necessary for the computation of the spectrum
of a LG vacuum \cite{v} are the anomalous dimensions of the scaling
fields
as well as the fact that in a configuration of weights
there exists a polynomial of appropriate degree with an
isolated singularity. To check whether there exists such a polynomial
in a configuration however is much easier than the actual construction
of such a potential. The reason is a theorem by Bertini \cite{algeom}
which  asserts that
if a polynomial does have an isolated singularity on the base locus
then
even though this potential may have worse singularities away from the
base locus there exists a deformation of the original polynomial that
only
admits an isolated singularity anywhere. Hence we only have to find
criteria
that guarantee at worst an isolated singularity on the base locus.
It is precisely this problem that was addressed in the mathematics
literature \cite{f} at the same time the explicit construction of LG
vacua  was started in ref. \cite{cls}.
We briefly review the main point of the argument in \cite{f} for the
sake
of  completeness.

Suppose we
wish to check whether a polynomial in $n$ variables $z_i$ with weights
$k_i$ has an isolated singularity, i.e. whether the condition
\beq
dp=\sum_i \frac{\del p}{\del z_i} dz_i = 0
\lleq{transv}
can be solved at the origin $z_1=\cdots = z_n=0$.  According to
Bertini's theorem
the singularities of a general element in $\IC_{(k_1,...,k_n)}[d]$
will lie
on the base locus, i.e. the intersection of the hypersurface and
all the
components of the base locus, described by coordinate planes of
dimension
$k=1,...,n$. Let $\cP_k$ such a $k$--plane which we may assume to be
described by setting to zero the coordinates $z_{k+1}=\cdots = z_n=0$.
Expand
the polynomials in terms of the nonvanishing coordinates $z_1,...,z_k$
\beq
p(z_1,...,z_n) =
q_0(z_1,...,z_k)~ +~ \sum_{j=k+1}^n q_j(z_1,...,_k)z_j + h.o.t.
\eeq
Clearly, if $q_0\neq 0$ then $\cP_k$ is not part of the base locus
and hence
the hypersurface is transverse. If on the other hand $q_j=0$ then
$\cP_k$ is part of the base locus and singular points
can occur on the intersection of the hypersurfaces defined by
$\cH_j=\{q_j=0\}$.
If however we can arrange this intersection to be empty then the
potentials is smooth on the base locus.

Thus we have found that the conditions for transversality in any
number of variables is the existence for any index set
$\cI_k=\{1,...,k\}$ of
\begin{itemize}
\begin{enumerate}
\item{either a monomial $z_1^{a_1}\cdots z_k^{a_k}$ of degree $d$}
\item{or of at least $k$ monomials $z_1^{a_1}\cdots z_k^{a_k}z_{e_i}$
with distinct $e_i$.}
\end{enumerate}
\end{itemize}

Assume on the other hand that neither of these conditions
hold  for
all index sets and let $\cI_k$ be the subset for which they fail. Then
the potential has the form
\beq
p(z_1,...,z_n) = \sum_{j=k+1}^n q_j(z_1,...,z_k)z_j ~+~ \cdots
\eeq
with at most $k-1$ nonvanishing $q_j$. In this case the intersection
of the hypersurfaces $\cH_j$ will be positive and hence the polynomial
$p$ is not transverse.

As an example for the considerable ease with which one can check whether
a given configuration allows for the existence of a
potential with an isolated singularity  consider the polynomial of Orlik
and Randall
\beq
p=z_1^3+z_1z_2^3+z_1z_3^5+z_4^{45}+z_2^2z_3^4z_4.
\eeq
Condition (\ref{transv}) is equivalent to  the  system of equations
\beq
\begin{array}{r l r l}
 0 &=~ 3z_1^2+z_2^3+z_3^5 , & 0 &=~ 3z_1z_2^2+2z_2z_3^4z_4  \\
 0 &=~ 5z_1z_3^4 + 4z_2^2z_3^3z_4, & 0 &=~ z_2^2z_3^4+45z_4^{44} .
\end{array}
\eeq
which, on the base locus, collapses to the trivial pair of
equations $z_2z_3=0=z_2^3+z_5^5$. Hence this configuration allows for
such a polynomial. To check the system away from the base locus
clearly is much more complicated.

By adding a fifth variable $z_5$ of weight 13 it is possible to define
a Calabi--Yau deformation class
$\IP_{(1,6,10,13,15)}[45]_{-72}^{(17,53)}$, a configuration not
considered in \cite{cls}.

\vskip .4truein
\noindent
\section{Finiteness Considerations.}

  \noindent
The problem of finiteness has two parts: first one has to put a
constraint
on the number of scaling fields that can appear in the LG theory
and then one has to determine limits on the exponents with which the
variables occur in the superpotential. Both of these constraints follow
>from the fact that the central charge of a Landau--Ginzburg theory with
fields of charge $q_i$
\beq
c=3\sum \left(1-2 q_i\right)=:\sum c_i
\lleq{cc}
has to be $c=9$ in order to describe a string vacuum.

It should be clear that without any additional input the number of
LG vacuum configurations one can exhibit is infinite. This is to be
expected simply because it is known from the construction of CICYs
\cite{cls} that it is often possible to rewrite a manifold in an infinite
number of ways and we ought to encounter similar things in the  LG
framework. A trivial way to do this is to simply add mass terms which
do not contribute to the central charge. Even though trivial such
mass terms
are important and necessary for LG theories, not only in orbifold
constructions \cite{kss} but also in order to relate them to  CY
manifolds. Consider e.g. the codimension four Calabi--Yau manifold
\beq
\matrix{\IP_1 \cr \IP_2\cr \IP_2 \cr \IP_2\cr}
\left[\matrix{2 &0 &0 &0\cr
              1 &2 &0 &0\cr
              0 &1 &2 &0\cr
              0 &0 &1 &2\cr}\right]
\lleq{boggle}
with the defining polynomials
\beq
\begin{array}{r l r l}
p_1 &=~ \sum_{i=1}^2 u_i^2v_i, & p_2 &=~ \sum_{i=1}^3 v_i^2w_i  \\
p_3 &=~ \sum_{i=1}^3 w_i^2x_i, & p_4 &=~ \sum_{i=1}^3 x_i^2
\end{array}
\lleq{bogglepollies}
the superpotential $W=\sum p_i$ of which lives in
\beq
\IC^{\star}_{(5,5,6,6,6,4,4,4,8,8,8)}[16]_{-32}^{10}
\eeq
and has an isolated  singularity at the origin. All eleven variables
are coupled and hence
this example appears to involve three fields with zero central charge
in a nontrivial way.

It turns out however that the manifold (\ref{boggle}) is equivalent
to a manifold with nine variables. One way to see this is by making
 use of some topological identities introduced in \cite{cdls}. First
consider the well known isomorphism
\beq
\IP_2[2] \equiv \IP_1,
\eeq
which allows to rewrite the manifold above as
\beq
\matrix{\IP_1 \cr \IP_2\cr \IP_2 \cr \IP_1\cr}
\left[\matrix{2 &0 &0\cr
              1 &2 &0\cr
              0 &1 &2\cr
              0 &0 &2\cr}\right].
\eeq
Using the surface identity \cite{cdls}
\beq
\matrix{\IP_1 \cr \IP_2\cr} \left[\matrix{ 2\cr 1\cr}\right]=
\matrix{\IP_1 \cr \IP_1\cr}
\eeq
applied via the rule
\beq
\matrix{\IP_1 \cr \IP_2\cr X\cr} \left[\matrix{ 2 &0\cr
                                                1 &a\cr
                                                0 &M\cr}\right]=
\matrix{\IP_1 \cr \IP_1\cr X\cr} \left[\matrix{ a\cr
                                                a\cr
                                                M\cr}\right]
\eeq
shows that this space in turn is equivalent to
\beq
\matrix{\IP_1 \cr \IP_1\cr \IP_1 \cr\IP_2\cr}
\left[\matrix{2 &0 \cr
              2 &0 \cr
              0 &2 \cr
              1 &2 \cr}\right],
\lleq{nonlg}
a manifold with only nine homogeneous coordinates.
It should be noted that the LG potential of (\ref{nonlg}) defined
by the sum of the two polynomials certainly does not
have an isolated singularity. Furthermore it is not possible to
 even assign weights to the fields such that the
central charge comes out to be nine!
It is thus possible, by applying topological identities, to extend
the applicability of Landau--Ginzburg theories to types of Calabi--Yau
manifolds that were hitherto completely inaccessible by the standard
formulation.

Further insight into the problem of redundancy in the construction of
LG potentials can be gained by an LG theoretic analysis of this example.
 From the weights of the scaling variables in the LG configuration
above it is clear that the spectrum of this LG configuration remains
the same if the last three coordinates are set to zero. In the
potential
\beq
W=\sum_{i=1}^2 \left(u_i^2v_i+v_i^2w_i+w_i^2x_i+x_i^2\right) +
   (v_3^2w_3+w_3^2x_3+x_3^2)
\lleq{bogglelg}
 described by the CICY polynomials (\ref{bogglepollies}) these variables
cannot be set to zero because they are coupled to other fields; hence it
seems impossible to reduce the number of fields. Consider however the
following change of variables
\beq
y_i=x_i+\frac{1}{2}w_i^2,~~i=1,2,3.
\eeq
 It follows from these transformations that the potential defined by
by (\ref{bogglelg}) is equivalent to
\beq
W=\sum_{i=1}^2(u_i^2v_i+v_i^2w_i-
      \frac{1}{4}w_i^4)+v_3^2w_3-\frac{1}{4}w_3^4
\eeq
 Adding a trivial factor and splitting this potential apart into three
separate polynomials
\beq
p_1=u_1^2v_1+u_2^2v_2,~~~p_2=v_1^2w_1+v_2^2w_2+v_3^2w_3,~~~
p_3=-\frac{1}{4}\left(w_1^4+w_2^4+w_3^4\right)+x^2
\eeq
we see that the original model is equivalent to a weighted
 configuration
\beq
\matrix{\IP_1 \hfill \cr \IP_2 \hfill \cr \IP_{(1,1,1,2)}\cr}
\left[\matrix{2 &0 &0\cr
              1 &2 &0\cr
              0 &1 &4\cr}\right],
\eeq
which again describes manifolds with nine variables.
Thus the original configuration (\ref{boggle}) is in fact equivalent
to two different (weighted) CICY respresentations
\beq
\matrix{\IP_1 \cr \IP_1\cr \IP_1 \cr\IP_2\cr}
\left[\matrix{2 &0 \cr
              2 &0 \cr
              0 &2 \cr
              1 &2 \cr}\right]
{}~~\equiv ~~
\matrix{\IP_1 \cr \IP_2\cr \IP_2 \cr \IP_2\cr}
\left[\matrix{2 &0 &0 &0\cr
              1 &2 &0 &0\cr
              0 &1 &2 &0\cr
              0 &0 &1 &2\cr}\right]
{}~~\equiv ~~
\matrix{\IP_1 \hfill \cr \IP_2 \hfill \cr \IP_{(1,1,1,2)}\cr}
\left[\matrix{2 &0 &0\cr
              1 &2 &0\cr
              0 &1 &4\cr}\right].
\eeq

To summarize the last few paragraphs, we have shown two things: first
that by adding trivial factors and coupling them to the remaining fields
we can give a Landau--Ginzburg description of a
larger class of Calabi--Yau manifolds than previously thought
possible. Furthermore we can use topological identities to obtain
an LG formulation of CY manifolds which do not admit a
canonical LG potential at all.
Incidentally we have also shown that it is possible to relate complete
intersection manifolds embedded in products of projective spaces to
 weighted complete intersection manifolds
embedded in products of weighted projective space.

Similarly the number of fields can grow without bound if we not only
allow
fields that do not contribute to the central charge but also fields
with
a negative contribution. Again such fields provide  redundant
descriptions
of simpler LG theories but are important for the LG/CY relation and
occur in the constructions of splitting and contraction introduced in
ref. \cite{cdls}. Even though these
constructions were discussed in \cite{cdls} only in the context of
Calabi--Yau
manifolds embedded in products of ordinary projective spaces they
readily generalize to the more general framework of weighted  projective
spaces.

In special circumstances the splitting or contraction process does not
change
the spectrum of the theory and hence it provides another tool to relate
 LG potentials with at most nine variables manifolds with more than
nine
homogenous coordinates.  Consider e.g. the manifold embedded in
\beq
\matrix{\IP_{(1,1)} \hfill \cr \IP_{(1,1,1,1,3)}\cr}
\left[\matrix{2&0\cr 1&6\cr}\right]
_{-252}^{(2,128)}
\eeq
which is described by the zero  locus of the two polynomials
\bea
p_1&=&x_1^2y_1+x_2^2y_2 \nn \\
p_2&=&y_1^6+y_2^6+y_3^6+y_4^6+y_5^2
\eea
which can be described by a superpotential $W=p_1+p_2$ which defines a
Landau--Ginzburg theory in
\beq
\IC^{\star}_{(5,5,2,2,2,2,6)}[12]^{(2,128)}_{-252}
\eeq
This manifold can be rewritten via an ineffective split in an infinite
sequence of different representations as
\beq
\matrix{\IP_{(1,1)} \hfill \cr \IP_{(1,1,1,1,3)}\cr}
\left[\matrix{2&0\cr 1&6\cr}\right]
\longrightarrow
\matrix{\IP_{(1,1)}\hfill\cr \IP_{(1,1)}\hfill\cr \IP_{(1,1,1,1,3)}\cr}
\left[\matrix{2&0&0\cr 1&1&0\cr 0&1&6\cr}\right]
\longrightarrow
\matrix{\IP_{(1,1)}\hfill\cr \IP_{(1,1)}\hfill\cr\IP_{(1,1)}\hfill \cr
        \IP_{(1,1,1,1,3)}\cr}
\left[\matrix{2&0&0&0\cr 1&1&0&0\cr 0&1&1&0\cr 0&0&1&6}\right]
\longrightarrow
\cdots
\eeq
which are described by LG potentials in the alternating classes
\beq
\IC^{\star 7}_{(5,5,2,2,2,2,6)}[12] \longrightarrow
\IC^{\star 9}_{(1,1,10,10,2,2,2,2,6)}[12] \longrightarrow
\IC^{\star 11}_{(5,5,2,2,10,10,2,2,2,2,6)}[12] \longrightarrow \cdots
\eeq
i.e. the infinite sequence belongs to the configurations
\beq
\IC^{\star 7+4k}_{(5,5,2,2,10,10,2,2,10,10,...,2,2,10,10,2,2,6)}[12]
\eeq
where the part $(2,2,10,10)$ occurs $k$ times, and
\beq
\IC^{\star 5+4k}_{(1,1,10,10,2,2,10,10,2,2,...10,10,2,2,2,2,6)}[12].
\eeq
where $(10,10,2,2)$ occurs $k$ times.

The construction above easily generalizes to a number of examples
which all belong to a class of spaces discussed in ref. \cite{rsres}.
Consider manifolds embedded in
\beq
\matrix{\IP_{(1,1)} \hfill\cr \IP_{(k_1,k_1,k_3,k_4,k_5)}\cr}
\left[\matrix{2&0\cr k_1&k\cr}\right].
\eeq
where $k=k_1+k_3+k_4+k_5$. These spaces can be split
into the infinite sequences
\beq
\matrix{\IP_{(1,1)} \hfill \cr \IP_{(k_1,k_1,k_3,k_4,k_5)}\cr}
\left[\matrix{2&0\cr k_1&k\cr}\right]
\longrightarrow
\matrix{\IP_{(1,1)} \hfill \cr \IP_{(1,1)}\hfill \cr
        \IP_{(k_1,k_1,k_3,k_4,k_5)}\cr}
\left[\matrix{2&0&0\cr 1&1&0\cr 0&k_1&k\cr}\right]
\longrightarrow
\matrix{\IP_{(1,1)} \hfill \cr \IP_{(1,1)}\hfill\cr\IP_{(1,1)}\hfill \cr
        \IP_{(k_1,k_1,k_3,k_4,k_5)}\cr}
\left[\matrix{2&0&0&0\cr 1&1&0&0\cr 0&1&1&0\cr 0&0&k_1&k}\right]
\longrightarrow
\cdots
\eeq
If the weights are such that $k/k_i$ is an integer
then it is easy to write down the tensor model that corresponds to
it (but this is not essential).
In such models the levels $l_i$ of the tensor model
 $l_1^2\cdot l_3\cdot l_4\cdot l_5$
in terms of the weights are given by
\beq
l_1=l_2=\frac{2k}{k_1}-2,~~~l_i=\frac{k}{k_i}-2,~i=3,4,5
\eeq
and the corresponding LG potentials live in
\bea
\IC^{\star 7}_{(k-k_1,k-k_1,2k_1,2k_1,2k_3,2k_4,2k_5)}[2k]
\longrightarrow
\IC^{\star 9}_{(k_1,k_1,2(k-k_1),2(k-k_1),2k_1,2k_1,2k_3,2k_4,2k_5)}[2k]
\longrightarrow \cdots
\eea
i.e. they belong to the sequences
\beq
\IC^{\star 7+4p}_{((k-k_1),(k-k_1),2k_1,2k_1,2(k-k_1),2(k-k_1),....,
                                         2k_1,2k_1,2k_3,2k_4,2k_5)}[2k]
\eeq
where the part $(2k_1,2k_1,2(k-k_1),2(k-k_1))$ occurs $p$ times, and
\beq
\IC^{\star 9+4k}_{(k_1,k_1,2(k-k_1),2(k-k_1),
2k_1,2k_1,...,2k_3,2k_4,2k_5)}[2k].
\eeq
where  $(2(k-k_1),2(k-k_1),2k_1,2k_1)$ occurs $p$ times.

All these models are constructed in such a way that they have central
charge
nine, but in contrast to the example discussed  previously,
there now appear fields with negative central charge.
In the case at hand however these dangerous fields
only occur in a {\it coupled} subpart of the theory and the smallest
subsystem which involves these fields and which one can isolate is in
fact a theory with positive central charge. In the series of splits
just described the fields with negative central charge appearing in the
first split e.g. always appear in the subsystem described by the
configurations
\beq
\IC^{\star}_{(k_1,2(k-k_1),2k_1)}[2k]
\eeq
with potentials of the form
\beq
x^2y +yz+z^{2k/k_1}.
\eeq
Thus the contribution to the central charge of this sector becomes
\beq
c=3\left[\left(1-\frac{k_1}{k}\right)+\left(1-\frac{2(k-k_1)}{k}\right)+
         \left(1-\frac{2k_1}{k}\right)\right]
\lleq{ccc}
which is always positive. This formula suggests that it ought
to be possible to dispense with the variables $y,z$ altogether as
their total contribution to the central charge adds up to zero and
that this theory is equivalent to that of a single monomial of degree
$k/k_1$.

More generally one may consider the Landau--Ginzburg theory defined by
the potential
\beq
x_1^ax_2+x_2x_3+\cdots + x_{n-1}x_n + x_n^b.
\lleq{triv}
 From the central charge
\beq
c= \left\{ \begin{array}{l l}
               6\left(1-\frac{1}{a}\right)\left(1-\frac{1}{b}\right),
                                                        & n~{\rm even} \\
               3\left(1-\frac{2}{ab}\right), &n~{\rm odd}
            \end{array} \right\}
\eeq as well as from the dimension of the chiral ring
\beq
dim~R_n = \prod \left(\frac{1}{q_i}-1\right)
        = \left\{ \begin{array}{l l}
                  ab-b-1,&n~{\rm even} \\
                   ab-1, &n~{\rm odd}
                  \end{array} \right\}
\eeq
we expect  this theory to be equivalent to
\beq
x_1^ax_2+x_2x_3+\cdots x_{n-1}x_n + x_n^b =
\left\{ \begin{array}{l l}
               x_1^ax_n+x_n^b,  &n~{\rm even} \\
               x_1^{ab},        &n~{\rm odd}
                  \end{array} \right\}
\lleq{ident}

Such identities are supported by the identification of rather different
 LG configurations such as
\beq
\IC^{\star}_{(1,1,1,6,6,6,3,3,3,3,3)}[9] \equiv \IC^{\star}_{(1,1,1,3,3)}[9]
\eeq
as well as many other identifications which are rather nontrivial
in the context of the associated manifolds.

It follows from the considerations above that in order to avoid redundant
reconstructions of LG theories we have to assume that the central charge
of all scaling fields of the potential should  be positive.
 In order to relate the potentials to manifolds we then may add
one or several trivial factors or  more complicated theories with zero
central charge.

Using the results above we derive in the following more detailed
finiteness conditions.  Observe first that from eqn. (\ref{cc})
written as
\begin{equation}
\sum_{i=1}^r q_i=\left( {r\over 2}-{c\over 6}\right):=\hat c\label{cbed}
\end{equation}
we obtain $r>c/3$.

Now let $p$ be a polynomial of degree $d$ in $r$ variables.
For the index set $\cI_1$ the conditions for transversality
implies the existence of $n_i\in \IN^+$, $i=1,\ldots,r$ and a map
$j:\cI_r \rightarrow \cI_r$  such that for all $i$ there exists
$j(i)$ such that
\begin{equation}
q_i={{1-q_{j(i)}}\over n_i}.
\lleq{qs1}
Let us first see how many non-trivial fields can occur at most.
Fields which have charge $q_i\le 1/3$
contribute $c_i\ge 1$ to the conformal anomaly.
Now consider fields with larger charge. Since we assume $c_i>0$ they are
in the range $1/3<q_i<1/2$.
Among these fields the transversality condition (1.) can not hold,
because two of them are not enough and three of them are too many
fields  in order to form a monomial of charge one. Transversality
condition (2.) implies that each of them has to occur together with a
partner field $z_{j(i)}$. These pairs contribute according to
(\ref{qs1},\ref{cc}) $c_i+c_{j(i)}>2$ to the conformal anomaly, so we
can conlude $r\le c$.

In order to construct all transversal
LG potentials for a given $c$
we choose a specific $r$ in the range obtained above and consider
all possible maps $j$ of which there are $r^r$.
Without restricting generality we may
{\sl then} assume the $n_i$ to be ordered
$n_1\le  \cdots  \le n_r$.
Starting with (\ref{cbed}) we obtain via eqn. (\ref{qs1}) and the
positivity of the charges a bound $n_1<r/c$. Now we choose $n_1$  in the
allowed range and use  (\ref{qs1}) in order to eliminate the $q_1$ if
necessary in favour of the $q_2,\ldots,q_r$. This yields
an equation of the general form $(p=2)$
\begin{equation}
\sum_{i=p}^r w^{(p)}_i q_i=\hat c^{(p)}.\label{cbed2}
\end{equation}
If $\hat c^{(p)}\neq 0$ eq. (\ref{cbed2}) is either in contradiction
with the positivity of the charges or one can derive a finite bound for
$n_{p}$. Assume $\hat c^{(p)}>0$ and let $\cI_+$ the
indices of the positive $w_i^{(p)}$ then one has
$n_{p}<(\sum_{i\in \cI_+} w_i^{(p)})/\hat c^{(p)}$, likewise if
$\hat c^{(p)}<0$ we have
$n_{p}<(\sum_{i\in\cI_-} w_i^{(p)})/\hat c^{(p)}$.

Consider the case $\hat c^{(p)}=0$.
If the $w_i^{(p)}$ are indefinite then
we get no bound from (\ref{cbed2}). However we will show that the
existence of certain monomials, which are required by the transversality
conditions implies a bound for $n_p$.
Let $\cI_a$ denote the indices of the already bounded
$n_i$ and $\cI_b$ the others. The charge of the field $z_a$ with
$a\in \cI_a$
will depend on the unknown charge of a field $z_{b(a)}$ with
$b\in \cI_b$ if
there is a chain of indices
$a_0=a,a_1=j(a),\dots,a_l=j(\ldots j(a)\ldots)=:b(a)$ linked by the
map $j$. The charge of $z_a$ is given by
\beq
q_a={1\over n_a}-{1\over n_a n_{a_1}}+
\ldots,-{(-1)^l\over n_a\ldots n_{l-1}}+
{(-1)^l q_{b(a)}\over n_a\ldots n_{l-1}}.\lleq{qform}
Indefiniteness of the $w_i^{(p)}$ can only occur if there are fields
$z_a$, $a\in \cI_a$, with odd $l$, i.e. the last term in (\ref{qform})
$s_a q_{b(a)}:=(-1)^l/(n_a\ldots n_{a_{l-1}} q_{b(a)})$ is negative.
Call the index set of these fields $\cI_a^-$.
Assume first transversality condition (1.) to hold.
This implies the existence of $m_i\in \IN^+$ ($m_i<2 n_i$)
such that $\sum_{i\in \cI^-_a}
m_i q_i=1$. For the unknown $q_i$, $i\in \cI_b$, we get an equation
of the form $\sum w_i q_i=\epsilon$, which yields a bound for the lowest
$n_i$, $i\in \cI_b$, since $w_i>0$.
The lowest possible value for $\epsilon>0$ can be readily calculated from
the denominators occurring in (\ref{qform}).
If transversality condition (2.) applies we have $|\cI_a^-|$ equations
of the form $\sum_{i\in \cI_a^-} m^{(j)}_i q_i=1-q_{e_j}$ which can be
rewritten as $\sum_{i\in \cI_b} w_i^{(j)} q_i=\epsilon^{(j)}$.
Only if all $w_i^{(j)}$ happen to be indefinite and all
$\epsilon^{(j)}$ are zero we get {\sl no bound} from this conditions.
 Assuming this to be true we have
\beq
\sum_{i\in \cI_b} m^{(j)} s_i q_i-s_{e_j} q_{b(e_j)}=0,
\lleq{g1}
where $s_{e_j}:=1$ and $b(e_j):=b$ if $e_j\in \cI_b$. Note that
$\sum_i m^{(j)}_i\ge 2$  in order to avoid quadratic mass terms.
Now we can rewrite (\ref{cbed2}) in the form
\beq
\sum s_i q_{b(i)}=0.
\lleq{g2}
If one uses now (\ref{g1}) and $\sum_i m^{(j)}_i\ge 2$
in order to eliminate the negative $s_i$ one finds
 $\sum_i w_i q_i\le 0$ with $w_i>0$ which is in contradiction with
the positivity of the charges, hence we get a bound in any case.

This procedure of restricting the bound for $n_{p}$ given
$n_i,\dots,n_{p-1}$
was implemented in a computer program. It allows to find all
configurations without testing unnecessary many combinations of
the $n_i$.
The actual upper bounds for the  $(n_i,\dots,n_r)$
in the four variable case are $(7,17,83,1805)$ and we have found $2390$
configurations which allows for transversal polynomials.
In the five variable case the bounds are $(4,6,14,62,923)$ and
$5165$ configurations exist.
By adding a trivial mass term $z_5^2$ in the four variable case these
configurations mentioned so far lead to three dimensional
Calabi-Yau manifolds described by one polynom constraint
in a four dimensional weighted projective space.
We list these $7555$ examples in table 3.

The same figures for the six variable case and seven variable case are
$(3,3,5,11,41,482)$, $2567$ and $(2,2,3,4,8,26,242)$, $669$ respectively.
The total of $3236$ combinations which we have found here are
displayed in table 4.

Likewise for eight and nine variables potentials we have bounds
$(2,2,2,2,2,3,3,5,14)$ with $47$ examples and  $(2,2,2,2,2,2,2,2,2)$
with
$1$ example respectively. Table 5 contains the few models of this type.

\vskip .4truein
\noindent
\section{Results and Comparisons.}

\noindent
We have constructed 10,839 Landau--Ginzburg theories with
2997 different spectra, i.e. pairs of generations and antigenerations.
The massless spectrum is very rough information about a theory and
it is likely that the degeneracy is lifted to a large degree when
additional information, such as the number of singlets and/or the
Yukawa couplings, becomes available. We expect the situation to
very similar to the class of CICYs \cite{cdls} which only leads to
some 250
different spectra \cite{ghl} but for which a detailed analysis of the
Yukawa couplings \cite{ch} shows that it contains several thousand
distinct theories.

It is clear however that there is in fact some redundancy in this
class of Landau--Ginzburg theories even beyond the one discussed in
the previous sections. In the list there appear e.g. two theories with
spectrum $(h^{(1,1)},h^{(2,1)},\chi)=(2,122,-240)$ involving five variables
\beq
\IP_{(2,2,2,1,7)}[14] \ni \{z_1^7+z_2^7+z_3^7+z_4^{14}+z_5^2=0\}
\eeq
and
\beq
\IP_{(1,1,1,1,3)}[7] \ni \{y_1^7+y_2^7+y_3^7+y_4^7+y_4y_5^2=0\}.
\eeq
Using the fractional transformations  introduced in ref. \cite{ls4}
it is
eeasy to show that these two models are equivalent even though
this is not obvious by just looking at the potentials. To see this
consider first the orbifold
\beq
\IP_{(2,2,2,1,7)}[14] {\large /} \ZZ_2: [~0~0~0~1~1]
\eeq
of the first model where the action indicated by
$\ZZ_2:~[~0~0~0~1~1] $
means that the first three coordinates remain invariant whereas the
latter
two variables are to be multiplied with the generator of the cylic
$\ZZ_2$, $\a=-1$.
Since the action of this $\ZZ_2$ on the weighted projective space
is part
of the  projective equivalence the orbifold is of course isomorphic
to the
original model.  On the other hand the fractional transformation \beq
z_i=y_i,~i=1,2,3;~~~ z_4=y_4^{1/2},~~~ z_5=y_4^{1/2}y_5
\eeq
associated with this symmetry \cite{ls4} define a 1--1 coordinate
transformation on the orbifold which transforms the first theory into
the second which  are therefore equivalent as well.

Similarly and can show the equivalences
\bea
\IP_{(2,2,2,3,9)}[18]_{-216}^{(4,112)}&=&\IP_{(1,1,1,3,3)}[9] \nonumber
\\
\IP_{(2,6,6,7,21)}[42]_{-96}^{(17,65)}&=&\IP_{(1,3,3,7,7)}[21]\nonumber
\\
\IP_{(2,5,14,14,35)}[70]_{-64}^{(27,59)} &=&\IP_{(1,5,7,7,15)}[35]
\eea
as well as a number of others.

It should be noted that even though we now have constructed
Landau--Ginzburg potentials with an arbitrary number of scaling fields
the basic range of the spectrum has not changed as compared with the
results of \cite{cls} where it was found that the spectra of all
6000 odd
theories constructed there lead to Euler numbers which fall into the
range
\fnote{6}{This should be compared with the result for the complete
   intersection Calabi--Yau manifolds where
\break $-200 \leq \chi \leq 0$ [12].}
\beq
-960  \leq \chi \leq  960.
\eeq
In fact not only
do all the LG spectra fall into this range, all known Calabi--Yau
spectra
and all the spectra from exactly solvable tensor models are contained
in this range as well! This suggests that perhaps the spectra of all
string vacua based on $c=9$
will be found within this range. Put differently we conjecture that
the Euler numbers of all Calabi--Yau manifolds are contained in the
range $-960 \leq \chi \leq 960$.

Similar to the  results in \cite{cls} the  Hodge pairs do not pair  up
completely.
In fact the mirror symmetry of the space of Landau--Ginzburg vacua is just
about $77\%$. As an illustration of this fact we have listed in table 1
the Euler numbers  that occur in the list and in Figs. 1  and 2  we have
plotted all the spectra and those without mirror partners respectively.
It thus  appears that orbifolding is an essential ingredient
in the construction of a mirror symmetric slice of the configuration space
of the Heterotic String. It is in fact easy to produce examples of
orbifolds whose spectrum does not appear in our list of LG vacua.
An example of a mirror pair is furnished by the orbifolds
\beq
\IP_{(1,1,1,1,2)}[6]^{(1,103)}_{-204} {\LARGE /} \ZZ_6:
\left[\matrix{3&2&1&0&0\cr}\right]
\eeq
which has the spectrum $(11,23,-24)$ and
\beq
\IP_{(1,1,1,1,2)}[6]^{(1,103)}_{-204} {\LARGE /} \ZZ_6\times \ZZ_3 :
\left[\matrix{3&2&1&0&0\cr
              0&0&1&0&2\cr}\right]
\eeq
which has the mirror flipped spectrum $(23,11,24)$.
The space of LG orbifolds that has been constructed sofar \cite{kss}
is indeed much closer to being mirror symmetric than the space of
LG theories itself. Even though the construction in \cite{kss} is
incomplete about 94\% of the Hodge numbers pair up.

By now there exist many different prescriptions to construct string
vacua and we would like to put the LG framework into the context of
other left--right symmetric constructions.
Among the more prominent ones other than Calabi--Yau manifolds which
are obviously closely related to LG theories are constructions which
have traditionally been called , somewhat misleading,
orbifolds (what is meant are
orbifolds of tori) \cite{dhvw}, free fermion constructions
\cite{freeferm}, lattice  constructions \cite{lls} and interacting
exactly solvable models \cite{g}\cite{ks}.

None of these classes are known completely even though much effort
has gone into the exploration of some of them. Because powerful
computational tools are available toroidal orbifolds have been analyzed
in some detail \cite{orbis}
and much attention has
focused on the explicit construction  of exactly solvable theories in
the
context of tensor models via  $N=2$ superconformal minimal theories
\cite{ls123} and Kazama--Suzuki models
\cite{fiq2}\cite{lvwg}\cite{ls5}\cite{s}\cite{fkss}\cite{fksv}
\cite{aafn} as well
as their in depth analysis \cite{lvwg}\cite{rs1a}\cite{fks}\cite{exyuk}.
In the context of
(2,2)--vacua orbifolds lead only to some tens distinct models  whereas
the known
classes of exactly solvable  theories lead only to a few hundred models
with distinct spectra. Similar results have been obtained sofar via the
covariant lattice approach \cite{lsw} and hence it is obvious that these
constructions do
not exhaust  the configuration space of Heterotic String by far. The
class of Landau--Ginzburg string vacua thus appears as a rather
extensive source of (2,2)--symmetric models.

Finally we should remark that the  (2,2) Landau--Ginzburg theories
we have constructed here can be used to build a probably much
larger class
of (2,0) models along the lines described in \cite{dg}, using an
appropriate adaption the work of \cite{cdgp}\cite{dave}
\cite{klth}\cite{anamaria} in order to determine the instantons
on which the existence of a certain split  of a vector bundle
has to be checked.


\def\plot#1#2{\vskip\parskip
                  \vbox{\hrule width\hsize
                        \hbox{\kern-0.2pt\vrule height#1
                              \vbox{\hfill}\kern-0.6pt
                              \vrule}\hrule width\hsize}
    \setbox0=\hbox{#2} \dimen0=\wd0 \divide\dimen0 by 2
    \setbox0=\hbox{\kern-\dimen0 #2}
    \dimen3=#1}

\def\hmark{\kern-0.2pt\lower10pt\hbox{\vrule height 5pt}}
\def\leftscalemark{\vbox{\hrule width5pt}}
\def\rightscalemark{\kern-5pt\vbox{\hrule width5pt}}

\def\Place#1#2#3{
    \count10=#1 \advance\count10 by 960
    \dimen1=\hsize \divide\dimen1 by 1920 \multiply\dimen1 by \count10
    \dimen2=\dimen3 \divide\dimen2 by 550 \multiply\dimen2 by #2
    \vbox to 0pt{\kern-\parskip\kern-18truept\kern-\dimen2
    \hbox{\kern\dimen1#3}\vss}\nointerlineskip}

\def\datum#1#2{\Place{#1}{#2}{\copy0}}

\begin{center}

\plot{8truein}{\tiny{$\bullet$}}
\nobreak
\Place{-960}{50}{\leftscalemark~~50}
\Place{-960}{100}{\leftscalemark~~100}
\Place{-960}{150}{\leftscalemark~~150}
\Place{-960}{200}{\leftscalemark~~200}
\Place{-960}{250}{\leftscalemark~~250}
\Place{-960}{300}{\leftscalemark~~300}
\Place{-960}{350}{\leftscalemark~~350}
\Place{-960}{400}{\leftscalemark~~400}
\Place{-960}{450}{\leftscalemark~~450}
\Place{-960}{500}{\leftscalemark~~500}
\Place{960}{50}{\rightscalemark\vphantom{0}}
\Place{960}{100}{\rightscalemark\vphantom{0}}
\Place{960}{150}{\rightscalemark\vphantom{0}}
\Place{960}{200}{\rightscalemark\vphantom{0}}
\Place{960}{250}{\rightscalemark\vphantom{0}}
\Place{960}{300}{\rightscalemark\vphantom{0}}
\Place{960}{350}{\rightscalemark\vphantom{0}}
\Place{960}{400}{\rightscalemark\vphantom{0}}
\Place{960}{450}{\rightscalemark\vphantom{0}}
\Place{960}{500}{\rightscalemark\vphantom{0}}
\Place{-960}{0}{\hmark\lower18pt\hbox{-960}}
\Place{-720}{0}{\hmark\lower18pt\hbox{-720}}
\Place{-480}{0}{\hmark\lower18pt\hbox{-480}}
\Place{-240}{0}{\hmark\lower18pt\hbox{-240}}
\Place{0}{0}{\hmark\lower18pt\hbox{0}}
\Place{240}{0}{\hmark\lower18pt\hbox{240}}
\Place{480}{0}{\hmark\lower18pt\hbox{480}}
\Place{720}{0}{\hmark\lower18pt\hbox{720}}
\Place{960}{0}{\hmark\lower18pt\hbox{960}}
\Place{-720}{550}{\hmark}
\Place{-480}{550}{\hmark}
\Place{-240}{550}{\hmark}
\Place{0}{550}{\hmark}
\Place{240}{550}{\hmark}
\Place{480}{550}{\hmark}
\Place{720}{550}{\hmark}
\Place{960}{550}{\hmark}
\nobreak
\begin{center}
\datum{           0}{          20}
\datum{           0}{          22}
\datum{           0}{          26}
\datum{           0}{          28}
\datum{           0}{          30}
\datum{           0}{          32}
\datum{           0}{          34}
\datum{           0}{          36}
\datum{           0}{          38}
\datum{           0}{          40}
\datum{           0}{          42}
\datum{           0}{          44}
\datum{           0}{          46}
\datum{           0}{          48}
\datum{           0}{          50}
\datum{           0}{          52}
\datum{           0}{          54}
\datum{           0}{          56}
\datum{           0}{          58}
\datum{           0}{          60}
\datum{           0}{          62}
\datum{           0}{          64}
\datum{           0}{          66}
\datum{           0}{          68}
\datum{           0}{          70}
\datum{           0}{          74}
\datum{           0}{          76}
\datum{           0}{          78}
\datum{           0}{          82}
\datum{           0}{          86}
\datum{           0}{          88}
\datum{           0}{          90}
\datum{           0}{          94}
\datum{           0}{          98}
\datum{           0}{         104}
\datum{           0}{         106}
\datum{           0}{         110}
\datum{           0}{         112}
\datum{           0}{         114}
\datum{           0}{         118}
\datum{           0}{         122}
\datum{           0}{         124}
\datum{           0}{         126}
\datum{           0}{         130}
\datum{           0}{         134}
\datum{           0}{         138}
\datum{           0}{         142}
\datum{           0}{         150}
\datum{           0}{         154}
\datum{           0}{         156}
\datum{           0}{         158}
\datum{           0}{         162}
\datum{           0}{         166}
\datum{           0}{         170}
\datum{           0}{         174}
\datum{           0}{         178}
\datum{           0}{         182}
\datum{           0}{         190}
\datum{           0}{         194}
\datum{           0}{         206}
\datum{           0}{         214}
\datum{           0}{         222}
\datum{           0}{         238}
\datum{           0}{         242}
\datum{           0}{         246}
\datum{           0}{         262}
\datum{           0}{         286}
\datum{           0}{         298}
\datum{           0}{         302}
\datum{           0}{         358}
\datum{           0}{         446}
\datum{           0}{         502}
\datum{           2}{          25}
\datum{           4}{          44}
\datum{           4}{          50}
\datum{           4}{          54}
\datum{           4}{          66}
\datum{           4}{          68}
\datum{           4}{          82}
\datum{           4}{          92}
\datum{           4}{         102}
\datum{           6}{          29}
\datum{           6}{          33}
\datum{           6}{          35}
\datum{           6}{          37}
\datum{           6}{          43}
\datum{           6}{          55}
\datum{           6}{          57}
\datum{           6}{          61}
\datum{           6}{          65}
\datum{           6}{          71}
\datum{           6}{          77}
\datum{           6}{          87}
\datum{           6}{          97}
\datum{           6}{         137}
\datum{           6}{         151}
\datum{           8}{          34}
\datum{           8}{          36}
\datum{           8}{          40}
\datum{           8}{          42}
\datum{           8}{          44}
\datum{           8}{          46}
\datum{           8}{          50}
\datum{           8}{          52}
\datum{           8}{          54}
\datum{           8}{          56}
\datum{           8}{          60}
\datum{           8}{          62}
\datum{           8}{          66}
\datum{           8}{          68}
\datum{           8}{          70}
\datum{           8}{          78}
\datum{           8}{          82}
\datum{           8}{          84}
\datum{           8}{          90}
\datum{           8}{         102}
\datum{           8}{         116}
\datum{          -4}{          38}
\datum{          -4}{          54}
\datum{          -4}{          64}
\datum{          -4}{          66}
\datum{          -4}{          82}
\datum{          -4}{          92}
\datum{          -4}{          96}
\datum{          -4}{         110}
\datum{          -4}{         156}
\datum{          -6}{          29}
\datum{          -6}{          37}
\datum{          -6}{          43}
\datum{          -6}{          45}
\datum{          -6}{          49}
\datum{          -6}{          55}
\datum{          -6}{          61}
\datum{          -6}{          67}
\datum{          -6}{          73}
\datum{          -6}{          77}
\datum{          -6}{          83}
\datum{          -6}{          97}
\datum{          -6}{          99}
\datum{          -6}{         117}
\datum{          -8}{          30}
\datum{          -8}{          34}
\datum{          -8}{          36}
\datum{          -8}{          38}
\datum{          -8}{          44}
\datum{          -8}{          50}
\datum{          -8}{          52}
\datum{          -8}{          54}
\datum{          -8}{          56}
\datum{          -8}{          58}
\datum{          -8}{          60}
\datum{          -8}{          62}
\datum{          -8}{          66}
\datum{          -8}{          70}
\datum{          -8}{          84}
\datum{          -8}{          90}
\datum{          -8}{          96}
\datum{          -8}{          98}
\datum{          -8}{         102}
\datum{          -8}{         116}
\datum{          10}{          53}
\datum{          10}{          83}
\datum{          10}{         141}
\datum{          12}{          22}
\datum{          12}{          26}
\datum{          12}{          30}
\datum{          12}{          32}
\datum{          12}{          34}
\datum{          12}{          36}
\datum{          12}{          38}
\datum{          12}{          42}
\datum{          12}{          44}
\datum{          12}{          48}
\datum{          12}{          50}
\datum{          12}{          52}
\datum{          12}{          54}
\datum{          12}{          56}
\datum{          12}{          58}
\datum{          12}{          64}
\datum{          12}{          66}
\datum{          12}{          70}
\datum{          12}{          72}
\datum{          12}{          76}
\datum{          12}{          80}
\datum{          12}{          82}
\datum{          12}{          86}
\datum{          12}{          88}
\datum{          12}{          94}
\datum{          12}{          96}
\datum{          12}{          98}
\datum{          12}{         100}
\datum{          12}{         108}
\datum{          12}{         116}
\datum{          12}{         122}
\datum{          12}{         128}
\datum{          12}{         130}
\datum{          12}{         146}
\datum{          12}{         156}
\datum{          14}{          45}
\datum{          14}{          61}
\datum{          16}{          26}
\datum{          16}{          34}
\datum{          16}{          36}
\datum{          16}{          38}
\datum{          16}{          42}
\datum{          16}{          44}
\datum{          16}{          46}
\datum{          16}{          48}
\datum{          16}{          50}
\datum{          16}{          52}
\datum{          16}{          54}
\datum{          16}{          56}
\datum{          16}{          58}
\datum{          16}{          62}
\datum{          16}{          64}
\datum{          16}{          66}
\datum{          16}{          74}
\datum{          16}{          76}
\datum{          16}{          78}
\datum{          16}{          82}
\datum{          16}{          84}
\datum{          16}{          86}
\datum{          16}{          92}
\datum{          16}{          94}
\datum{          16}{         122}
\datum{          18}{          45}
\datum{          18}{          47}
\datum{          18}{          49}
\datum{          18}{          55}
\datum{          18}{          61}
\datum{          18}{          63}
\datum{          18}{          67}
\datum{          18}{          69}
\datum{          18}{          79}
\datum{          18}{          81}
\datum{          18}{          85}
\datum{          18}{          89}
\datum{          18}{          91}
\datum{          18}{          97}
\datum{          18}{         103}
\datum{          18}{         105}
\datum{          20}{          46}
\datum{          20}{          48}
\datum{          20}{          58}
\datum{          20}{          84}
\datum{          20}{          90}
\datum{          20}{         108}
\datum{          20}{         114}
\datum{          20}{         198}
\datum{          20}{         234}
\datum{          24}{          26}
\datum{          24}{          28}
\datum{          24}{          30}
\datum{          24}{          32}
\datum{          24}{          34}
\datum{          24}{          36}
\datum{          24}{          38}
\datum{          24}{          40}
\datum{          24}{          42}
\datum{          24}{          44}
\datum{          24}{          46}
\datum{          24}{          48}
\datum{          24}{          50}
\datum{          24}{          52}
\datum{          24}{          54}
\datum{          24}{          56}
\datum{          24}{          58}
\datum{          24}{          60}
\datum{          24}{          62}
\datum{          24}{          64}
\datum{          24}{          66}
\datum{          24}{          68}
\datum{          24}{          70}
\datum{          24}{          72}
\datum{          24}{          74}
\datum{          24}{          76}
\datum{          24}{          80}
\datum{          24}{          82}
\datum{          24}{          84}
\datum{          24}{          86}
\datum{          24}{          88}
\datum{          24}{          90}
\datum{          24}{          92}
\datum{          24}{          94}
\datum{          24}{          98}
\datum{          24}{         100}
\datum{          24}{         102}
\datum{          24}{         104}
\datum{          24}{         112}
\datum{          24}{         114}
\datum{          24}{         116}
\datum{          24}{         120}
\datum{          24}{         128}
\datum{          24}{         130}
\datum{          24}{         132}
\datum{          24}{         134}
\datum{          24}{         148}
\datum{          24}{         160}
\datum{          24}{         164}
\datum{          24}{         166}
\datum{          24}{         232}
\datum{          28}{          38}
\datum{          28}{          54}
\datum{          28}{          58}
\datum{          28}{          62}
\datum{          28}{          98}
\datum{          28}{         146}
\datum{          30}{          35}
\datum{          30}{          43}
\datum{          30}{          49}
\datum{          30}{          51}
\datum{          30}{          53}
\datum{          30}{          63}
\datum{          30}{          73}
\datum{          30}{          77}
\datum{          30}{          79}
\datum{          30}{          83}
\datum{          30}{         103}
\datum{          30}{         163}
\datum{          32}{          30}
\datum{          32}{          38}
\datum{          32}{          42}
\datum{          32}{          46}
\datum{          32}{          50}
\datum{          32}{          52}
\datum{          32}{          54}
\datum{          32}{          60}
\datum{          32}{          62}
\datum{          32}{          66}
\datum{          32}{          70}
\datum{          32}{          74}
\datum{          32}{          76}
\datum{          32}{          78}
\datum{          32}{          82}
\datum{          32}{          84}
\datum{          32}{          94}
\datum{          32}{         104}
\datum{          32}{         110}
\datum{          32}{         114}
\datum{          32}{         118}
\datum{          32}{         190}
\datum{          34}{          45}
\datum{          34}{          69}
\datum{          34}{          83}
\datum{          36}{          30}
\datum{          36}{          34}
\datum{          36}{          36}
\datum{          36}{          38}
\datum{          36}{          40}
\datum{          36}{          42}
\datum{          36}{          46}
\datum{          36}{          50}
\datum{          36}{          52}
\datum{          36}{          54}
\datum{          36}{          56}
\datum{          36}{          58}
\datum{          36}{          60}
\datum{          36}{          62}
\datum{          36}{          66}
\datum{          36}{          68}
\datum{          36}{          70}
\datum{          36}{          74}
\datum{          36}{          76}
\datum{          36}{          78}
\datum{          36}{          82}
\datum{          36}{          84}
\datum{          36}{          86}
\datum{          36}{          88}
\datum{          36}{          94}
\datum{          36}{         100}
\datum{          36}{         102}
\datum{          36}{         106}
\datum{          36}{         108}
\datum{          36}{         110}
\datum{          36}{         112}
\datum{          36}{         114}
\datum{          36}{         118}
\datum{          36}{         126}
\datum{          36}{         130}
\datum{          36}{         134}
\datum{          36}{         142}
\datum{          36}{         144}
\datum{          36}{         156}
\datum{          36}{         162}
\datum{          36}{         172}
\datum{          36}{         184}
\datum{          36}{         202}
\datum{          36}{         212}
\datum{          36}{         214}
\datum{          36}{         222}
\datum{          36}{         314}
\datum{          40}{          30}
\datum{          40}{          38}
\datum{          40}{          44}
\datum{          40}{          46}
\datum{          40}{          48}
\datum{          40}{          50}
\datum{          40}{          52}
\datum{          40}{          54}
\datum{          40}{          56}
\datum{          40}{          58}
\datum{          40}{          62}
\datum{          40}{          68}
\datum{          40}{          70}
\datum{          40}{          72}
\datum{          40}{          78}
\datum{          40}{          80}
\datum{          40}{          86}
\datum{          40}{          90}
\datum{          40}{          94}
\datum{          40}{         110}
\datum{          40}{         116}
\datum{          40}{         118}
\datum{          40}{         120}
\datum{          40}{         130}
\datum{          40}{         148}
\datum{          42}{          43}
\datum{          42}{          49}
\datum{          42}{          53}
\datum{          42}{          55}
\datum{          42}{          61}
\datum{          42}{          79}
\datum{          42}{          81}
\datum{          42}{          89}
\datum{          42}{          91}
\datum{          42}{          93}
\datum{          42}{         115}
\datum{          44}{          44}
\datum{          44}{          52}
\datum{          44}{          56}
\datum{          44}{          58}
\datum{          44}{          64}
\datum{          44}{          66}
\datum{          44}{          78}
\datum{          44}{          80}
\datum{          44}{          94}
\datum{          44}{         108}
\datum{          48}{          30}
\datum{          48}{          34}
\datum{          48}{          38}
\datum{          48}{          40}
\datum{          48}{          42}
\datum{          48}{          44}
\datum{          48}{          46}
\datum{          48}{          48}
\datum{          48}{          50}
\datum{          48}{          52}
\datum{          48}{          54}
\datum{          48}{          56}
\datum{          48}{          58}
\datum{          48}{          60}
\datum{          48}{          62}
\datum{          48}{          64}
\datum{          48}{          66}
\datum{          48}{          68}
\datum{          48}{          70}
\datum{          48}{          72}
\datum{          48}{          74}
\datum{          48}{          76}
\datum{          48}{          78}
\datum{          48}{          80}
\datum{          48}{          82}
\datum{          48}{          84}
\datum{          48}{          86}
\datum{          48}{          88}
\datum{          48}{          92}
\datum{          48}{          94}
\datum{          48}{          96}
\datum{          48}{          98}
\datum{          48}{         100}
\datum{          48}{         102}
\datum{          48}{         106}
\datum{          48}{         108}
\datum{          48}{         110}
\datum{          48}{         112}
\datum{          48}{         118}
\datum{          48}{         122}
\datum{          48}{         124}
\datum{          48}{         134}
\datum{          48}{         138}
\datum{          48}{         142}
\datum{          48}{         158}
\datum{          48}{         166}
\datum{          48}{         178}
\datum{          48}{         202}
\datum{          48}{         230}
\datum{          48}{         266}
\datum{          50}{          45}
\datum{          50}{          63}
\datum{          50}{          93}
\datum{          52}{          64}
\datum{          52}{         108}
\datum{          52}{         116}
\datum{          54}{          41}
\datum{          54}{          43}
\datum{          54}{          47}
\datum{          54}{          49}
\datum{          54}{          57}
\datum{          54}{          59}
\datum{          54}{          61}
\datum{          54}{          73}
\datum{          54}{          77}
\datum{          54}{          79}
\datum{          54}{          83}
\datum{          54}{          85}
\datum{          54}{         103}
\datum{          54}{         117}
\datum{          54}{         133}
\datum{          54}{         157}
\datum{          56}{          38}
\datum{          56}{          52}
\datum{          56}{          54}
\datum{          56}{          58}
\datum{          56}{          62}
\datum{          56}{          66}
\datum{          56}{          68}
\datum{          56}{          70}
\datum{          56}{          72}
\datum{          56}{          74}
\datum{          56}{          84}
\datum{          56}{          86}
\datum{          56}{          88}
\datum{          56}{          90}
\datum{          56}{         108}
\datum{          56}{         114}
\datum{          56}{         124}
\datum{          56}{         126}
\datum{          56}{         150}
\datum{          56}{         166}
\datum{          56}{         180}
\datum{          56}{         214}
\datum{          60}{          36}
\datum{          60}{          40}
\datum{          60}{          42}
\datum{          60}{          44}
\datum{          60}{          46}
\datum{          60}{          48}
\datum{          60}{          52}
\datum{          60}{          54}
\datum{          60}{          58}
\datum{          60}{          60}
\datum{          60}{          62}
\datum{          60}{          64}
\datum{          60}{          66}
\datum{          60}{          68}
\datum{          60}{          70}
\datum{          60}{          74}
\datum{          60}{          78}
\datum{          60}{          82}
\datum{          60}{          84}
\datum{          60}{          88}
\datum{          60}{          96}
\datum{          60}{          98}
\datum{          60}{         110}
\datum{          60}{         112}
\datum{          60}{         120}
\datum{          60}{         122}
\datum{          60}{         132}
\datum{          60}{         144}
\datum{          60}{         166}
\datum{          60}{         178}
\datum{          60}{         196}
\datum{          60}{         218}
\datum{          60}{         222}
\datum{          60}{         234}
\datum{          60}{         358}
\datum{          60}{         474}
\datum{          64}{          50}
\datum{          64}{          54}
\datum{          64}{          60}
\datum{          64}{          62}
\datum{          64}{          66}
\datum{          64}{          70}
\datum{          64}{          76}
\datum{          64}{          80}
\datum{          64}{          82}
\datum{          64}{          86}
\datum{          64}{          90}
\datum{          64}{         102}
\datum{          64}{         106}
\datum{          64}{         110}
\datum{          64}{         134}
\datum{          64}{         154}
\datum{          64}{         166}
\datum{          66}{          61}
\datum{          66}{          63}
\datum{          66}{          65}
\datum{          66}{          69}
\datum{          66}{          73}
\datum{          66}{          75}
\datum{          66}{          79}
\datum{          66}{          87}
\datum{          66}{          97}
\datum{          66}{         129}
\datum{          66}{         137}
\datum{          66}{         159}
\datum{          68}{          64}
\datum{          68}{          78}
\datum{          68}{         110}
\datum{          68}{         150}
\datum{          70}{          75}
\datum{          70}{          89}
\datum{          70}{          95}
\datum{          70}{         101}
\datum{          70}{         135}
\datum{          72}{          40}
\datum{          72}{          44}
\datum{          72}{          46}
\datum{          72}{          48}
\datum{          72}{          50}
\datum{          72}{          52}
\datum{          72}{          54}
\datum{          72}{          56}
\datum{          72}{          58}
\datum{          72}{          60}
\datum{          72}{          62}
\datum{          72}{          64}
\datum{          72}{          66}
\datum{          72}{          68}
\datum{          72}{          70}
\datum{          72}{          74}
\datum{          72}{          76}
\datum{          72}{          78}
\datum{          72}{          80}
\datum{          72}{          82}
\datum{          72}{          84}
\datum{          72}{          86}
\datum{          72}{          88}
\datum{          72}{          90}
\datum{          72}{          92}
\datum{          72}{          94}
\datum{          72}{          96}
\datum{          72}{          98}
\datum{          72}{         100}
\datum{          72}{         102}
\datum{          72}{         104}
\datum{          72}{         112}
\datum{          72}{         116}
\datum{          72}{         118}
\datum{          72}{         120}
\datum{          72}{         124}
\datum{          72}{         126}
\datum{          72}{         128}
\datum{          72}{         132}
\datum{          72}{         136}
\datum{          72}{         140}
\datum{          72}{         142}
\datum{          72}{         144}
\datum{          72}{         148}
\datum{          72}{         150}
\datum{          72}{         158}
\datum{          72}{         160}
\datum{          72}{         164}
\datum{          72}{         182}
\datum{          72}{         198}
\datum{          72}{         212}
\datum{          74}{         121}
\datum{          76}{          68}
\datum{          76}{          94}
\datum{          76}{         100}
\datum{          76}{         124}
\datum{          78}{          61}
\datum{          78}{          79}
\datum{          78}{          83}
\datum{          78}{          89}
\datum{          78}{          91}
\datum{          78}{          93}
\datum{          78}{          95}
\datum{          78}{         109}
\datum{          78}{         111}
\datum{          78}{         153}
\datum{          78}{         163}
\datum{          80}{          54}
\datum{          80}{          58}
\datum{          80}{          60}
\datum{          80}{          62}
\datum{          80}{          66}
\datum{          80}{          68}
\datum{          80}{          72}
\datum{          80}{          74}
\datum{          80}{          76}
\datum{          80}{          78}
\datum{          80}{          94}
\datum{          80}{          96}
\datum{          80}{          98}
\datum{          80}{         100}
\datum{          80}{         102}
\datum{          80}{         106}
\datum{          80}{         118}
\datum{          84}{          50}
\datum{          84}{          52}
\datum{          84}{          54}
\datum{          84}{          58}
\datum{          84}{          64}
\datum{          84}{          66}
\datum{          84}{          68}
\datum{          84}{          70}
\datum{          84}{          72}
\datum{          84}{          74}
\datum{          84}{          78}
\datum{          84}{          80}
\datum{          84}{          82}
\datum{          84}{          84}
\datum{          84}{          86}
\datum{          84}{          88}
\datum{          84}{          90}
\datum{          84}{          94}
\datum{          84}{          96}
\datum{          84}{          98}
\datum{          84}{         102}
\datum{          84}{         106}
\datum{          84}{         124}
\datum{          84}{         130}
\datum{          84}{         134}
\datum{          84}{         138}
\datum{          84}{         154}
\datum{          84}{         162}
\datum{          84}{         164}
\datum{          84}{         166}
\datum{          84}{         172}
\datum{          84}{         174}
\datum{          84}{         178}
\datum{          84}{         190}
\datum{          84}{         194}
\datum{          84}{         262}
\datum{          84}{         322}
\datum{          86}{          57}
\datum{          86}{          87}
\datum{          86}{         127}
\datum{          88}{          54}
\datum{          88}{          60}
\datum{          88}{          62}
\datum{          88}{          68}
\datum{          88}{          70}
\datum{          88}{          82}
\datum{          88}{          84}
\datum{          88}{         102}
\datum{          88}{         104}
\datum{          88}{         128}
\datum{          88}{         158}
\datum{          90}{          53}
\datum{          90}{          63}
\datum{          90}{          65}
\datum{          90}{          67}
\datum{          90}{          71}
\datum{          90}{          73}
\datum{          90}{          79}
\datum{          90}{          83}
\datum{          90}{          93}
\datum{          90}{          95}
\datum{          90}{          97}
\datum{          90}{         105}
\datum{          90}{         131}
\datum{          90}{         133}
\datum{          90}{         171}
\datum{          92}{          94}
\datum{          92}{         150}
\datum{          96}{          50}
\datum{          96}{          54}
\datum{          96}{          58}
\datum{          96}{          60}
\datum{          96}{          62}
\datum{          96}{          64}
\datum{          96}{          66}
\datum{          96}{          70}
\datum{          96}{          74}
\datum{          96}{          76}
\datum{          96}{          78}
\datum{          96}{          80}
\datum{          96}{          82}
\datum{          96}{          84}
\datum{          96}{          86}
\datum{          96}{          88}
\datum{          96}{          90}
\datum{          96}{          92}
\datum{          96}{          94}
\datum{          96}{          98}
\datum{          96}{         102}
\datum{          96}{         106}
\datum{          96}{         108}
\datum{          96}{         110}
\datum{          96}{         112}
\datum{          96}{         114}
\datum{          96}{         118}
\datum{          96}{         120}
\datum{          96}{         122}
\datum{          96}{         126}
\datum{          96}{         134}
\datum{          96}{         138}
\datum{          96}{         142}
\datum{          96}{         146}
\datum{          96}{         154}
\datum{          96}{         158}
\datum{          96}{         166}
\datum{          96}{         186}
\datum{          96}{         190}
\datum{          96}{         204}
\datum{          96}{         212}
\datum{          96}{         250}
\datum{          96}{         286}
\datum{          96}{         342}
\datum{          96}{         402}
\datum{         -10}{          43}
\datum{         -10}{          53}
\datum{         -10}{          63}
\datum{         -10}{          73}
\datum{         -12}{          28}
\datum{         -12}{          32}
\datum{         -12}{          34}
\datum{         -12}{          36}
\datum{         -12}{          38}
\datum{         -12}{          40}
\datum{         -12}{          42}
\datum{         -12}{          44}
\datum{         -12}{          46}
\datum{         -12}{          48}
\datum{         -12}{          50}
\datum{         -12}{          52}
\datum{         -12}{          54}
\datum{         -12}{          56}
\datum{         -12}{          58}
\datum{         -12}{          64}
\datum{         -12}{          66}
\datum{         -12}{          70}
\datum{         -12}{          72}
\datum{         -12}{          76}
\datum{         -12}{          78}
\datum{         -12}{          80}
\datum{         -12}{          82}
\datum{         -12}{          86}
\datum{         -12}{          90}
\datum{         -12}{          94}
\datum{         -12}{         100}
\datum{         -12}{         104}
\datum{         -12}{         108}
\datum{         -12}{         116}
\datum{         -12}{         128}
\datum{         -12}{         146}
\datum{         -12}{         156}
\datum{         -14}{          47}
\datum{         -14}{         137}
\datum{         -16}{          28}
\datum{         -16}{          30}
\datum{         -16}{          36}
\datum{         -16}{          38}
\datum{         -16}{          42}
\datum{         -16}{          44}
\datum{         -16}{          46}
\datum{         -16}{          48}
\datum{         -16}{          52}
\datum{         -16}{          54}
\datum{         -16}{          56}
\datum{         -16}{          58}
\datum{         -16}{          60}
\datum{         -16}{          62}
\datum{         -16}{          64}
\datum{         -16}{          66}
\datum{         -16}{          70}
\datum{         -16}{          72}
\datum{         -16}{          74}
\datum{         -16}{          76}
\datum{         -16}{          78}
\datum{         -16}{          82}
\datum{         -16}{          94}
\datum{         -16}{         150}
\datum{         -18}{          25}
\datum{         -18}{          29}
\datum{         -18}{          37}
\datum{         -18}{          39}
\datum{         -18}{          47}
\datum{         -18}{          49}
\datum{         -18}{          55}
\datum{         -18}{          61}
\datum{         -18}{          63}
\datum{         -18}{          65}
\datum{         -18}{          79}
\datum{         -18}{          83}
\datum{         -18}{          85}
\datum{         -18}{          97}
\datum{         -18}{         109}
\datum{         -18}{         115}
\datum{         -18}{         149}
\datum{         -20}{          26}
\datum{         -20}{          30}
\datum{         -20}{          40}
\datum{         -20}{          44}
\datum{         -20}{          48}
\datum{         -20}{          56}
\datum{         -20}{          58}
\datum{         -20}{          64}
\datum{         -20}{          68}
\datum{         -20}{          74}
\datum{         -20}{         108}
\datum{         -20}{         114}
\datum{         -20}{         116}
\datum{         -20}{         198}
\datum{         -22}{          41}
\datum{         -24}{          26}
\datum{         -24}{          28}
\datum{         -24}{          30}
\datum{         -24}{          32}
\datum{         -24}{          34}
\datum{         -24}{          36}
\datum{         -24}{          38}
\datum{         -24}{          40}
\datum{         -24}{          42}
\datum{         -24}{          44}
\datum{         -24}{          46}
\datum{         -24}{          48}
\datum{         -24}{          50}
\datum{         -24}{          52}
\datum{         -24}{          54}
\datum{         -24}{          56}
\datum{         -24}{          58}
\datum{         -24}{          60}
\datum{         -24}{          62}
\datum{         -24}{          64}
\datum{         -24}{          66}
\datum{         -24}{          68}
\datum{         -24}{          70}
\datum{         -24}{          72}
\datum{         -24}{          74}
\datum{         -24}{          76}
\datum{         -24}{          78}
\datum{         -24}{          80}
\datum{         -24}{          82}
\datum{         -24}{          84}
\datum{         -24}{          86}
\datum{         -24}{          88}
\datum{         -24}{          90}
\datum{         -24}{          92}
\datum{         -24}{          94}
\datum{         -24}{          96}
\datum{         -24}{          98}
\datum{         -24}{         100}
\datum{         -24}{         102}
\datum{         -24}{         104}
\datum{         -24}{         112}
\datum{         -24}{         114}
\datum{         -24}{         116}
\datum{         -24}{         120}
\datum{         -24}{         128}
\datum{         -24}{         130}
\datum{         -24}{         132}
\datum{         -24}{         134}
\datum{         -24}{         142}
\datum{         -24}{         144}
\datum{         -24}{         148}
\datum{         -24}{         160}
\datum{         -24}{         162}
\datum{         -24}{         164}
\datum{         -24}{         166}
\datum{         -24}{         184}
\datum{         -24}{         232}
\datum{         -26}{          63}
\datum{         -26}{          89}
\datum{         -28}{          36}
\datum{         -28}{          48}
\datum{         -28}{          54}
\datum{         -28}{          62}
\datum{         -28}{          72}
\datum{         -28}{         186}
\datum{         -30}{          33}
\datum{         -30}{          43}
\datum{         -30}{          49}
\datum{         -30}{          53}
\datum{         -30}{          61}
\datum{         -30}{          63}
\datum{         -30}{          69}
\datum{         -30}{          71}
\datum{         -30}{          73}
\datum{         -30}{          79}
\datum{         -30}{          83}
\datum{         -30}{          91}
\datum{         -30}{          93}
\datum{         -30}{         133}
\datum{         -30}{         163}
\datum{         -32}{          30}
\datum{         -32}{          32}
\datum{         -32}{          36}
\datum{         -32}{          38}
\datum{         -32}{          42}
\datum{         -32}{          44}
\datum{         -32}{          46}
\datum{         -32}{          48}
\datum{         -32}{          50}
\datum{         -32}{          54}
\datum{         -32}{          56}
\datum{         -32}{          62}
\datum{         -32}{          66}
\datum{         -32}{          70}
\datum{         -32}{          74}
\datum{         -32}{          76}
\datum{         -32}{          78}
\datum{         -32}{          86}
\datum{         -32}{         104}
\datum{         -32}{         118}
\datum{         -32}{         190}
\datum{         -36}{          28}
\datum{         -36}{          34}
\datum{         -36}{          36}
\datum{         -36}{          38}
\datum{         -36}{          40}
\datum{         -36}{          46}
\datum{         -36}{          48}
\datum{         -36}{          50}
\datum{         -36}{          52}
\datum{         -36}{          54}
\datum{         -36}{          58}
\datum{         -36}{          62}
\datum{         -36}{          66}
\datum{         -36}{          70}
\datum{         -36}{          72}
\datum{         -36}{          74}
\datum{         -36}{          76}
\datum{         -36}{          78}
\datum{         -36}{          82}
\datum{         -36}{          84}
\datum{         -36}{          86}
\datum{         -36}{          88}
\datum{         -36}{          94}
\datum{         -36}{         100}
\datum{         -36}{         102}
\datum{         -36}{         104}
\datum{         -36}{         106}
\datum{         -36}{         108}
\datum{         -36}{         112}
\datum{         -36}{         118}
\datum{         -36}{         120}
\datum{         -36}{         126}
\datum{         -36}{         134}
\datum{         -36}{         142}
\datum{         -36}{         144}
\datum{         -36}{         156}
\datum{         -36}{         162}
\datum{         -36}{         172}
\datum{         -36}{         184}
\datum{         -36}{         202}
\datum{         -36}{         212}
\datum{         -36}{         214}
\datum{         -36}{         222}
\datum{         -36}{         314}
\datum{         -38}{          45}
\datum{         -40}{          34}
\datum{         -40}{          36}
\datum{         -40}{          38}
\datum{         -40}{          44}
\datum{         -40}{          46}
\datum{         -40}{          48}
\datum{         -40}{          50}
\datum{         -40}{          52}
\datum{         -40}{          54}
\datum{         -40}{          58}
\datum{         -40}{          62}
\datum{         -40}{          68}
\datum{         -40}{          70}
\datum{         -40}{          74}
\datum{         -40}{          78}
\datum{         -40}{          86}
\datum{         -40}{          90}
\datum{         -40}{          92}
\datum{         -40}{         110}
\datum{         -40}{         116}
\datum{         -40}{         118}
\datum{         -40}{         138}
\datum{         -40}{         148}
\datum{         -40}{         160}
\datum{         -42}{          35}
\datum{         -42}{          43}
\datum{         -42}{          49}
\datum{         -42}{          55}
\datum{         -42}{          61}
\datum{         -42}{          65}
\datum{         -42}{          67}
\datum{         -42}{          73}
\datum{         -42}{          79}
\datum{         -42}{          83}
\datum{         -42}{          85}
\datum{         -42}{          91}
\datum{         -42}{          97}
\datum{         -42}{         137}
\datum{         -44}{          44}
\datum{         -44}{          78}
\datum{         -44}{          80}
\datum{         -44}{          92}
\datum{         -44}{         134}
\datum{         -46}{         153}
\datum{         -48}{          34}
\datum{         -48}{          38}
\datum{         -48}{          42}
\datum{         -48}{          44}
\datum{         -48}{          46}
\datum{         -48}{          48}
\datum{         -48}{          50}
\datum{         -48}{          52}
\datum{         -48}{          54}
\datum{         -48}{          56}
\datum{         -48}{          58}
\datum{         -48}{          60}
\datum{         -48}{          62}
\datum{         -48}{          64}
\datum{         -48}{          66}
\datum{         -48}{          68}
\datum{         -48}{          70}
\datum{         -48}{          72}
\datum{         -48}{          74}
\datum{         -48}{          76}
\datum{         -48}{          78}
\datum{         -48}{          80}
\datum{         -48}{          82}
\datum{         -48}{          86}
\datum{         -48}{          88}
\datum{         -48}{          90}
\datum{         -48}{          92}
\datum{         -48}{          94}
\datum{         -48}{          96}
\datum{         -48}{          98}
\datum{         -48}{         102}
\datum{         -48}{         106}
\datum{         -48}{         108}
\datum{         -48}{         110}
\datum{         -48}{         112}
\datum{         -48}{         118}
\datum{         -48}{         122}
\datum{         -48}{         124}
\datum{         -48}{         134}
\datum{         -48}{         138}
\datum{         -48}{         142}
\datum{         -48}{         158}
\datum{         -48}{         166}
\datum{         -48}{         168}
\datum{         -48}{         178}
\datum{         -48}{         202}
\datum{         -48}{         230}
\datum{         -48}{         266}
\datum{         -50}{          53}
\datum{         -50}{          73}
\datum{         -52}{          64}
\datum{         -54}{          37}
\datum{         -54}{          43}
\datum{         -54}{          47}
\datum{         -54}{          49}
\datum{         -54}{          61}
\datum{         -54}{          63}
\datum{         -54}{          71}
\datum{         -54}{          77}
\datum{         -54}{          79}
\datum{         -54}{          85}
\datum{         -54}{          87}
\datum{         -54}{         103}
\datum{         -54}{         133}
\datum{         -54}{         147}
\datum{         -56}{          38}
\datum{         -56}{          46}
\datum{         -56}{          50}
\datum{         -56}{          54}
\datum{         -56}{          62}
\datum{         -56}{          66}
\datum{         -56}{          68}
\datum{         -56}{          72}
\datum{         -56}{          74}
\datum{         -56}{          82}
\datum{         -56}{          86}
\datum{         -56}{          88}
\datum{         -56}{          96}
\datum{         -56}{         114}
\datum{         -56}{         116}
\datum{         -56}{         126}
\datum{         -56}{         134}
\datum{         -56}{         150}
\datum{         -56}{         166}
\datum{         -56}{         180}
\datum{         -56}{         214}
\datum{         -58}{          43}
\datum{         -58}{          49}
\datum{         -58}{         129}
\datum{         -60}{          42}
\datum{         -60}{          48}
\datum{         -60}{          52}
\datum{         -60}{          54}
\datum{         -60}{          56}
\datum{         -60}{          58}
\datum{         -60}{          60}
\datum{         -60}{          62}
\datum{         -60}{          64}
\datum{         -60}{          66}
\datum{         -60}{          68}
\datum{         -60}{          70}
\datum{         -60}{          74}
\datum{         -60}{          78}
\datum{         -60}{          82}
\datum{         -60}{          84}
\datum{         -60}{          88}
\datum{         -60}{          90}
\datum{         -60}{          94}
\datum{         -60}{          96}
\datum{         -60}{          98}
\datum{         -60}{         106}
\datum{         -60}{         110}
\datum{         -60}{         112}
\datum{         -60}{         122}
\datum{         -60}{         124}
\datum{         -60}{         126}
\datum{         -60}{         132}
\datum{         -60}{         144}
\datum{         -60}{         166}
\datum{         -60}{         178}
\datum{         -60}{         196}
\datum{         -60}{         218}
\datum{         -60}{         222}
\datum{         -60}{         234}
\datum{         -60}{         358}
\datum{         -60}{         474}
\datum{         -62}{          89}
\datum{         -62}{         105}
\datum{         -62}{         147}
\datum{         -64}{          42}
\datum{         -64}{          46}
\datum{         -64}{          48}
\datum{         -64}{          52}
\datum{         -64}{          54}
\datum{         -64}{          60}
\datum{         -64}{          62}
\datum{         -64}{          66}
\datum{         -64}{          72}
\datum{         -64}{          76}
\datum{         -64}{          78}
\datum{         -64}{          80}
\datum{         -64}{          82}
\datum{         -64}{          86}
\datum{         -64}{          90}
\datum{         -64}{         102}
\datum{         -64}{         106}
\datum{         -64}{         110}
\datum{         -64}{         118}
\datum{         -64}{         124}
\datum{         -64}{         134}
\datum{         -64}{         166}
\datum{         -66}{          65}
\datum{         -66}{          75}
\datum{         -66}{          79}
\datum{         -66}{          87}
\datum{         -66}{          89}
\datum{         -66}{          97}
\datum{         -68}{          42}
\datum{         -68}{          54}
\datum{         -68}{          58}
\datum{         -68}{          86}
\datum{         -68}{         126}
\datum{         -70}{          51}
\datum{         -70}{          53}
\datum{         -70}{          73}
\datum{         -72}{          40}
\datum{         -72}{          44}
\datum{         -72}{          46}
\datum{         -72}{          48}
\datum{         -72}{          50}
\datum{         -72}{          52}
\datum{         -72}{          54}
\datum{         -72}{          56}
\datum{         -72}{          58}
\datum{         -72}{          60}
\datum{         -72}{          62}
\datum{         -72}{          64}
\datum{         -72}{          66}
\datum{         -72}{          68}
\datum{         -72}{          70}
\datum{         -72}{          74}
\datum{         -72}{          76}
\datum{         -72}{          78}
\datum{         -72}{          80}
\datum{         -72}{          82}
\datum{         -72}{          84}
\datum{         -72}{          86}
\datum{         -72}{          88}
\datum{         -72}{          92}
\datum{         -72}{          94}
\datum{         -72}{          96}
\datum{         -72}{          98}
\datum{         -72}{         100}
\datum{         -72}{         102}
\datum{         -72}{         104}
\datum{         -72}{         112}
\datum{         -72}{         116}
\datum{         -72}{         118}
\datum{         -72}{         120}
\datum{         -72}{         124}
\datum{         -72}{         126}
\datum{         -72}{         128}
\datum{         -72}{         132}
\datum{         -72}{         136}
\datum{         -72}{         140}
\datum{         -72}{         142}
\datum{         -72}{         144}
\datum{         -72}{         148}
\datum{         -72}{         150}
\datum{         -72}{         158}
\datum{         -72}{         160}
\datum{         -72}{         164}
\datum{         -72}{         174}
\datum{         -72}{         182}
\datum{         -72}{         198}
\datum{         -72}{         212}
\datum{         -74}{         177}
\datum{         -76}{          74}
\datum{         -76}{          94}
\datum{         -76}{         112}
\datum{         -76}{         124}
\datum{         -76}{         142}
\datum{         -78}{          61}
\datum{         -78}{          91}
\datum{         -78}{         109}
\datum{         -78}{         175}
\datum{         -80}{          46}
\datum{         -80}{          50}
\datum{         -80}{          54}
\datum{         -80}{          58}
\datum{         -80}{          60}
\datum{         -80}{          62}
\datum{         -80}{          66}
\datum{         -80}{          68}
\datum{         -80}{          70}
\datum{         -80}{          72}
\datum{         -80}{          74}
\datum{         -80}{          78}
\datum{         -80}{          94}
\datum{         -80}{          96}
\datum{         -80}{          98}
\datum{         -80}{         102}
\datum{         -80}{         106}
\datum{         -80}{         118}
\datum{         -80}{         122}
\datum{         -80}{         138}
\datum{         -80}{         146}
\datum{         -84}{          50}
\datum{         -84}{          52}
\datum{         -84}{          54}
\datum{         -84}{          56}
\datum{         -84}{          58}
\datum{         -84}{          60}
\datum{         -84}{          64}
\datum{         -84}{          66}
\datum{         -84}{          68}
\datum{         -84}{          70}
\datum{         -84}{          72}
\datum{         -84}{          78}
\datum{         -84}{          82}
\datum{         -84}{          84}
\datum{         -84}{          86}
\datum{         -84}{          88}
\datum{         -84}{          90}
\datum{         -84}{          94}
\datum{         -84}{          96}
\datum{         -84}{          98}
\datum{         -84}{         100}
\datum{         -84}{         102}
\datum{         -84}{         114}
\datum{         -84}{         124}
\datum{         -84}{         130}
\datum{         -84}{         132}
\datum{         -84}{         134}
\datum{         -84}{         138}
\datum{         -84}{         154}
\datum{         -84}{         164}
\datum{         -84}{         166}
\datum{         -84}{         172}
\datum{         -84}{         174}
\datum{         -84}{         178}
\datum{         -84}{         190}
\datum{         -84}{         194}
\datum{         -84}{         262}
\datum{         -84}{         322}
\datum{         -86}{          57}
\datum{         -86}{          77}
\datum{         -86}{         137}
\datum{         -88}{          52}
\datum{         -88}{          54}
\datum{         -88}{          60}
\datum{         -88}{          62}
\datum{         -88}{          66}
\datum{         -88}{          68}
\datum{         -88}{          70}
\datum{         -88}{          74}
\datum{         -88}{          78}
\datum{         -88}{          82}
\datum{         -88}{          84}
\datum{         -88}{         102}
\datum{         -88}{         108}
\datum{         -88}{         132}
\datum{         -88}{         158}
\datum{         -90}{          61}
\datum{         -90}{          63}
\datum{         -90}{          71}
\datum{         -90}{          73}
\datum{         -90}{          79}
\datum{         -90}{          93}
\datum{         -90}{          99}
\datum{         -90}{         109}
\datum{         -90}{         133}
\datum{         -90}{         141}
\datum{         -92}{          56}
\datum{         -92}{          58}
\datum{         -92}{          64}
\datum{         -92}{          86}
\datum{         -92}{          94}
\datum{         -92}{          98}
\datum{         -92}{         150}
\datum{         -94}{          87}
\datum{         -96}{          54}
\datum{         -96}{          56}
\datum{         -96}{          58}
\datum{         -96}{          60}
\datum{         -96}{          62}
\datum{         -96}{          64}
\datum{         -96}{          66}
\datum{         -96}{          70}
\datum{         -96}{          74}
\datum{         -96}{          76}
\datum{         -96}{          78}
\datum{         -96}{          80}
\datum{         -96}{          82}
\datum{         -96}{          84}
\datum{         -96}{          86}
\datum{         -96}{          88}
\datum{         -96}{          90}
\datum{         -96}{          92}
\datum{         -96}{          94}
\datum{         -96}{          96}
\datum{         -96}{          98}
\datum{         -96}{         102}
\datum{         -96}{         106}
\datum{         -96}{         108}
\datum{         -96}{         110}
\datum{         -96}{         112}
\datum{         -96}{         114}
\datum{         -96}{         118}
\datum{         -96}{         120}
\datum{         -96}{         122}
\datum{         -96}{         126}
\datum{         -96}{         134}
\datum{         -96}{         138}
\datum{         -96}{         142}
\datum{         -96}{         146}
\datum{         -96}{         154}
\datum{         -96}{         158}
\datum{         -96}{         166}
\datum{         -96}{         186}
\datum{         -96}{         190}
\datum{         -96}{         204}
\datum{         -96}{         212}
\datum{         -96}{         250}
\datum{         -96}{         286}
\datum{         -96}{         342}
\datum{         -96}{         402}
\datum{         -98}{          87}
\datum{         100}{          62}
\datum{         100}{          70}
\datum{         100}{          80}
\datum{         100}{          88}
\datum{         100}{          96}
\datum{         100}{          98}
\datum{         100}{         118}
\datum{         100}{         168}
\datum{         102}{          61}
\datum{         102}{          67}
\datum{         102}{          93}
\datum{         102}{          97}
\datum{         102}{          99}
\datum{         104}{          58}
\datum{         104}{          62}
\datum{         104}{          68}
\datum{         104}{          74}
\datum{         104}{          76}
\datum{         104}{          78}
\datum{         104}{          80}
\datum{         104}{          82}
\datum{         104}{          84}
\datum{         104}{          86}
\datum{         104}{          90}
\datum{         104}{         100}
\datum{         104}{         102}
\datum{         104}{         106}
\datum{         104}{         108}
\datum{         104}{         132}
\datum{         104}{         154}
\datum{         106}{          69}
\datum{         106}{         129}
\datum{         108}{          58}
\datum{         108}{          66}
\datum{         108}{          74}
\datum{         108}{          76}
\datum{         108}{          78}
\datum{         108}{          80}
\datum{         108}{          84}
\datum{         108}{          86}
\datum{         108}{          88}
\datum{         108}{          90}
\datum{         108}{          94}
\datum{         108}{          96}
\datum{         108}{         100}
\datum{         108}{         106}
\datum{         108}{         108}
\datum{         108}{         112}
\datum{         108}{         118}
\datum{         108}{         120}
\datum{         108}{         122}
\datum{         108}{         130}
\datum{         108}{         150}
\datum{         108}{         166}
\datum{         108}{         178}
\datum{         108}{         190}
\datum{         108}{         212}
\datum{         110}{          73}
\datum{         112}{          62}
\datum{         112}{          66}
\datum{         112}{          70}
\datum{         112}{          76}
\datum{         112}{          78}
\datum{         112}{          80}
\datum{         112}{          86}
\datum{         112}{          88}
\datum{         112}{          90}
\datum{         112}{          92}
\datum{         112}{          94}
\datum{         112}{          96}
\datum{         112}{         108}
\datum{         112}{         110}
\datum{         112}{         118}
\datum{         112}{         124}
\datum{         112}{         126}
\datum{         112}{         138}
\datum{         112}{         150}
\datum{         114}{          79}
\datum{         114}{          85}
\datum{         114}{          91}
\datum{         114}{          97}
\datum{         114}{         103}
\datum{         114}{         113}
\datum{         114}{         115}
\datum{         114}{         127}
\datum{         114}{         147}
\datum{         114}{         169}
\datum{         116}{          72}
\datum{         116}{         108}
\datum{         116}{         124}
\datum{         120}{          62}
\datum{         120}{          64}
\datum{         120}{          66}
\datum{         120}{          68}
\datum{         120}{          70}
\datum{         120}{          72}
\datum{         120}{          74}
\datum{         120}{          76}
\datum{         120}{          78}
\datum{         120}{          80}
\datum{         120}{          82}
\datum{         120}{          86}
\datum{         120}{          88}
\datum{         120}{          90}
\datum{         120}{          92}
\datum{         120}{          94}
\datum{         120}{          96}
\datum{         120}{          98}
\datum{         120}{         100}
\datum{         120}{         102}
\datum{         120}{         104}
\datum{         120}{         106}
\datum{         120}{         108}
\datum{         120}{         110}
\datum{         120}{         112}
\datum{         120}{         114}
\datum{         120}{         116}
\datum{         120}{         118}
\datum{         120}{         122}
\datum{         120}{         124}
\datum{         120}{         128}
\datum{         120}{         130}
\datum{         120}{         132}
\datum{         120}{         134}
\datum{         120}{         136}
\datum{         120}{         138}
\datum{         120}{         140}
\datum{         120}{         142}
\datum{         120}{         146}
\datum{         120}{         148}
\datum{         120}{         150}
\datum{         120}{         152}
\datum{         120}{         154}
\datum{         120}{         156}
\datum{         120}{         158}
\datum{         120}{         168}
\datum{         120}{         178}
\datum{         120}{         190}
\datum{         120}{         196}
\datum{         120}{         212}
\datum{         120}{         248}
\datum{         120}{         276}
\datum{         120}{         278}
\datum{         124}{          94}
\datum{         124}{         106}
\datum{         124}{         162}
\datum{         126}{          69}
\datum{         126}{          77}
\datum{         126}{          79}
\datum{         126}{          85}
\datum{         126}{          87}
\datum{         126}{          89}
\datum{         126}{         101}
\datum{         126}{         109}
\datum{         126}{         115}
\datum{         126}{         117}
\datum{         126}{         133}
\datum{         126}{         143}
\datum{         128}{          74}
\datum{         128}{          78}
\datum{         128}{          82}
\datum{         128}{          86}
\datum{         128}{          90}
\datum{         128}{          92}
\datum{         128}{          94}
\datum{         128}{         102}
\datum{         128}{         108}
\datum{         128}{         110}
\datum{         128}{         118}
\datum{         128}{         124}
\datum{         128}{         126}
\datum{         128}{         142}
\datum{         130}{         109}
\datum{         132}{          72}
\datum{         132}{          76}
\datum{         132}{          78}
\datum{         132}{          84}
\datum{         132}{          86}
\datum{         132}{          90}
\datum{         132}{          92}
\datum{         132}{          94}
\datum{         132}{          96}
\datum{         132}{         100}
\datum{         132}{         102}
\datum{         132}{         104}
\datum{         132}{         114}
\datum{         132}{         128}
\datum{         132}{         150}
\datum{         132}{         172}
\datum{         132}{         178}
\datum{         132}{         238}
\datum{         132}{         302}
\datum{         136}{          78}
\datum{         136}{          92}
\datum{         136}{         102}
\datum{         136}{         104}
\datum{         136}{         112}
\datum{         136}{         126}
\datum{         136}{         136}
\datum{         136}{         138}
\datum{         136}{         182}
\datum{         138}{          79}
\datum{         138}{          85}
\datum{         138}{          97}
\datum{         138}{         103}
\datum{         138}{         151}
\datum{         140}{          84}
\datum{         140}{          88}
\datum{         140}{          94}
\datum{         140}{          96}
\datum{         140}{          98}
\datum{         140}{         114}
\datum{         140}{         130}
\datum{         140}{         138}
\datum{         140}{         148}
\datum{         140}{         158}
\datum{         142}{         129}
\datum{         144}{          74}
\datum{         144}{          76}
\datum{         144}{          78}
\datum{         144}{          80}
\datum{         144}{          82}
\datum{         144}{          84}
\datum{         144}{          86}
\datum{         144}{          88}
\datum{         144}{          90}
\datum{         144}{          92}
\datum{         144}{          94}
\datum{         144}{          96}
\datum{         144}{          98}
\datum{         144}{         100}
\datum{         144}{         102}
\datum{         144}{         104}
\datum{         144}{         106}
\datum{         144}{         110}
\datum{         144}{         112}
\datum{         144}{         114}
\datum{         144}{         118}
\datum{         144}{         122}
\datum{         144}{         124}
\datum{         144}{         126}
\datum{         144}{         130}
\datum{         144}{         134}
\datum{         144}{         138}
\datum{         144}{         140}
\datum{         144}{         148}
\datum{         144}{         150}
\datum{         144}{         152}
\datum{         144}{         154}
\datum{         144}{         158}
\datum{         144}{         164}
\datum{         144}{         172}
\datum{         144}{         174}
\datum{         144}{         182}
\datum{         144}{         188}
\datum{         144}{         190}
\datum{         144}{         206}
\datum{         144}{         214}
\datum{         146}{         193}
\datum{         148}{          92}
\datum{         148}{         106}
\datum{         150}{          93}
\datum{         150}{          97}
\datum{         150}{         101}
\datum{         150}{         123}
\datum{         150}{         153}
\datum{         150}{         181}
\datum{         152}{          78}
\datum{         152}{          84}
\datum{         152}{          86}
\datum{         152}{          96}
\datum{         152}{         102}
\datum{         152}{         108}
\datum{         152}{         110}
\datum{         152}{         134}
\datum{         154}{          89}
\datum{         154}{          97}
\datum{         156}{          86}
\datum{         156}{          88}
\datum{         156}{          92}
\datum{         156}{          94}
\datum{         156}{         102}
\datum{         156}{         106}
\datum{         156}{         108}
\datum{         156}{         110}
\datum{         156}{         112}
\datum{         156}{         114}
\datum{         156}{         120}
\datum{         156}{         122}
\datum{         156}{         124}
\datum{         156}{         126}
\datum{         156}{         132}
\datum{         156}{         148}
\datum{         156}{         154}
\datum{         156}{         178}
\datum{         156}{         210}
\datum{         156}{         232}
\datum{         156}{         234}
\datum{         156}{         430}
\datum{         160}{          86}
\datum{         160}{          90}
\datum{         160}{          94}
\datum{         160}{          98}
\datum{         160}{         102}
\datum{         160}{         108}
\datum{         160}{         110}
\datum{         160}{         114}
\datum{         160}{         124}
\datum{         160}{         126}
\datum{         160}{         150}
\datum{         160}{         156}
\datum{         160}{         170}
\datum{         160}{         178}
\datum{         162}{          97}
\datum{         162}{         115}
\datum{         162}{         131}
\datum{         162}{         133}
\datum{         164}{          94}
\datum{         168}{          84}
\datum{         168}{          86}
\datum{         168}{          88}
\datum{         168}{          90}
\datum{         168}{          94}
\datum{         168}{          96}
\datum{         168}{          98}
\datum{         168}{         100}
\datum{         168}{         102}
\datum{         168}{         106}
\datum{         168}{         108}
\datum{         168}{         110}
\datum{         168}{         112}
\datum{         168}{         114}
\datum{         168}{         116}
\datum{         168}{         118}
\datum{         168}{         120}
\datum{         168}{         122}
\datum{         168}{         124}
\datum{         168}{         126}
\datum{         168}{         128}
\datum{         168}{         134}
\datum{         168}{         138}
\datum{         168}{         140}
\datum{         168}{         142}
\datum{         168}{         144}
\datum{         168}{         148}
\datum{         168}{         152}
\datum{         168}{         164}
\datum{         168}{         168}
\datum{         168}{         178}
\datum{         168}{         184}
\datum{         168}{         256}
\datum{         170}{          93}
\datum{         170}{         117}
\datum{         172}{         114}
\datum{         172}{         122}
\datum{         174}{          97}
\datum{         174}{         143}
\datum{         174}{         157}
\datum{         174}{         167}
\datum{         176}{          92}
\datum{         176}{          98}
\datum{         176}{         100}
\datum{         176}{         102}
\datum{         176}{         106}
\datum{         176}{         108}
\datum{         176}{         114}
\datum{         176}{         126}
\datum{         176}{         132}
\datum{         176}{         174}
\datum{         180}{          90}
\datum{         180}{          98}
\datum{         180}{         100}
\datum{         180}{         104}
\datum{         180}{         106}
\datum{         180}{         108}
\datum{         180}{         110}
\datum{         180}{         112}
\datum{         180}{         114}
\datum{         180}{         118}
\datum{         180}{         120}
\datum{         180}{         124}
\datum{         180}{         126}
\datum{         180}{         134}
\datum{         180}{         136}
\datum{         180}{         138}
\datum{         180}{         142}
\datum{         180}{         144}
\datum{         180}{         148}
\datum{         180}{         154}
\datum{         180}{         158}
\datum{         180}{         168}
\datum{         180}{         174}
\datum{         180}{         184}
\datum{         180}{         226}
\datum{         180}{         228}
\datum{         180}{         286}
\datum{         180}{         366}
\datum{         184}{         102}
\datum{         184}{         106}
\datum{         184}{         110}
\datum{         184}{         116}
\datum{         184}{         122}
\datum{         184}{         134}
\datum{         184}{         136}
\datum{         184}{         148}
\datum{         184}{         180}
\datum{         186}{          97}
\datum{         186}{         115}
\datum{         186}{         123}
\datum{         186}{         167}
\datum{         192}{         102}
\datum{         192}{         106}
\datum{         192}{         108}
\datum{         192}{         110}
\datum{         192}{         112}
\datum{         192}{         114}
\datum{         192}{         116}
\datum{         192}{         118}
\datum{         192}{         122}
\datum{         192}{         124}
\datum{         192}{         126}
\datum{         192}{         128}
\datum{         192}{         130}
\datum{         192}{         134}
\datum{         192}{         138}
\datum{         192}{         142}
\datum{         192}{         150}
\datum{         192}{         154}
\datum{         192}{         162}
\datum{         192}{         166}
\datum{         192}{         170}
\datum{         192}{         178}
\datum{         192}{         184}
\datum{         192}{         190}
\datum{         192}{         198}
\datum{         192}{         206}
\datum{         192}{         218}
\datum{         192}{         226}
\datum{         192}{         250}
\datum{         196}{         106}
\datum{         196}{         150}
\datum{         196}{         172}
\datum{         198}{         125}
\datum{         198}{         131}
\datum{         200}{         102}
\datum{         200}{         106}
\datum{         200}{         108}
\datum{         200}{         116}
\datum{         200}{         118}
\datum{         200}{         128}
\datum{         200}{         134}
\datum{         200}{         136}
\datum{         200}{         148}
\datum{         200}{         150}
\datum{         200}{         156}
\datum{         200}{         168}
\datum{         204}{         104}
\datum{         204}{         108}
\datum{         204}{         114}
\datum{         204}{         118}
\datum{         204}{         120}
\datum{         204}{         124}
\datum{         204}{         126}
\datum{         204}{         130}
\datum{         204}{         134}
\datum{         204}{         142}
\datum{         204}{         164}
\datum{         204}{         168}
\datum{         204}{         190}
\datum{         208}{         108}
\datum{         208}{         118}
\datum{         208}{         126}
\datum{         208}{         138}
\datum{         210}{         113}
\datum{         210}{         117}
\datum{         210}{         133}
\datum{         210}{         143}
\datum{         210}{         145}
\datum{         210}{         151}
\datum{         210}{         171}
\datum{         210}{         221}
\datum{         212}{         150}
\datum{         212}{         158}
\datum{         216}{         112}
\datum{         216}{         116}
\datum{         216}{         120}
\datum{         216}{         124}
\datum{         216}{         126}
\datum{         216}{         128}
\datum{         216}{         132}
\datum{         216}{         134}
\datum{         216}{         136}
\datum{         216}{         138}
\datum{         216}{         140}
\datum{         216}{         142}
\datum{         216}{         144}
\datum{         216}{         150}
\datum{         216}{         152}
\datum{         216}{         154}
\datum{         216}{         158}
\datum{         216}{         164}
\datum{         216}{         168}
\datum{         216}{         174}
\datum{         216}{         180}
\datum{         216}{         182}
\datum{         216}{         188}
\datum{         216}{         192}
\datum{         216}{         212}
\datum{         216}{         240}
\datum{         216}{         244}
\datum{         216}{         274}
\datum{         216}{         292}
\datum{         220}{         126}
\datum{         220}{         158}
\datum{         220}{         218}
\datum{         222}{         133}
\datum{         222}{         135}
\datum{         224}{         122}
\datum{         224}{         124}
\datum{         224}{         126}
\datum{         224}{         134}
\datum{         224}{         138}
\datum{         224}{         142}
\datum{         224}{         178}
\datum{         228}{         126}
\datum{         228}{         130}
\datum{         228}{         132}
\datum{         228}{         138}
\datum{         228}{         140}
\datum{         228}{         142}
\datum{         228}{         146}
\datum{         228}{         170}
\datum{         228}{         176}
\datum{         228}{         178}
\datum{         228}{         196}
\datum{         228}{         202}
\datum{         230}{         129}
\datum{         232}{         126}
\datum{         232}{         134}
\datum{         232}{         158}
\datum{         234}{         131}
\datum{         234}{         133}
\datum{         234}{         141}
\datum{         234}{         217}
\datum{         236}{         172}
\datum{         240}{         124}
\datum{         240}{         126}
\datum{         240}{         130}
\datum{         240}{         134}
\datum{         240}{         138}
\datum{         240}{         140}
\datum{         240}{         142}
\datum{         240}{         144}
\datum{         240}{         146}
\datum{         240}{         150}
\datum{         240}{         158}
\datum{         240}{         162}
\datum{         240}{         166}
\datum{         240}{         174}
\datum{         240}{         178}
\datum{         240}{         188}
\datum{         240}{         206}
\datum{         240}{         218}
\datum{         240}{         226}
\datum{         240}{         232}
\datum{         240}{         268}
\datum{         240}{         394}
\datum{         242}{         139}
\datum{         242}{         189}
\datum{         246}{         167}
\datum{         246}{         185}
\datum{         246}{         237}
\datum{         248}{         134}
\datum{         248}{         174}
\datum{         252}{         130}
\datum{         252}{         138}
\datum{         252}{         142}
\datum{         252}{         158}
\datum{         252}{         162}
\datum{         252}{         166}
\datum{         252}{         174}
\datum{         252}{         178}
\datum{         252}{         202}
\datum{         252}{         206}
\datum{         252}{         278}
\datum{         256}{         134}
\datum{         256}{         156}
\datum{         256}{         158}
\datum{         256}{         166}
\datum{         256}{         174}
\datum{         256}{         234}
\datum{         258}{         151}
\datum{         258}{         163}
\datum{         260}{         134}
\datum{         260}{         150}
\datum{         264}{         144}
\datum{         264}{         148}
\datum{         264}{         150}
\datum{         264}{         154}
\datum{         264}{         158}
\datum{         264}{         160}
\datum{         264}{         162}
\datum{         264}{         164}
\datum{         264}{         172}
\datum{         264}{         178}
\datum{         264}{         184}
\datum{         264}{         188}
\datum{         264}{         214}
\datum{         264}{         228}
\datum{         264}{         262}
\datum{         266}{         169}
\datum{         270}{         223}
\datum{         272}{         148}
\datum{         272}{         150}
\datum{         272}{         166}
\datum{         272}{         178}
\datum{         276}{         150}
\datum{         276}{         154}
\datum{         276}{         162}
\datum{         276}{         172}
\datum{         276}{         174}
\datum{         276}{         186}
\datum{         276}{         192}
\datum{         276}{         212}
\datum{         276}{         214}
\datum{         276}{         222}
\datum{         276}{         234}
\datum{         276}{         262}
\datum{         276}{         330}
\datum{         280}{         148}
\datum{         280}{         150}
\datum{         280}{         166}
\datum{         286}{         177}
\datum{         288}{         146}
\datum{         288}{         152}
\datum{         288}{         158}
\datum{         288}{         162}
\datum{         288}{         166}
\datum{         288}{         170}
\datum{         288}{         182}
\datum{         288}{         186}
\datum{         288}{         188}
\datum{         288}{         190}
\datum{         288}{         206}
\datum{         288}{         214}
\datum{         288}{         222}
\datum{         288}{         226}
\datum{         288}{         270}
\datum{         288}{         374}
\datum{         292}{         158}
\datum{         294}{         159}
\datum{         294}{         187}
\datum{         296}{         150}
\datum{         296}{         166}
\datum{         296}{         174}
\datum{         300}{         168}
\datum{         300}{         178}
\datum{         300}{         180}
\datum{         300}{         194}
\datum{         300}{         198}
\datum{         300}{         218}
\datum{         300}{         226}
\datum{         300}{         230}
\datum{         304}{         176}
\datum{         306}{         169}
\datum{         306}{         217}
\datum{         312}{         166}
\datum{         312}{         172}
\datum{         312}{         174}
\datum{         312}{         178}
\datum{         312}{         180}
\datum{         312}{         190}
\datum{         312}{         192}
\datum{         312}{         196}
\datum{         312}{         224}
\datum{         316}{         262}
\datum{         318}{         197}
\datum{         320}{         170}
\datum{         320}{         174}
\datum{         320}{         190}
\datum{         320}{         198}
\datum{         320}{         206}
\datum{         320}{         222}
\datum{         322}{         193}
\datum{         324}{         168}
\datum{         324}{         182}
\datum{         324}{         184}
\datum{         324}{         222}
\datum{         324}{         230}
\datum{         324}{         232}
\datum{         324}{         262}
\datum{         330}{         181}
\datum{         330}{         221}
\datum{         330}{         261}
\datum{         336}{         178}
\datum{         336}{         188}
\datum{         336}{         194}
\datum{         336}{         198}
\datum{         336}{         202}
\datum{         336}{         206}
\datum{         336}{         222}
\datum{         336}{         230}
\datum{         336}{         312}
\datum{         336}{         358}
\datum{         340}{         198}
\datum{         342}{         233}
\datum{         348}{         186}
\datum{         348}{         198}
\datum{         348}{         226}
\datum{         348}{         238}
\datum{         352}{         190}
\datum{         354}{         259}
\datum{         356}{         202}
\datum{         356}{         220}
\datum{         360}{         190}
\datum{         360}{         192}
\datum{         360}{         194}
\datum{         360}{         206}
\datum{         360}{         212}
\datum{         360}{         228}
\datum{         360}{         258}
\datum{         360}{         306}
\datum{         364}{         194}
\datum{         368}{         204}
\datum{         372}{         194}
\datum{         372}{         202}
\datum{         372}{         226}
\datum{         372}{         258}
\datum{         372}{         262}
\datum{         372}{         346}
\datum{         376}{         214}
\datum{         380}{         198}
\datum{         384}{         204}
\datum{         384}{         218}
\datum{         384}{         222}
\datum{         384}{         232}
\datum{         384}{         234}
\datum{         384}{         242}
\datum{         384}{         250}
\datum{         384}{         262}
\datum{         396}{         214}
\datum{         396}{         222}
\datum{         396}{         262}
\datum{         396}{         340}
\datum{         408}{         212}
\datum{         408}{         224}
\datum{         408}{         232}
\datum{         408}{         240}
\datum{         408}{         268}
\datum{         420}{         218}
\datum{         420}{         230}
\datum{         420}{         248}
\datum{         420}{         250}
\datum{         420}{         306}
\datum{         420}{         334}
\datum{         426}{         265}
\datum{         432}{         238}
\datum{         432}{         242}
\datum{         432}{         266}
\datum{         432}{         274}
\datum{         432}{         334}
\datum{         444}{         234}
\datum{         450}{         303}
\datum{         456}{         234}
\datum{         456}{         248}
\datum{         456}{         256}
\datum{         456}{         262}
\datum{         456}{         264}
\datum{         456}{         272}
\datum{         456}{         302}
\datum{         468}{         286}
\datum{         468}{         306}
\datum{         476}{         270}
\datum{         480}{         246}
\datum{         480}{         262}
\datum{         480}{         278}
\datum{         480}{         286}
\datum{         480}{         334}
\datum{         492}{         256}
\datum{         504}{         276}
\datum{         504}{         312}
\datum{         510}{         331}
\datum{         512}{         286}
\datum{         516}{         302}
\datum{         516}{         330}
\datum{         528}{         278}
\datum{         528}{         286}
\datum{         528}{         318}
\datum{         528}{         334}
\datum{         540}{         274}
\datum{         540}{         298}
\datum{         540}{         334}
\datum{         552}{         306}
\datum{         564}{         322}
\datum{         564}{         330}
\datum{         564}{         340}
\datum{         576}{         302}
\datum{         576}{         314}
\datum{         588}{         346}
\datum{         612}{         330}
\datum{         612}{         346}
\datum{         624}{         330}
\datum{         624}{         358}
\datum{         636}{         342}
\datum{         648}{         358}
\datum{         660}{         366}
\datum{         672}{         374}
\datum{         720}{         394}
\datum{         732}{         386}
\datum{         744}{         402}
\datum{         804}{         430}
\datum{         840}{         446}
\datum{         900}{         474}
\datum{         960}{         502}
\datum{        -100}{          62}
\datum{        -100}{          66}
\datum{        -100}{          70}
\datum{        -100}{          76}
\datum{        -100}{          80}
\datum{        -100}{          98}
\datum{        -100}{         118}
\datum{        -100}{         120}
\datum{        -102}{          61}
\datum{        -102}{          67}
\datum{        -102}{          73}
\datum{        -102}{          81}
\datum{        -102}{          83}
\datum{        -102}{          93}
\datum{        -102}{         113}
\datum{        -104}{          54}
\datum{        -104}{          58}
\datum{        -104}{          62}
\datum{        -104}{          68}
\datum{        -104}{          74}
\datum{        -104}{          78}
\datum{        -104}{          80}
\datum{        -104}{          84}
\datum{        -104}{          86}
\datum{        -104}{          92}
\datum{        -104}{         100}
\datum{        -104}{         104}
\datum{        -104}{         106}
\datum{        -104}{         126}
\datum{        -104}{         140}
\datum{        -104}{         154}
\datum{        -106}{         117}
\datum{        -106}{         133}
\datum{        -108}{          58}
\datum{        -108}{          60}
\datum{        -108}{          66}
\datum{        -108}{          68}
\datum{        -108}{          70}
\datum{        -108}{          72}
\datum{        -108}{          74}
\datum{        -108}{          76}
\datum{        -108}{          80}
\datum{        -108}{          82}
\datum{        -108}{          84}
\datum{        -108}{          86}
\datum{        -108}{          88}
\datum{        -108}{          90}
\datum{        -108}{          94}
\datum{        -108}{         100}
\datum{        -108}{         106}
\datum{        -108}{         108}
\datum{        -108}{         112}
\datum{        -108}{         120}
\datum{        -108}{         122}
\datum{        -108}{         130}
\datum{        -108}{         150}
\datum{        -108}{         166}
\datum{        -108}{         178}
\datum{        -108}{         186}
\datum{        -108}{         190}
\datum{        -108}{         212}
\datum{        -110}{          83}
\datum{        -110}{         119}
\datum{        -110}{         141}
\datum{        -110}{         193}
\datum{        -112}{          60}
\datum{        -112}{          62}
\datum{        -112}{          66}
\datum{        -112}{          70}
\datum{        -112}{          74}
\datum{        -112}{          76}
\datum{        -112}{          78}
\datum{        -112}{          88}
\datum{        -112}{          90}
\datum{        -112}{          92}
\datum{        -112}{          94}
\datum{        -112}{          96}
\datum{        -112}{         102}
\datum{        -112}{         108}
\datum{        -112}{         110}
\datum{        -112}{         118}
\datum{        -112}{         124}
\datum{        -112}{         126}
\datum{        -112}{         160}
\datum{        -114}{          67}
\datum{        -114}{          77}
\datum{        -114}{          85}
\datum{        -114}{          97}
\datum{        -114}{         127}
\datum{        -114}{         145}
\datum{        -114}{         147}
\datum{        -114}{         187}
\datum{        -114}{         197}
\datum{        -116}{          72}
\datum{        -116}{         124}
\datum{        -120}{          62}
\datum{        -120}{          64}
\datum{        -120}{          66}
\datum{        -120}{          68}
\datum{        -120}{          70}
\datum{        -120}{          72}
\datum{        -120}{          76}
\datum{        -120}{          78}
\datum{        -120}{          80}
\datum{        -120}{          82}
\datum{        -120}{          86}
\datum{        -120}{          88}
\datum{        -120}{          90}
\datum{        -120}{          92}
\datum{        -120}{          94}
\datum{        -120}{          96}
\datum{        -120}{          98}
\datum{        -120}{         100}
\datum{        -120}{         102}
\datum{        -120}{         104}
\datum{        -120}{         108}
\datum{        -120}{         110}
\datum{        -120}{         112}
\datum{        -120}{         116}
\datum{        -120}{         118}
\datum{        -120}{         122}
\datum{        -120}{         124}
\datum{        -120}{         128}
\datum{        -120}{         130}
\datum{        -120}{         132}
\datum{        -120}{         134}
\datum{        -120}{         136}
\datum{        -120}{         138}
\datum{        -120}{         142}
\datum{        -120}{         146}
\datum{        -120}{         148}
\datum{        -120}{         150}
\datum{        -120}{         152}
\datum{        -120}{         154}
\datum{        -120}{         156}
\datum{        -120}{         158}
\datum{        -120}{         168}
\datum{        -120}{         178}
\datum{        -120}{         190}
\datum{        -120}{         196}
\datum{        -120}{         212}
\datum{        -120}{         248}
\datum{        -120}{         276}
\datum{        -120}{         278}
\datum{        -124}{          74}
\datum{        -124}{          84}
\datum{        -124}{         106}
\datum{        -124}{         136}
\datum{        -124}{         150}
\datum{        -126}{          69}
\datum{        -126}{          79}
\datum{        -126}{          87}
\datum{        -126}{          89}
\datum{        -126}{          91}
\datum{        -126}{          99}
\datum{        -126}{         115}
\datum{        -126}{         143}
\datum{        -126}{         179}
\datum{        -128}{          78}
\datum{        -128}{          82}
\datum{        -128}{          86}
\datum{        -128}{          90}
\datum{        -128}{          92}
\datum{        -128}{          94}
\datum{        -128}{          98}
\datum{        -128}{         102}
\datum{        -128}{         108}
\datum{        -128}{         110}
\datum{        -128}{         114}
\datum{        -128}{         142}
\datum{        -130}{          93}
\datum{        -132}{          72}
\datum{        -132}{          74}
\datum{        -132}{          76}
\datum{        -132}{          78}
\datum{        -132}{          80}
\datum{        -132}{          86}
\datum{        -132}{          94}
\datum{        -132}{          96}
\datum{        -132}{         100}
\datum{        -132}{         102}
\datum{        -132}{         104}
\datum{        -132}{         114}
\datum{        -132}{         126}
\datum{        -132}{         148}
\datum{        -132}{         178}
\datum{        -132}{         202}
\datum{        -132}{         238}
\datum{        -132}{         302}
\datum{        -134}{          87}
\datum{        -136}{          72}
\datum{        -136}{          78}
\datum{        -136}{          82}
\datum{        -136}{          86}
\datum{        -136}{          94}
\datum{        -136}{         102}
\datum{        -136}{         108}
\datum{        -136}{         112}
\datum{        -136}{         118}
\datum{        -136}{         126}
\datum{        -136}{         136}
\datum{        -138}{          85}
\datum{        -138}{          97}
\datum{        -138}{         105}
\datum{        -138}{         125}
\datum{        -138}{         151}
\datum{        -140}{          88}
\datum{        -140}{          96}
\datum{        -140}{          98}
\datum{        -140}{         110}
\datum{        -140}{         158}
\datum{        -142}{         127}
\datum{        -144}{          74}
\datum{        -144}{          76}
\datum{        -144}{          78}
\datum{        -144}{          80}
\datum{        -144}{          82}
\datum{        -144}{          84}
\datum{        -144}{          86}
\datum{        -144}{          88}
\datum{        -144}{          90}
\datum{        -144}{          92}
\datum{        -144}{          94}
\datum{        -144}{          96}
\datum{        -144}{          98}
\datum{        -144}{         100}
\datum{        -144}{         102}
\datum{        -144}{         104}
\datum{        -144}{         106}
\datum{        -144}{         110}
\datum{        -144}{         112}
\datum{        -144}{         114}
\datum{        -144}{         118}
\datum{        -144}{         122}
\datum{        -144}{         124}
\datum{        -144}{         126}
\datum{        -144}{         130}
\datum{        -144}{         134}
\datum{        -144}{         138}
\datum{        -144}{         140}
\datum{        -144}{         148}
\datum{        -144}{         150}
\datum{        -144}{         154}
\datum{        -144}{         158}
\datum{        -144}{         164}
\datum{        -144}{         172}
\datum{        -144}{         174}
\datum{        -144}{         182}
\datum{        -144}{         188}
\datum{        -144}{         190}
\datum{        -144}{         206}
\datum{        -144}{         214}
\datum{        -148}{          92}
\datum{        -148}{          96}
\datum{        -148}{         158}
\datum{        -150}{          93}
\datum{        -150}{          97}
\datum{        -150}{         103}
\datum{        -150}{         109}
\datum{        -150}{         133}
\datum{        -150}{         151}
\datum{        -152}{          78}
\datum{        -152}{          84}
\datum{        -152}{          86}
\datum{        -152}{          96}
\datum{        -152}{         108}
\datum{        -152}{         110}
\datum{        -152}{         130}
\datum{        -152}{         134}
\datum{        -152}{         150}
\datum{        -156}{          86}
\datum{        -156}{          88}
\datum{        -156}{          92}
\datum{        -156}{          94}
\datum{        -156}{         102}
\datum{        -156}{         106}
\datum{        -156}{         108}
\datum{        -156}{         110}
\datum{        -156}{         112}
\datum{        -156}{         114}
\datum{        -156}{         118}
\datum{        -156}{         120}
\datum{        -156}{         122}
\datum{        -156}{         124}
\datum{        -156}{         132}
\datum{        -156}{         150}
\datum{        -156}{         210}
\datum{        -156}{         232}
\datum{        -156}{         234}
\datum{        -156}{         430}
\datum{        -160}{          90}
\datum{        -160}{          96}
\datum{        -160}{          98}
\datum{        -160}{         102}
\datum{        -160}{         108}
\datum{        -160}{         110}
\datum{        -160}{         124}
\datum{        -160}{         126}
\datum{        -160}{         150}
\datum{        -160}{         156}
\datum{        -160}{         178}
\datum{        -160}{         198}
\datum{        -160}{         314}
\datum{        -162}{         121}
\datum{        -162}{         131}
\datum{        -162}{         133}
\datum{        -162}{         137}
\datum{        -162}{         185}
\datum{        -164}{          94}
\datum{        -166}{         127}
\datum{        -168}{          84}
\datum{        -168}{          86}
\datum{        -168}{          88}
\datum{        -168}{          90}
\datum{        -168}{          94}
\datum{        -168}{          96}
\datum{        -168}{          98}
\datum{        -168}{         100}
\datum{        -168}{         102}
\datum{        -168}{         106}
\datum{        -168}{         108}
\datum{        -168}{         110}
\datum{        -168}{         112}
\datum{        -168}{         114}
\datum{        -168}{         116}
\datum{        -168}{         118}
\datum{        -168}{         120}
\datum{        -168}{         122}
\datum{        -168}{         124}
\datum{        -168}{         128}
\datum{        -168}{         134}
\datum{        -168}{         138}
\datum{        -168}{         140}
\datum{        -168}{         142}
\datum{        -168}{         144}
\datum{        -168}{         148}
\datum{        -168}{         152}
\datum{        -168}{         168}
\datum{        -168}{         174}
\datum{        -168}{         178}
\datum{        -168}{         184}
\datum{        -168}{         256}
\datum{        -172}{         106}
\datum{        -172}{         144}
\datum{        -174}{          97}
\datum{        -174}{         167}
\datum{        -176}{          92}
\datum{        -176}{         102}
\datum{        -176}{         106}
\datum{        -176}{         108}
\datum{        -176}{         114}
\datum{        -176}{         126}
\datum{        -176}{         150}
\datum{        -176}{         174}
\datum{        -180}{          90}
\datum{        -180}{          98}
\datum{        -180}{         108}
\datum{        -180}{         112}
\datum{        -180}{         118}
\datum{        -180}{         120}
\datum{        -180}{         124}
\datum{        -180}{         126}
\datum{        -180}{         136}
\datum{        -180}{         138}
\datum{        -180}{         142}
\datum{        -180}{         148}
\datum{        -180}{         154}
\datum{        -180}{         158}
\datum{        -180}{         168}
\datum{        -180}{         174}
\datum{        -180}{         184}
\datum{        -180}{         194}
\datum{        -180}{         226}
\datum{        -180}{         228}
\datum{        -180}{         286}
\datum{        -180}{         366}
\datum{        -184}{         102}
\datum{        -184}{         110}
\datum{        -184}{         112}
\datum{        -184}{         120}
\datum{        -184}{         134}
\datum{        -184}{         148}
\datum{        -184}{         168}
\datum{        -184}{         174}
\datum{        -186}{          97}
\datum{        -186}{         117}
\datum{        -186}{         127}
\datum{        -186}{         151}
\datum{        -190}{         129}
\datum{        -192}{         102}
\datum{        -192}{         106}
\datum{        -192}{         108}
\datum{        -192}{         110}
\datum{        -192}{         112}
\datum{        -192}{         114}
\datum{        -192}{         116}
\datum{        -192}{         118}
\datum{        -192}{         122}
\datum{        -192}{         124}
\datum{        -192}{         126}
\datum{        -192}{         128}
\datum{        -192}{         130}
\datum{        -192}{         134}
\datum{        -192}{         138}
\datum{        -192}{         150}
\datum{        -192}{         154}
\datum{        -192}{         162}
\datum{        -192}{         166}
\datum{        -192}{         170}
\datum{        -192}{         178}
\datum{        -192}{         184}
\datum{        -192}{         190}
\datum{        -192}{         198}
\datum{        -192}{         206}
\datum{        -192}{         218}
\datum{        -192}{         226}
\datum{        -192}{         250}
\datum{        -194}{         113}
\datum{        -194}{         119}
\datum{        -196}{         106}
\datum{        -196}{         172}
\datum{        -198}{         113}
\datum{        -198}{         139}
\datum{        -200}{         102}
\datum{        -200}{         106}
\datum{        -200}{         116}
\datum{        -200}{         130}
\datum{        -200}{         134}
\datum{        -200}{         150}
\datum{        -200}{         168}
\datum{        -202}{         133}
\datum{        -204}{         104}
\datum{        -204}{         108}
\datum{        -204}{         114}
\datum{        -204}{         118}
\datum{        -204}{         120}
\datum{        -204}{         126}
\datum{        -204}{         130}
\datum{        -204}{         134}
\datum{        -204}{         142}
\datum{        -204}{         164}
\datum{        -204}{         178}
\datum{        -204}{         222}
\datum{        -208}{         108}
\datum{        -208}{         126}
\datum{        -208}{         138}
\datum{        -208}{         178}
\datum{        -210}{         113}
\datum{        -210}{         131}
\datum{        -210}{         145}
\datum{        -210}{         151}
\datum{        -210}{         171}
\datum{        -210}{         221}
\datum{        -210}{         241}
\datum{        -216}{         112}
\datum{        -216}{         116}
\datum{        -216}{         120}
\datum{        -216}{         124}
\datum{        -216}{         126}
\datum{        -216}{         128}
\datum{        -216}{         132}
\datum{        -216}{         134}
\datum{        -216}{         136}
\datum{        -216}{         142}
\datum{        -216}{         144}
\datum{        -216}{         150}
\datum{        -216}{         152}
\datum{        -216}{         154}
\datum{        -216}{         158}
\datum{        -216}{         164}
\datum{        -216}{         174}
\datum{        -216}{         180}
\datum{        -216}{         182}
\datum{        -216}{         188}
\datum{        -216}{         192}
\datum{        -216}{         212}
\datum{        -216}{         240}
\datum{        -216}{         244}
\datum{        -216}{         274}
\datum{        -216}{         292}
\datum{        -220}{         158}
\datum{        -222}{         133}
\datum{        -224}{         122}
\datum{        -224}{         126}
\datum{        -224}{         142}
\datum{        -224}{         146}
\datum{        -224}{         152}
\datum{        -224}{         170}
\datum{        -228}{         126}
\datum{        -228}{         130}
\datum{        -228}{         132}
\datum{        -228}{         138}
\datum{        -228}{         140}
\datum{        -228}{         142}
\datum{        -228}{         146}
\datum{        -228}{         170}
\datum{        -228}{         176}
\datum{        -228}{         196}
\datum{        -228}{         346}
\datum{        -232}{         126}
\datum{        -232}{         134}
\datum{        -232}{         158}
\datum{        -234}{         217}
\datum{        -236}{         162}
\datum{        -236}{         172}
\datum{        -236}{         182}
\datum{        -240}{         124}
\datum{        -240}{         126}
\datum{        -240}{         134}
\datum{        -240}{         138}
\datum{        -240}{         140}
\datum{        -240}{         142}
\datum{        -240}{         144}
\datum{        -240}{         146}
\datum{        -240}{         150}
\datum{        -240}{         158}
\datum{        -240}{         162}
\datum{        -240}{         166}
\datum{        -240}{         174}
\datum{        -240}{         178}
\datum{        -240}{         188}
\datum{        -240}{         206}
\datum{        -240}{         218}
\datum{        -240}{         226}
\datum{        -240}{         232}
\datum{        -240}{         268}
\datum{        -240}{         394}
\datum{        -244}{         162}
\datum{        -246}{         167}
\datum{        -246}{         237}
\datum{        -252}{         130}
\datum{        -252}{         138}
\datum{        -252}{         142}
\datum{        -252}{         158}
\datum{        -252}{         162}
\datum{        -252}{         178}
\datum{        -252}{         202}
\datum{        -252}{         206}
\datum{        -252}{         278}
\datum{        -256}{         134}
\datum{        -256}{         166}
\datum{        -256}{         174}
\datum{        -258}{         161}
\datum{        -260}{         134}
\datum{        -260}{         150}
\datum{        -264}{         144}
\datum{        -264}{         148}
\datum{        -264}{         150}
\datum{        -264}{         154}
\datum{        -264}{         158}
\datum{        -264}{         162}
\datum{        -264}{         164}
\datum{        -264}{         172}
\datum{        -264}{         184}
\datum{        -264}{         188}
\datum{        -264}{         214}
\datum{        -264}{         228}
\datum{        -264}{         262}
\datum{        -272}{         150}
\datum{        -272}{         178}
\datum{        -276}{         150}
\datum{        -276}{         154}
\datum{        -276}{         172}
\datum{        -276}{         174}
\datum{        -276}{         192}
\datum{        -276}{         214}
\datum{        -276}{         234}
\datum{        -276}{         262}
\datum{        -276}{         330}
\datum{        -280}{         148}
\datum{        -280}{         166}
\datum{        -284}{         166}
\datum{        -286}{         163}
\datum{        -288}{         146}
\datum{        -288}{         152}
\datum{        -288}{         158}
\datum{        -288}{         162}
\datum{        -288}{         166}
\datum{        -288}{         170}
\datum{        -288}{         182}
\datum{        -288}{         186}
\datum{        -288}{         188}
\datum{        -288}{         190}
\datum{        -288}{         206}
\datum{        -288}{         214}
\datum{        -288}{         222}
\datum{        -288}{         226}
\datum{        -288}{         258}
\datum{        -288}{         270}
\datum{        -288}{         374}
\datum{        -292}{         158}
\datum{        -294}{         259}
\datum{        -296}{         150}
\datum{        -296}{         162}
\datum{        -296}{         166}
\datum{        -300}{         168}
\datum{        -300}{         178}
\datum{        -300}{         180}
\datum{        -300}{         198}
\datum{        -300}{         218}
\datum{        -300}{         222}
\datum{        -300}{         226}
\datum{        -300}{         230}
\datum{        -304}{         176}
\datum{        -306}{         169}
\datum{        -306}{         177}
\datum{        -306}{         217}
\datum{        -312}{         166}
\datum{        -312}{         172}
\datum{        -312}{         178}
\datum{        -312}{         180}
\datum{        -312}{         190}
\datum{        -312}{         192}
\datum{        -312}{         196}
\datum{        -312}{         224}
\datum{        -312}{         318}
\datum{        -320}{         174}
\datum{        -324}{         168}
\datum{        -324}{         182}
\datum{        -324}{         184}
\datum{        -324}{         232}
\datum{        -324}{         262}
\datum{        -330}{         181}
\datum{        -330}{         221}
\datum{        -330}{         261}
\datum{        -336}{         178}
\datum{        -336}{         188}
\datum{        -336}{         194}
\datum{        -336}{         198}
\datum{        -336}{         202}
\datum{        -336}{         206}
\datum{        -336}{         222}
\datum{        -336}{         230}
\datum{        -336}{         312}
\datum{        -336}{         358}
\datum{        -340}{         198}
\datum{        -342}{         233}
\datum{        -344}{         224}
\datum{        -348}{         186}
\datum{        -348}{         198}
\datum{        -348}{         226}
\datum{        -348}{         238}
\datum{        -348}{         258}
\datum{        -356}{         234}
\datum{        -360}{         190}
\datum{        -360}{         192}
\datum{        -360}{         212}
\datum{        -360}{         228}
\datum{        -364}{         204}
\datum{        -368}{         204}
\datum{        -372}{         194}
\datum{        -372}{         202}
\datum{        -372}{         226}
\datum{        -372}{         258}
\datum{        -372}{         262}
\datum{        -372}{         306}
\datum{        -372}{         346}
\datum{        -376}{         214}
\datum{        -384}{         204}
\datum{        -384}{         218}
\datum{        -384}{         232}
\datum{        -384}{         242}
\datum{        -384}{         250}
\datum{        -390}{         303}
\datum{        -396}{         222}
\datum{        -396}{         262}
\datum{        -396}{         340}
\datum{        -408}{         212}
\datum{        -408}{         224}
\datum{        -408}{         232}
\datum{        -408}{         240}
\datum{        -408}{         268}
\datum{        -416}{         262}
\datum{        -420}{         218}
\datum{        -420}{         230}
\datum{        -420}{         248}
\datum{        -420}{         250}
\datum{        -420}{         306}
\datum{        -420}{         334}
\datum{        -426}{         265}
\datum{        -432}{         238}
\datum{        -432}{         242}
\datum{        -432}{         266}
\datum{        -432}{         274}
\datum{        -432}{         334}
\datum{        -444}{         234}
\datum{        -444}{         330}
\datum{        -450}{         331}
\datum{        -456}{         234}
\datum{        -456}{         248}
\datum{        -456}{         256}
\datum{        -456}{         262}
\datum{        -456}{         264}
\datum{        -456}{         272}
\datum{        -468}{         286}
\datum{        -480}{         246}
\datum{        -480}{         262}
\datum{        -480}{         278}
\datum{        -480}{         286}
\datum{        -480}{         306}
\datum{        -480}{         334}
\datum{        -492}{         256}
\datum{        -504}{         276}
\datum{        -504}{         312}
\datum{        -516}{         302}
\datum{        -528}{         278}
\datum{        -528}{         286}
\datum{        -528}{         334}
\datum{        -540}{         274}
\datum{        -540}{         334}
\datum{        -552}{         306}
\datum{        -564}{         322}
\datum{        -564}{         330}
\datum{        -564}{         340}
\datum{        -576}{         314}
\datum{        -588}{         346}
\datum{        -612}{         330}
\datum{        -624}{         330}
\datum{        -624}{         358}
\datum{        -636}{         342}
\datum{        -660}{         366}
\datum{        -672}{         374}
\datum{        -720}{         394}
\datum{        -732}{         386}
\datum{        -744}{         402}
\datum{        -804}{         430}
\datum{        -900}{         474}
\datum{        -960}{         502}
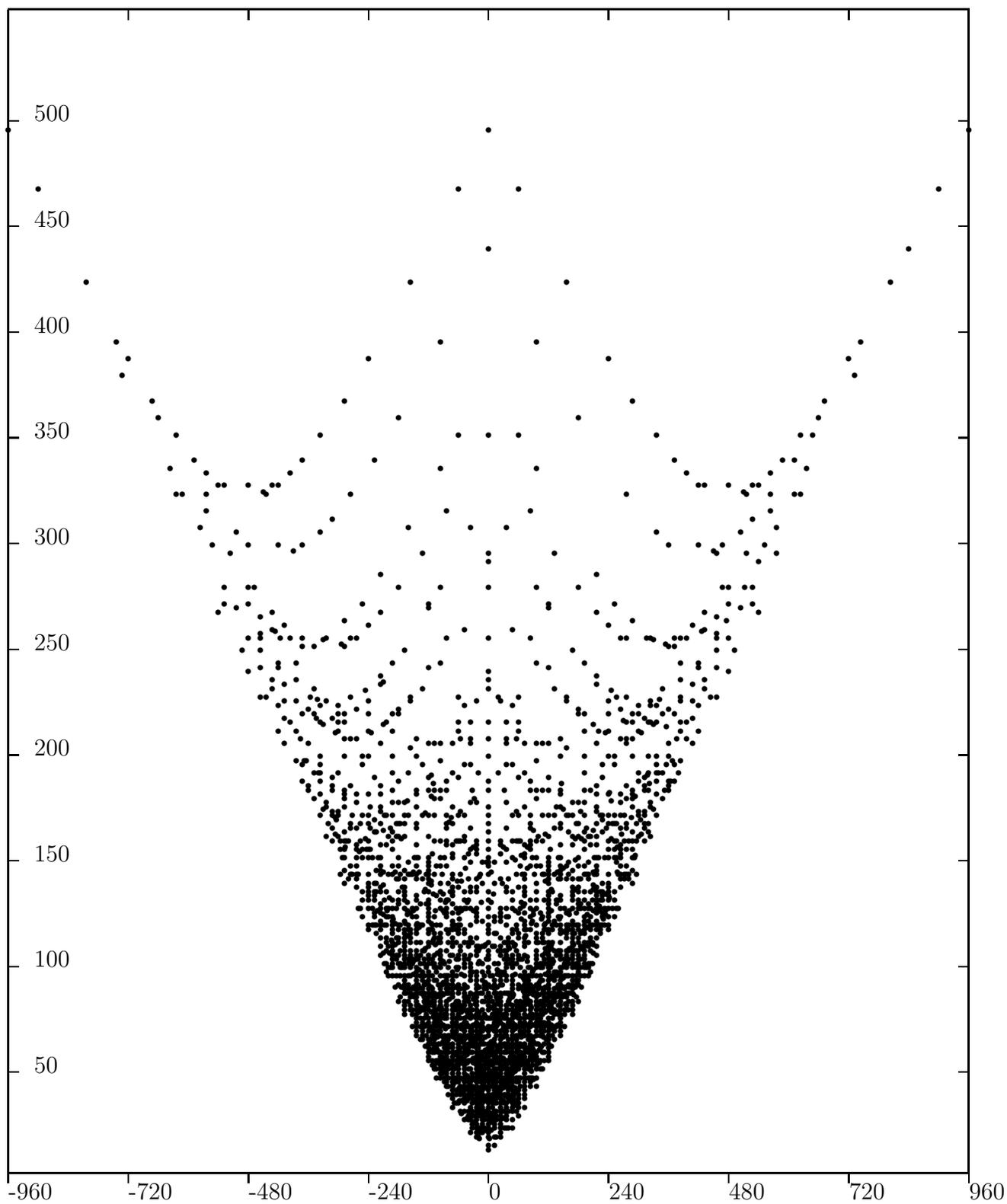
\parbox{6.4truein}{\noindent {\bf Fig. 1}~~{\it A plot of Euler numbers against
 ${\bar n}_g+n_g$ for the 2997 spectra of all the LG potentials.
}}
\end{center}
\end{center}


\def\plot#1#2{\vskip\parskip
                  \vbox{\hrule width\hsize
                        \hbox{\kern-0.2pt\vrule height#1
                              \vbox{\hfill}\kern-0.6pt
                              \vrule}\hrule width\hsize}
    \setbox0=\hbox{#2} \dimen0=\wd0 \divide\dimen0 by 2
    \setbox0=\hbox{\kern-\dimen0 #2}
    \dimen3=#1}

\def\hmark{\kern-0.2pt\lower10pt\hbox{\vrule height 5pt}}
\def\leftscalemark{\vbox{\hrule width5pt}}
\def\rightscalemark{\kern-5pt\vbox{\hrule width5pt}}

\def\Place#1#2#3{
    \count10=#1 \advance\count10 by 960
    \dimen1=\hsize \divide\dimen1 by 1920 \multiply\dimen1 by \count10
    \dimen2=\dimen3 \divide\dimen2 by 550 \multiply\dimen2 by #2
    \vbox to 0pt{\kern-\parskip\kern-18truept\kern-\dimen2
    \hbox{\kern\dimen1#3}\vss}\nointerlineskip}

\def\datum#1#2{\Place{#1}{#2}{\copy0}}

\begin{center}
\plot{8truein}{\tiny{$\bullet$}}
\nobreak
\Place{-960}{50}{\leftscalemark~~50}
\Place{-960}{100}{\leftscalemark~~100}
\Place{-960}{150}{\leftscalemark~~150}
\Place{-960}{200}{\leftscalemark~~200}
\Place{-960}{250}{\leftscalemark~~250}
\Place{-960}{300}{\leftscalemark~~300}
\Place{-960}{350}{\leftscalemark~~350}
\Place{-960}{400}{\leftscalemark~~400}
\Place{-960}{450}{\leftscalemark~~450}
\Place{-960}{500}{\leftscalemark~~500}
\Place{960}{50}{\rightscalemark\vphantom{0}}
\Place{960}{100}{\rightscalemark\vphantom{0}}
\Place{960}{150}{\rightscalemark\vphantom{0}}
\Place{960}{200}{\rightscalemark\vphantom{0}}
\Place{960}{250}{\rightscalemark\vphantom{0}}
\Place{960}{300}{\rightscalemark\vphantom{0}}
\Place{960}{350}{\rightscalemark\vphantom{0}}
\Place{960}{400}{\rightscalemark\vphantom{0}}
\Place{960}{450}{\rightscalemark\vphantom{0}}
\Place{960}{500}{\rightscalemark\vphantom{0}}
\Place{-960}{0}{\hmark\lower18pt\hbox{-960}}
\Place{-720}{0}{\hmark\lower18pt\hbox{-720}}
\Place{-480}{0}{\hmark\lower18pt\hbox{-480}}
\Place{-240}{0}{\hmark\lower18pt\hbox{-240}}
\Place{0}{0}{\hmark\lower18pt\hbox{0}}
\Place{240}{0}{\hmark\lower18pt\hbox{240}}
\Place{480}{0}{\hmark\lower18pt\hbox{480}}
\Place{720}{0}{\hmark\lower18pt\hbox{720}}
\Place{960}{0}{\hmark\lower18pt\hbox{960}}
\Place{-720}{550}{\hmark}
\Place{-480}{550}{\hmark}
\Place{-240}{550}{\hmark}
\Place{0}{550}{\hmark}
\Place{240}{550}{\hmark}
\Place{480}{550}{\hmark}
\Place{720}{550}{\hmark}
\Place{960}{550}{\hmark}
\nobreak
\begin{center}
\datum{        -104}{          54}
\datum{        -112}{          60}
\datum{        -136}{          72}
\datum{         -80}{          46}
\datum{        -108}{          60}
\datum{         -68}{          42}
\datum{         -88}{          52}
\datum{         -96}{          56}
\datum{        -132}{          74}
\datum{         -36}{          28}
\datum{         -54}{          37}
\datum{         -64}{          42}
\datum{         -80}{          50}
\datum{         -92}{          56}
\datum{        -114}{          67}
\datum{         -92}{          58}
\datum{        -124}{          74}
\datum{         -40}{          34}
\datum{         -42}{          35}
\datum{         -58}{          43}
\datum{         -64}{          46}
\datum{         -84}{          56}
\datum{        -108}{          68}
\datum{        -132}{          80}
\datum{        -136}{          82}
\datum{        -198}{         113}
\datum{        -296}{         162}
\datum{         -18}{          25}
\datum{         -20}{          26}
\datum{         -32}{          32}
\datum{         -40}{          36}
\datum{         -64}{          48}
\datum{         -70}{          51}
\datum{         -90}{          61}
\datum{        -100}{          66}
\datum{        -108}{          70}
\datum{        -160}{          96}
\datum{        -194}{         113}
\datum{         -30}{          33}
\datum{         -56}{          46}
\datum{         -70}{          53}
\datum{         -84}{          60}
\datum{         -92}{          64}
\datum{        -108}{          72}
\datum{        -112}{          74}
\datum{        -136}{          86}
\datum{         -16}{          28}
\datum{         -18}{          29}
\datum{         -20}{          30}
\datum{         -32}{          36}
\datum{         -58}{          49}
\datum{         -64}{          52}
\datum{         -68}{          54}
\datum{        -114}{          77}
\datum{        -134}{          87}
\datum{        -172}{         106}
\datum{        -184}{         112}
\datum{        -286}{         163}
\datum{         -12}{          28}
\datum{         -16}{          30}
\datum{         -28}{          36}
\datum{         -56}{          50}
\datum{         -88}{          66}
\datum{        -102}{          73}
\datum{        -124}{          84}
\datum{        -148}{          96}
\datum{        -194}{         119}
\datum{        -364}{         204}
\datum{         -68}{          58}
\datum{        -186}{         117}
\datum{        -284}{         166}
\datum{        -306}{         177}
\datum{           2}{          25}
\datum{          -8}{          30}
\datum{         -38}{          45}
\datum{         -60}{          56}
\datum{        -100}{          76}
\datum{        -136}{          94}
\datum{        -210}{         131}
\datum{          12}{          22}
\datum{         -18}{          37}
\datum{         -32}{          44}
\datum{         -50}{          53}
\datum{        -108}{          82}
\datum{        -110}{          83}
\datum{        -126}{          91}
\datum{        -130}{          93}
\datum{        -150}{         103}
\datum{        -184}{         120}
\datum{         -18}{          39}
\datum{         -20}{          40}
\datum{         -22}{          41}
\datum{         -36}{          48}
\datum{         -80}{          70}
\datum{         -88}{          74}
\datum{        -102}{          81}
\datum{        -200}{         130}
\datum{          12}{          26}
\datum{         -32}{          48}
\datum{        -102}{          83}
\datum{        -202}{         133}
\datum{        -258}{         161}
\datum{          16}{          26}
\datum{          -8}{          38}
\datum{         -12}{          40}
\datum{         -20}{          44}
\datum{         -28}{          48}
\datum{         -86}{          77}
\datum{         -88}{          78}
\datum{        -128}{          98}
\datum{        -150}{         109}
\datum{        -186}{         127}
\datum{        -190}{         129}
\datum{        -224}{         146}
\datum{          12}{          30}
\datum{           6}{          33}
\datum{          -4}{          38}
\datum{         -54}{          63}
\datum{         -76}{          74}
\datum{        -126}{          99}
\datum{        -138}{         105}
\datum{           6}{          35}
\datum{         -10}{          43}
\datum{         -70}{          73}
\datum{         -98}{          87}
\datum{         -12}{          46}
\datum{         -14}{          47}
\datum{         -32}{          56}
\datum{         -64}{          72}
\datum{         -92}{          86}
\datum{         -94}{          87}
\datum{        -104}{          92}
\datum{        -136}{         108}
\datum{        -140}{         110}
\datum{        -156}{         118}
\datum{        -162}{         121}
\datum{        -198}{         139}
\datum{        -224}{         152}
\datum{        -244}{         162}
\datum{          16}{          34}
\datum{          -6}{          45}
\datum{           8}{          40}
\datum{         -42}{          65}
\datum{         -54}{          71}
\datum{        -166}{         127}
\datum{        -236}{         162}
\datum{           8}{          42}
\datum{           4}{          44}
\datum{          -6}{          49}
\datum{         -20}{          56}
\datum{         -30}{          61}
\datum{         -42}{          67}
\datum{         -64}{          78}
\datum{        -112}{         102}
\datum{          36}{          30}
\datum{         -50}{          73}
\datum{         -96}{          96}
\datum{          40}{          30}
\datum{          30}{          35}
\datum{           8}{          46}
\datum{         -26}{          63}
\datum{        -128}{         114}
\datum{        -136}{         118}
\datum{          28}{          38}
\datum{          14}{          45}
\datum{           4}{          50}
\datum{         -16}{          60}
\datum{         -42}{          73}
\datum{         -68}{          86}
\datum{         -92}{          98}
\datum{        -104}{         104}
\datum{        -344}{         224}
\datum{          48}{          30}
\datum{          18}{          45}
\datum{          -8}{          58}
\datum{         -20}{          64}
\datum{         -30}{          69}
\datum{         -36}{          72}
\datum{         -40}{          74}
\datum{         -56}{          82}
\datum{         -90}{          99}
\datum{        -152}{         130}
\datum{        -416}{         262}
\datum{          20}{          46}
\datum{         -18}{          65}
\datum{         -30}{          71}
\datum{         -66}{          89}
\datum{        -138}{         125}
\datum{        -142}{         127}
\datum{        -162}{         137}
\datum{        -356}{         234}
\datum{          16}{          50}
\datum{         -10}{          63}
\datum{         -20}{          68}
\datum{         -28}{          72}
\datum{         -62}{          89}
\datum{         -84}{         100}
\datum{        -150}{         133}
\datum{        -172}{         144}
\datum{        -186}{         151}
\datum{        -224}{         170}
\datum{          36}{          42}
\datum{           6}{          57}
\datum{         -54}{          87}
\datum{         -60}{          90}
\datum{        -132}{         126}
\datum{          34}{          45}
\datum{          -4}{          64}
\datum{         -16}{          70}
\datum{         -42}{          83}
\datum{        -102}{         113}
\datum{        -176}{         150}
\datum{          48}{          40}
\datum{          -6}{          67}
\datum{         -16}{          72}
\datum{         -20}{          74}
\datum{         -42}{          85}
\datum{         -60}{          94}
\datum{         -88}{         108}
\datum{         -90}{         109}
\datum{        -106}{         117}
\datum{        -110}{         119}
\datum{        -236}{         182}
\datum{          60}{          36}
\datum{          30}{          51}
\datum{         -24}{          78}
\datum{         -48}{          90}
\datum{        -480}{         306}
\datum{          54}{          41}
\datum{          32}{          52}
\datum{          14}{          61}
\datum{           6}{          65}
\datum{         -10}{          73}
\datum{         -56}{          96}
\datum{          60}{          40}
\datum{          50}{          45}
\datum{           4}{          68}
\datum{          -6}{          73}
\datum{         -32}{          86}
\datum{         -44}{          92}
\datum{        -100}{         120}
\datum{          28}{          58}
\datum{           8}{          68}
\datum{         -12}{          78}
\datum{         -40}{          92}
\datum{         -84}{         114}
\datum{        -156}{         150}
\datum{        -300}{         222}
\datum{          60}{          44}
\datum{          44}{          52}
\datum{          42}{          53}
\datum{          36}{          56}
\datum{           6}{          71}
\datum{         -18}{          83}
\datum{         -62}{         105}
\datum{         -76}{         112}
\datum{        -104}{         126}
\datum{        -124}{         136}
\datum{        -152}{         150}
\datum{        -208}{         178}
\datum{          60}{          46}
\datum{          40}{          56}
\datum{          32}{          60}
\datum{          18}{          67}
\datum{         -26}{          89}
\datum{         -30}{          91}
\datum{         -42}{          97}
\datum{         -60}{         106}
\datum{        -150}{         151}
\datum{        -184}{         168}
\datum{        -204}{         178}
\datum{          44}{          56}
\datum{          36}{          60}
\datum{          18}{          69}
\datum{         -30}{          93}
\datum{          56}{          52}
\datum{          44}{          58}
\datum{          -6}{          83}
\datum{        -106}{         133}
\datum{          64}{          50}
\datum{           8}{          78}
\datum{         -80}{         122}
\datum{        -132}{         148}
\datum{        -184}{         174}
\datum{          54}{          57}
\datum{         -12}{          90}
\datum{         -24}{          96}
\datum{        -148}{         158}
\datum{        -348}{         258}
\datum{          56}{          58}
\datum{          54}{          59}
\datum{          44}{          64}
\datum{          36}{          68}
\datum{          34}{          69}
\datum{           8}{          82}
\datum{         -36}{         104}
\datum{         -64}{         118}
\datum{        -110}{         141}
\datum{          50}{          63}
\datum{          44}{          66}
\datum{          10}{          83}
\datum{         -56}{         116}
\datum{         -88}{         132}
\datum{        -104}{         140}
\datum{        -114}{         145}
\datum{        -124}{         150}
\datum{          18}{          81}
\datum{           6}{          87}
\datum{         -84}{         132}
\datum{        -168}{         174}
\datum{          40}{          72}
\datum{          30}{          77}
\datum{          16}{          84}
\datum{          -8}{          96}
\datum{         -64}{         124}
\datum{         -68}{         126}
\datum{          66}{          61}
\datum{          20}{          84}
\datum{          16}{          86}
\datum{          12}{          88}
\datum{          -4}{          96}
\datum{          -8}{          98}
\datum{         -60}{         124}
\datum{         -86}{         137}
\datum{          66}{          63}
\datum{          -6}{          99}
\datum{         -60}{         126}
\datum{         -90}{         141}
\datum{          96}{          50}
\datum{          90}{          53}
\datum{          68}{          64}
\datum{          56}{          70}
\datum{          32}{          82}
\datum{          18}{          89}
\datum{         -12}{         104}
\datum{         -80}{         138}
\datum{          54}{          73}
\datum{          40}{          80}
\datum{          34}{          83}
\datum{          32}{          84}
\datum{          20}{          90}
\datum{          18}{          91}
\datum{          16}{          92}
\datum{         -18}{         109}
\datum{         -58}{         129}
\datum{          66}{          69}
\datum{          64}{          70}
\datum{          42}{          81}
\datum{          12}{          96}
\datum{         -36}{         120}
\datum{          12}{          98}
\datum{           4}{         102}
\datum{         -76}{         142}
\datum{        -112}{         160}
\datum{        -162}{         185}
\datum{        -180}{         194}
\datum{          76}{          68}
\datum{          66}{          73}
\datum{         -18}{         115}
\datum{         -20}{         116}
\datum{         -56}{         134}
\datum{         -80}{         146}
\datum{        -450}{         331}
\datum{          48}{          84}
\datum{          -4}{         110}
\datum{        -390}{         303}
\datum{        -444}{         330}
\datum{          90}{          65}
\datum{          70}{          75}
\datum{          54}{          83}
\datum{          42}{          89}
\datum{          32}{          94}
\datum{          90}{          67}
\datum{          68}{          78}
\datum{          56}{          84}
\datum{          28}{          98}
\datum{          18}{         103}
\datum{         -44}{         134}
\datum{        -294}{         259}
\datum{          42}{          93}
\datum{          40}{          94}
\datum{          18}{         105}
\datum{          -6}{         117}
\datum{        -288}{         258}
\datum{          84}{          74}
\datum{          80}{          76}
\datum{          44}{          94}
\datum{         -42}{         137}
\datum{         -62}{         147}
\datum{        -126}{         179}
\datum{          78}{          79}
\datum{          56}{          90}
\datum{          50}{          93}
\datum{          30}{         103}
\datum{         -30}{         133}
\datum{         -40}{         138}
\datum{        -160}{         198}
\datum{         -54}{         147}
\datum{        -204}{         222}
\datum{        -372}{         306}
\datum{         106}{          69}
\datum{          84}{          80}
\datum{          78}{          83}
\datum{          70}{          89}
\datum{          48}{         100}
\datum{          72}{          90}
\datum{          32}{         110}
\datum{         110}{          73}
\datum{         104}{          76}
\datum{          90}{          83}
\datum{          78}{          89}
\datum{          36}{         110}
\datum{          12}{         122}
\datum{          86}{          87}
\datum{          70}{          95}
\datum{          44}{         108}
\datum{          32}{         114}
\datum{          16}{         122}
\datum{         -14}{         137}
\datum{         -24}{         142}
\datum{         -46}{         153}
\datum{        -114}{         187}
\datum{         108}{          78}
\datum{          78}{          93}
\datum{          36}{         114}
\datum{         -24}{         144}
\datum{        -108}{         186}
\datum{         120}{          74}
\datum{         104}{          82}
\datum{          78}{          95}
\datum{          52}{         108}
\datum{         114}{          79}
\datum{         112}{          80}
\datum{          70}{         101}
\datum{          56}{         108}
\datum{          42}{         115}
\datum{          12}{         130}
\datum{         -78}{         175}
\datum{        -132}{         202}
\datum{        -210}{         241}
\datum{         128}{          74}
\datum{         100}{          88}
\datum{          76}{         100}
\datum{         -72}{         174}
\datum{        -110}{         193}
\datum{         126}{          77}
\datum{          90}{          95}
\datum{          80}{         100}
\datum{          40}{         120}
\datum{           6}{         137}
\datum{         -18}{         149}
\datum{         -40}{         160}
\datum{         -74}{         177}
\datum{        -114}{         197}
\datum{         112}{          86}
\datum{         104}{          90}
\datum{          90}{          97}
\datum{          52}{         116}
\datum{         -16}{         150}
\datum{          68}{         110}
\datum{          54}{         117}
\datum{         -48}{         168}
\datum{         100}{          96}
\datum{          10}{         141}
\datum{         138}{          79}
\datum{         126}{          85}
\datum{         114}{          91}
\datum{         102}{          97}
\datum{          88}{         104}
\datum{          84}{         106}
\datum{          36}{         130}
\datum{         132}{          84}
\datum{         108}{          96}
\datum{         102}{          99}
\datum{          90}{         105}
\datum{          78}{         111}
\datum{          60}{         120}
\datum{          40}{         130}
\datum{         -24}{         162}
\datum{          56}{         124}
\datum{         140}{          84}
\datum{         104}{         102}
\datum{           6}{         151}
\datum{          -4}{         156}
\datum{         132}{          90}
\datum{         124}{          94}
\datum{         132}{          92}
\datum{          74}{         121}
\datum{         136}{          92}
\datum{         114}{         103}
\datum{         104}{         108}
\datum{          28}{         146}
\datum{          66}{         129}
\datum{        -312}{         318}
\datum{         140}{          94}
\datum{         126}{         101}
\datum{         160}{          86}
\datum{         154}{          89}
\datum{         120}{         106}
\datum{         116}{         108}
\datum{         114}{         113}
\datum{          86}{         127}
\datum{          70}{         135}
\datum{          66}{         137}
\datum{         138}{         103}
\datum{         136}{         104}
\datum{         126}{         109}
\datum{         114}{         115}
\datum{         108}{         118}
\datum{          88}{         128}
\datum{         -24}{         184}
\datum{         -28}{         186}
\datum{         160}{          94}
\datum{         154}{          97}
\datum{         130}{         109}
\datum{         120}{         114}
\datum{         150}{         101}
\datum{          90}{         131}
\datum{         170}{          93}
\datum{         162}{          97}
\datum{         152}{         102}
\datum{         148}{         106}
\datum{         126}{         117}
\datum{         128}{         118}
\datum{         106}{         129}
\datum{         140}{         114}
\datum{         104}{         132}
\datum{          68}{         150}
\datum{          54}{         157}
\datum{         176}{          98}
\datum{          64}{         154}
\datum{         176}{         100}
\datum{         128}{         124}
\datum{         180}{         100}
\datum{         128}{         126}
\datum{          78}{         153}
\datum{          66}{         159}
\datum{         180}{         104}
\datum{         160}{         114}
\datum{         132}{         128}
\datum{         112}{         138}
\datum{         180}{         106}
\datum{         162}{         115}
\datum{         126}{         133}
\datum{         184}{         106}
\datum{         150}{         123}
\datum{         180}{         110}
\datum{         172}{         114}
\datum{         142}{         129}
\datum{         140}{         130}
\datum{         120}{         140}
\datum{         170}{         117}
\datum{          78}{         163}
\datum{         180}{         114}
\datum{         156}{         126}
\datum{          84}{         162}
\datum{         136}{         138}
\datum{         112}{         150}
\datum{         200}{         108}
\datum{         186}{         115}
\datum{         184}{         116}
\datum{         172}{         122}
\datum{         140}{         138}
\datum{         168}{         126}
\datum{         184}{         122}
\datum{         186}{         123}
\datum{         132}{         150}
\datum{          90}{         171}
\datum{         200}{         118}
\datum{         140}{         148}
\datum{         100}{         168}
\datum{         176}{         132}
\datum{         210}{         117}
\datum{         208}{         118}
\datum{         198}{         125}
\datum{         180}{         134}
\datum{         144}{         152}
\datum{         124}{         162}
\datum{         204}{         124}
\datum{         156}{         148}
\datum{         114}{         169}
\datum{         200}{         128}
\datum{         184}{         136}
\datum{         150}{         153}
\datum{         198}{         131}
\datum{         174}{         143}
\datum{         156}{         154}
\datum{        -228}{         346}
\datum{         180}{         144}
\datum{        -160}{         314}
\datum{         224}{         124}
\datum{         220}{         126}
\datum{         200}{         136}
\datum{         210}{         133}
\datum{         192}{         142}
\datum{         132}{         172}
\datum{         230}{         129}
\datum{         174}{         157}
\datum{          20}{         234}
\datum{         224}{         134}
\datum{         222}{         135}
\datum{         216}{         138}
\datum{         234}{         131}
\datum{         216}{         140}
\datum{         210}{         143}
\datum{         200}{         148}
\datum{         196}{         150}
\datum{         168}{         164}
\datum{         240}{         130}
\datum{         234}{         133}
\datum{         224}{         138}
\datum{         160}{         170}
\datum{         136}{         182}
\datum{         212}{         150}
\datum{         200}{         156}
\datum{         156}{         178}
\datum{         150}{         181}
\datum{         248}{         134}
\datum{         234}{         141}
\datum{         242}{         139}
\datum{         186}{         167}
\datum{         212}{         158}
\datum{         146}{         193}
\datum{         204}{         168}
\datum{         184}{         180}
\datum{         216}{         168}
\datum{         258}{         151}
\datum{         272}{         148}
\datum{         256}{         156}
\datum{         256}{         158}
\datum{         280}{         150}
\datum{         224}{         178}
\datum{         264}{         160}
\datum{         258}{         163}
\datum{         252}{         166}
\datum{         228}{         178}
\datum{         204}{         190}
\datum{         248}{         174}
\datum{         276}{         162}
\datum{         252}{         174}
\datum{         272}{         166}
\datum{         266}{         169}
\datum{         294}{         159}
\datum{         246}{         185}
\datum{         264}{         178}
\datum{         242}{         189}
\datum{         228}{         202}
\datum{         286}{         177}
\datum{         296}{         174}
\datum{         276}{         186}
\datum{         220}{         218}
\datum{         320}{         170}
\datum{         312}{         174}
\datum{         294}{         187}
\datum{         300}{         194}
\datum{         320}{         190}
\datum{         276}{         212}
\datum{         322}{         193}
\datum{         318}{         197}
\datum{         320}{         198}
\datum{         270}{         223}
\datum{         276}{         222}
\datum{         256}{         234}
\datum{         352}{         190}
\datum{         320}{         206}
\datum{         360}{         194}
\datum{         364}{         194}
\datum{         356}{         202}
\datum{         320}{         222}
\datum{         324}{         222}
\datum{         360}{         206}
\datum{         380}{         198}
\datum{         324}{         230}
\datum{         356}{         220}
\datum{         396}{         214}
\datum{         384}{         222}
\datum{         316}{         262}
\datum{         384}{         234}
\datum{         354}{         259}
\datum{         360}{         258}
\datum{         384}{         262}
\datum{         360}{         306}
\datum{         476}{         270}
\datum{         450}{         303}
\datum{         456}{         302}
\datum{         468}{         306}
\datum{         512}{         286}
\datum{         540}{         298}
\datum{         528}{         318}
\datum{         510}{         331}
\datum{         516}{         330}
\datum{         576}{         302}
\datum{         612}{         346}
\datum{         648}{         358}
\datum{         840}{         446}
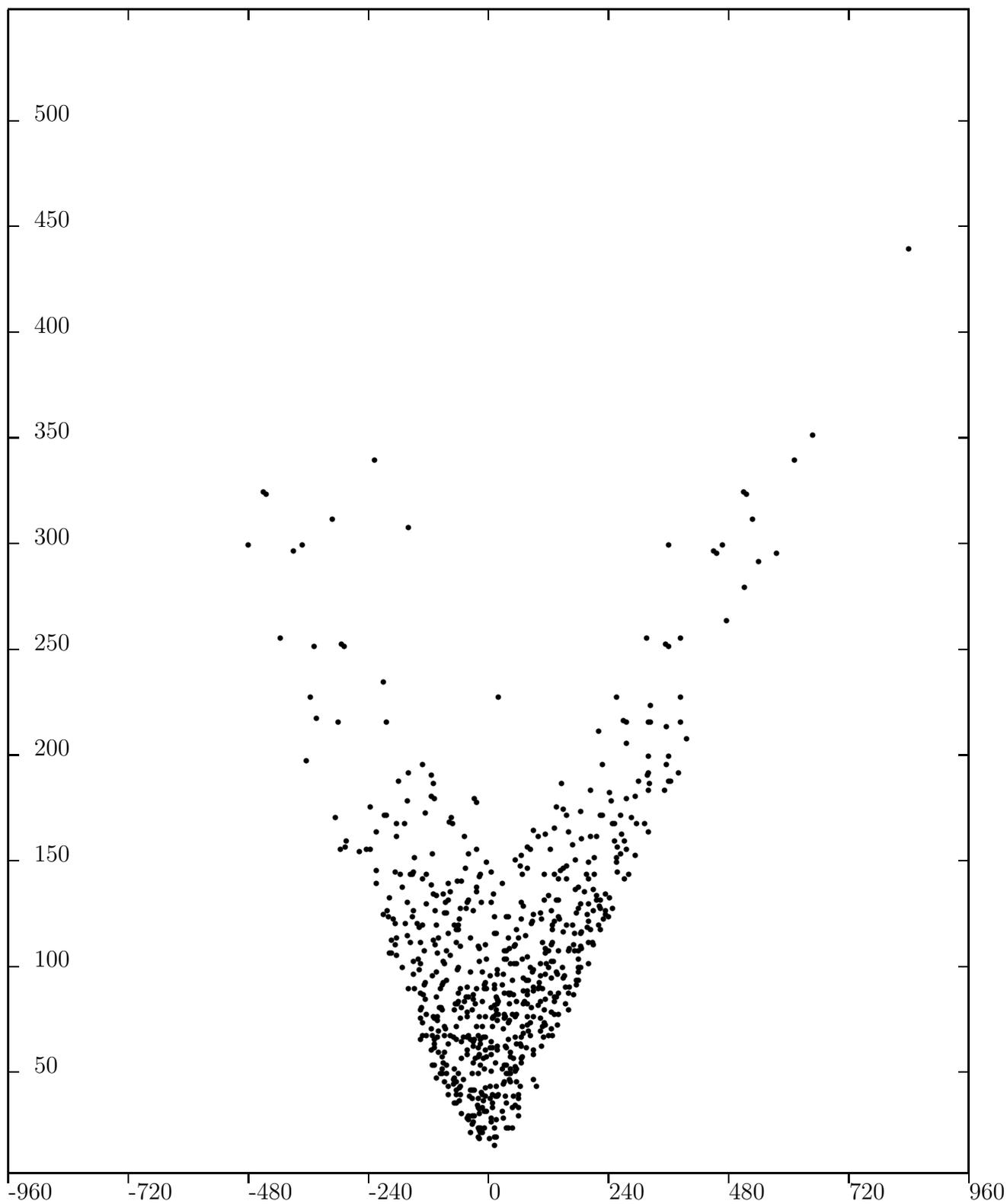
\parbox{6.4truein}{\noindent {\bf Fig. 2}~~{\it A plot of Euler numbers against
 ${\bar n}_g+n_g$ for the 695 spectra of the LG potentials without a mirror
candidate in the list.}}
\end{center}
\end{center}


\vfill \eject

\begin{center}
\parbox{6.4truein}{\noindent {\bf Table 1 }~~{\it A list of all the Euler
numbers that arise via Landau--Ginzburg vacua }.\smallskip}
\end{center}

\vskip .2truein

\begin{center}
\begin{small}
\begin{tabular}{c c c c c c}
\begin{tabular}{|c|}
\hline
{}~0\\
\hline
\makebox[.4truein]{\hfill \hphantom{$-$}  $\star$} \vrule
\makebox[.4truein]{\hfill \hphantom{$-$}~         2}\\
\hline
\makebox[.4truein]{\hfill $-$         4} \vrule
\makebox[.4truein]{\hfill \hphantom{$-$}~         4}\\
\hline
\makebox[.4truein]{\hfill $-$         6} \vrule
\makebox[.4truein]{\hfill \hphantom{$-$}~         6}\\
\hline
\makebox[.4truein]{\hfill $-$         8} \vrule
\makebox[.4truein]{\hfill \hphantom{$-$}~         8}\\
\hline
\makebox[.4truein]{\hfill $-$        10} \vrule
\makebox[.4truein]{\hfill \hphantom{$-$}~         10}\\
\hline
\makebox[.4truein]{\hfill $-$        12} \vrule
\makebox[.4truein]{\hfill \hphantom{$-$}~        12}\\
\hline
\makebox[.4truein]{\hfill $-$        14} \vrule
\makebox[.4truein]{\hfill \hphantom{$-$}~        14}\\
\hline
\makebox[.4truein]{\hfill $-$        16} \vrule
\makebox[.4truein]{\hfill \hphantom{$-$}~        16}\\
\hline
\makebox[.4truein]{\hfill $-$        18} \vrule
\makebox[.4truein]{\hfill \hphantom{$-$}~        18}\\
\hline
\makebox[.4truein]{\hfill $-$        20} \vrule
\makebox[.4truein]{\hfill \hphantom{$-$}~        20}\\
\hline
\makebox[.4truein]{\hfill $-$        22} \vrule
\makebox[.4truein]{\hfill \hphantom{$-$}~        $\star$}\\
\hline
\makebox[.4truein]{\hfill $-$        24} \vrule
\makebox[.4truein]{\hfill \hphantom{$-$}~        24}\\
\hline
\makebox[.4truein]{\hfill $-$        26} \vrule
\makebox[.4truein]{\hfill \hphantom{$-$}~        $\star$}\\
\hline
\makebox[.4truein]{\hfill $-$        28} \vrule
\makebox[.4truein]{\hfill \hphantom{$-$}~        28}\\
\hline
\makebox[.4truein]{\hfill $-$        30} \vrule
\makebox[.4truein]{\hfill \hphantom{$-$}~        30}\\
\hline
\makebox[.4truein]{\hfill $-$        32} \vrule
\makebox[.4truein]{\hfill \hphantom{$-$}~        32}\\
\hline
\makebox[.4truein]{\hfill    $\star$} \vrule
\makebox[.4truein]{\hfill \hphantom{$-$}~        34}\\
\hline
\makebox[.4truein]{\hfill $-$        36} \vrule
\makebox[.4truein]{\hfill \hphantom{$-$}~        36}\\
\hline
\makebox[.4truein]{\hfill $-$        38} \vrule
\makebox[.4truein]{\hfill \hphantom{$-$}~       $\star$}\\
\hline
\makebox[.4truein]{\hfill $-$        40} \vrule
\makebox[.4truein]{\hfill \hphantom{$-$}~        40}\\
\hline
\makebox[.4truein]{\hfill $-$        42} \vrule
\makebox[.4truein]{\hfill \hphantom{$-$}~        42}\\
\hline
\makebox[.4truein]{\hfill $-$        44} \vrule
\makebox[.4truein]{\hfill \hphantom{$-$}~        44}\\
\hline
\makebox[.4truein]{\hfill $-$        46} \vrule
\makebox[.4truein]{\hfill \hphantom{$-$}~       $\star$}\\
\hline
\makebox[.4truein]{\hfill $-$        48} \vrule
\makebox[.4truein]{\hfill \hphantom{$-$}~        48}\\
\hline
\makebox[.4truein]{\hfill $-$        50} \vrule
\makebox[.4truein]{\hfill \hphantom{$-$}~        50}\\
\hline
\makebox[.4truein]{\hfill $-$        52} \vrule
\makebox[.4truein]{\hfill \hphantom{$-$}~        52}\\
\hline
\makebox[.4truein]{\hfill $-$        54} \vrule
\makebox[.4truein]{\hfill \hphantom{$-$}~        54}\\
\hline
\makebox[.4truein]{\hfill $-$        56} \vrule
\makebox[.4truein]{\hfill \hphantom{$-$}~        $\star $}\\
\hline
\makebox[.4truein]{\hfill $-$        58} \vrule
\makebox[.4truein]{\hfill \hphantom{$-$}~       $\star $}\\
\hline
\makebox[.4truein]{\hfill $-$        60} \vrule
\makebox[.4truein]{\hfill \hphantom{$-$}~        60}\\
\hline
\makebox[.4truein]{\hfill $-$        62} \vrule
\makebox[.4truein]{\hfill \hphantom{$-$}~        $\star$}\\
\hline
\makebox[.4truein]{\hfill $-$        64} \vrule
\makebox[.4truein]{\hfill \hphantom{$-$}~        64}\\
\hline
\end{tabular}

\begin{tabular}{|c|}
\hline
\makebox[.4truein]{\hfill $-$        66} \vrule
\makebox[.4truein]{\hfill \hphantom{$-$}~        66}\\
\hline
\makebox[.4truein]{\hfill $-$        68} \vrule
\makebox[.4truein]{\hfill \hphantom{$-$}~        $\star$}\\
\hline
\makebox[.4truein]{\hfill $-$        70} \vrule
\makebox[.4truein]{\hfill \hphantom{$-$}~        70}\\
\hline
\makebox[.4truein]{\hfill $-$        72} \vrule
\makebox[.4truein]{\hfill \hphantom{$-$}~        72}\\
\hline
\makebox[.4truein]{\hfill $-$        74} \vrule
\makebox[.4truein]{\hfill \hphantom{$-$}~        74}\\
\hline
\makebox[.4truein]{\hfill $-$        76} \vrule
\makebox[.4truein]{\hfill \hphantom{$-$}~        76}\\
\hline
\makebox[.4truein]{\hfill $-$        78} \vrule
\makebox[.4truein]{\hfill \hphantom{$-$}~        78}\\
\hline
\makebox[.4truein]{\hfill $-$        80} \vrule
\makebox[.4truein]{\hfill \hphantom{$-$}~        80}\\
\hline
\makebox[.4truein]{\hfill $-$        84} \vrule
\makebox[.4truein]{\hfill \hphantom{$-$}~        84}\\
\hline
\makebox[.4truein]{\hfill $-$        86} \vrule
\makebox[.4truein]{\hfill \hphantom{$-$}~        86}\\
\hline
\makebox[.4truein]{\hfill $-$        88} \vrule
\makebox[.4truein]{\hfill \hphantom{$-$}~        88}\\
\hline
\makebox[.4truein]{\hfill $-$        90} \vrule
\makebox[.4truein]{\hfill \hphantom{$-$}~        90}\\
\hline
\makebox[.4truein]{\hfill $-$        92} \vrule
\makebox[.4truein]{\hfill \hphantom{$-$}~        92}\\
\hline
\makebox[.4truein]{\hfill $-$        94} \vrule
\makebox[.4truein]{\hfill \hphantom{$-$}~       $\star$}\\
\hline
\makebox[.4truein]{\hfill $-$        96} \vrule
\makebox[.4truein]{\hfill \hphantom{$-$}~        96}\\
\hline
\makebox[.4truein]{\hfill $-$        98} \vrule
\makebox[.4truein]{\hfill \hphantom{$-$}~       $\star $}\\
\hline
\makebox[.4truein]{\hfill $-$       100} \vrule
\makebox[.4truein]{\hfill \hphantom{$-$}~       100}\\
\hline
\makebox[.4truein]{\hfill $-$       102} \vrule
\makebox[.4truein]{\hfill \hphantom{$-$}~       102}\\
\hline
\makebox[.4truein]{\hfill $-$       104} \vrule
\makebox[.4truein]{\hfill \hphantom{$-$}~       104}\\
\hline
\makebox[.4truein]{\hfill $-$       106} \vrule
\makebox[.4truein]{\hfill \hphantom{$-$}~       106}\\
\hline
\makebox[.4truein]{\hfill $-$       108} \vrule
\makebox[.4truein]{\hfill \hphantom{$-$}~       108}\\
\hline
\makebox[.4truein]{\hfill $-$       110} \vrule
\makebox[.4truein]{\hfill \hphantom{$-$}~       110}\\
\hline
\makebox[.4truein]{\hfill $-$       112} \vrule
\makebox[.4truein]{\hfill \hphantom{$-$}~       112}\\
\hline
\makebox[.4truein]{\hfill $-$       114} \vrule
\makebox[.4truein]{\hfill \hphantom{$-$}~       114}\\
\hline
\makebox[.4truein]{\hfill $-$       116} \vrule
\makebox[.4truein]{\hfill \hphantom{$-$}~       116}\\
\hline
\makebox[.4truein]{\hfill $-$       118} \vrule
\makebox[.4truein]{\hfill \hphantom{$-$}~      $\star$}\\
\hline
\makebox[.4truein]{\hfill $-$       120} \vrule
\makebox[.4truein]{\hfill \hphantom{$-$}~       120}\\
\hline
\makebox[.4truein]{\hfill $-$       124} \vrule
\makebox[.4truein]{\hfill \hphantom{$-$}~       124}\\
\hline
\makebox[.4truein]{\hfill $-$       126} \vrule
\makebox[.4truein]{\hfill \hphantom{$-$}~       126}\\
\hline
\makebox[.4truein]{\hfill $-$       128} \vrule
\makebox[.4truein]{\hfill \hphantom{$-$}~       128}\\
\hline
\makebox[.4truein]{\hfill $-$       130} \vrule
\makebox[.4truein]{\hfill \hphantom{$-$}~       130}\\
\hline
\makebox[.4truein]{\hfill $-$       132} \vrule
\makebox[.4truein]{\hfill \hphantom{$-$}~       132}\\
\hline
\makebox[.4truein]{\hfill $-$       134} \vrule
\makebox[.4truein]{\hfill \hphantom{$-$}~       $\star$}\\
\hline
\end{tabular}

\begin{tabular}{|c|}
\hline
\makebox[.4truein]{\hfill $-$       136} \vrule
\makebox[.4truein]{\hfill \hphantom{$-$}~       136}\\
\hline
\makebox[.4truein]{\hfill $-$       138} \vrule
\makebox[.4truein]{\hfill \hphantom{$-$}~       138}\\
\hline
\makebox[.4truein]{\hfill $-$       140} \vrule
\makebox[.4truein]{\hfill \hphantom{$-$}~       140}\\
\hline
\makebox[.4truein]{\hfill $-$       142} \vrule
\makebox[.4truein]{\hfill \hphantom{$-$}~       142}\\
\hline
\makebox[.4truein]{\hfill $-$       144} \vrule
\makebox[.4truein]{\hfill \hphantom{$-$}~       144}\\
\hline
\makebox[.4truein]{\hfill       $\star$} \vrule
\makebox[.4truein]{\hfill \hphantom{$-$}~       144}\\
\hline
\makebox[.4truein]{\hfill          $\star$} \vrule
\makebox[.4truein]{\hfill \hphantom{$-$}~       146}\\
\hline
\makebox[.4truein]{\hfill $-$       148} \vrule
\makebox[.4truein]{\hfill \hphantom{$-$}~       148}\\
\hline
\makebox[.4truein]{\hfill $-$       150} \vrule
\makebox[.4truein]{\hfill \hphantom{$-$}~       150}\\
\hline
\makebox[.4truein]{\hfill $-$       152} \vrule
\makebox[.4truein]{\hfill \hphantom{$-$}~       152}\\
\hline
\makebox[.4truein]{\hfill          $\star$} \vrule
\makebox[.4truein]{\hfill \hphantom{$-$}~       154}\\
\hline
\makebox[.4truein]{\hfill $-$       156} \vrule
\makebox[.4truein]{\hfill \hphantom{$-$}~       156}\\
\hline
\makebox[.4truein]{\hfill $-$       160} \vrule
\makebox[.4truein]{\hfill \hphantom{$-$}~       160}\\
\hline
\makebox[.4truein]{\hfill $-$       162} \vrule
\makebox[.4truein]{\hfill \hphantom{$-$}~       162}\\
\hline
\makebox[.4truein]{\hfill $-$       164} \vrule
\makebox[.4truein]{\hfill \hphantom{$-$}~       164}\\
\hline
\makebox[.4truein]{\hfill $-$       166} \vrule
\makebox[.4truein]{\hfill \hphantom{$-$}~      $\star$}\\
\hline
\makebox[.4truein]{\hfill $-$       168} \vrule
\makebox[.4truein]{\hfill \hphantom{$-$}~       168}\\
\hline
\makebox[.4truein]{\hfill          $\star$} \vrule
\makebox[.4truein]{\hfill \hphantom{$-$}~       170}\\
\hline
\makebox[.4truein]{\hfill $-$       172} \vrule
\makebox[.4truein]{\hfill \hphantom{$-$}~       172}\\
\hline
\makebox[.4truein]{\hfill $-$       174} \vrule
\makebox[.4truein]{\hfill \hphantom{$-$}~       174}\\
\hline
\makebox[.4truein]{\hfill $-$       176} \vrule
\makebox[.4truein]{\hfill \hphantom{$-$}~       176}\\
\hline
\makebox[.4truein]{\hfill $-$       180} \vrule
\makebox[.4truein]{\hfill \hphantom{$-$}~       180}\\
\hline
\makebox[.4truein]{\hfill $-$       184} \vrule
\makebox[.4truein]{\hfill \hphantom{$-$}~       184}\\
\hline
\makebox[.4truein]{\hfill $-$       186} \vrule
\makebox[.4truein]{\hfill \hphantom{$-$}~       186}\\
\hline
\makebox[.4truein]{\hfill $-$       190} \vrule
\makebox[.4truein]{\hfill \hphantom{$-$}~       $\star$}\\
\hline
\makebox[.4truein]{\hfill $-$       192} \vrule
\makebox[.4truein]{\hfill \hphantom{$-$}~       192}\\
\hline
\makebox[.4truein]{\hfill $-$       194} \vrule
\makebox[.4truein]{\hfill \hphantom{$-$}~       $\star$}\\
\hline
\makebox[.4truein]{\hfill $-$       196} \vrule
\makebox[.4truein]{\hfill \hphantom{$-$}~       196}\\
\hline
\makebox[.4truein]{\hfill $-$       198} \vrule
\makebox[.4truein]{\hfill \hphantom{$-$}~       198}\\
\hline
\makebox[.4truein]{\hfill $-$       200} \vrule
\makebox[.4truein]{\hfill \hphantom{$-$}~       200}\\
\hline
\makebox[.4truein]{\hfill $-$       202} \vrule
\makebox[.4truein]{\hfill \hphantom{$-$}~       $\star$}\\
\hline
\makebox[.4truein]{\hfill $-$       204} \vrule
\makebox[.4truein]{\hfill \hphantom{$-$}~       204}\\
\hline
\makebox[.4truein]{\hfill $-$       208} \vrule
\makebox[.4truein]{\hfill \hphantom{$-$}~       208}\\
\hline
\end{tabular}

\begin{tabular}{|c|}
\hline
\makebox[.4truein]{\hfill $-$       210} \vrule
\makebox[.4truein]{\hfill \hphantom{$-$}~       210}\\
\hline
\makebox[.4truein]{\hfill       $\star $} \vrule
\makebox[.4truein]{\hfill \hphantom{$-$}~       212}\\
\hline
\makebox[.4truein]{\hfill $-$       216} \vrule
\makebox[.4truein]{\hfill \hphantom{$-$}~       216}\\
\hline
\makebox[.4truein]{\hfill $-$       220} \vrule
\makebox[.4truein]{\hfill \hphantom{$-$}~       220}\\
\hline
\makebox[.4truein]{\hfill $-$       222} \vrule
\makebox[.4truein]{\hfill \hphantom{$-$}~       222}\\
\hline
\makebox[.4truein]{\hfill $-$       224} \vrule
\makebox[.4truein]{\hfill \hphantom{$-$}~       224}\\
\hline
\makebox[.4truein]{\hfill $-$       228} \vrule
\makebox[.4truein]{\hfill \hphantom{$-$}~       228}\\
\hline
\makebox[.4truein]{\hfill          $\star$} \vrule
\makebox[.4truein]{\hfill \hphantom{$-$}~       230}\\
\hline
\makebox[.4truein]{\hfill $-$       232} \vrule
\makebox[.4truein]{\hfill \hphantom{$-$}~       232}\\
\hline
\makebox[.4truein]{\hfill $-$       234} \vrule
\makebox[.4truein]{\hfill \hphantom{$-$}~       234}\\
\hline
\makebox[.4truein]{\hfill $-$       236} \vrule
\makebox[.4truein]{\hfill \hphantom{$-$}~       236}\\
\hline
\makebox[.4truein]{\hfill $-$       240} \vrule
\makebox[.4truein]{\hfill \hphantom{$-$}~       240}\\
\hline
\makebox[.4truein]{\hfill          $\star$} \vrule
\makebox[.4truein]{\hfill \hphantom{$-$}~       242}\\
\hline
\makebox[.4truein]{\hfill $-$       244} \vrule
\makebox[.4truein]{\hfill \hphantom{$-$}~       $\star$}\\
\hline
\makebox[.4truein]{\hfill $-$       246} \vrule
\makebox[.4truein]{\hfill \hphantom{$-$}~       246}\\
\hline
\makebox[.4truein]{\hfill          $\star$} \vrule
\makebox[.4truein]{\hfill \hphantom{$-$}~       248}\\
\hline
\makebox[.4truein]{\hfill $-$       252} \vrule
\makebox[.4truein]{\hfill \hphantom{$-$}~       252}\\
\hline
\makebox[.4truein]{\hfill $-$       256} \vrule
\makebox[.4truein]{\hfill \hphantom{$-$}~       256}\\
\hline
\makebox[.4truein]{\hfill $-$       258} \vrule
\makebox[.4truein]{\hfill \hphantom{$-$}~       258}\\
\hline
\makebox[.4truein]{\hfill $-$       260} \vrule
\makebox[.4truein]{\hfill \hphantom{$-$}~       260}\\
\hline
\makebox[.4truein]{\hfill $-$       264} \vrule
\makebox[.4truein]{\hfill \hphantom{$-$}~       264}\\
\hline
\makebox[.4truein]{\hfill          $\star$} \vrule
\makebox[.4truein]{\hfill \hphantom{$-$}~       266}\\
\hline
\makebox[.4truein]{\hfill          $\star$} \vrule
\makebox[.4truein]{\hfill \hphantom{$-$}~       270}\\
\hline
\makebox[.4truein]{\hfill $-$       272} \vrule
\makebox[.4truein]{\hfill \hphantom{$-$}~       272}\\
\hline
\makebox[.4truein]{\hfill $-$       276} \vrule
\makebox[.4truein]{\hfill \hphantom{$-$}~       276}\\
\hline
\makebox[.4truein]{\hfill $-$       280} \vrule
\makebox[.4truein]{\hfill \hphantom{$-$}~       280}\\
\hline
\makebox[.4truein]{\hfill $-$       284} \vrule
\makebox[.4truein]{\hfill \hphantom{$-$}~       $\star$}\\
\hline
\makebox[.4truein]{\hfill $-$       286} \vrule
\makebox[.4truein]{\hfill \hphantom{$-$}~       286}\\
\hline
\makebox[.4truein]{\hfill $-$       288} \vrule
\makebox[.4truein]{\hfill \hphantom{$-$}~       288}\\
\hline
\makebox[.4truein]{\hfill $-$       292} \vrule
\makebox[.4truein]{\hfill \hphantom{$-$}~       292}\\
\hline
\makebox[.4truein]{\hfill $-$       294} \vrule
\makebox[.4truein]{\hfill \hphantom{$-$}~       294}\\
\hline
\makebox[.4truein]{\hfill $-$       296} \vrule
\makebox[.4truein]{\hfill \hphantom{$-$}~       296}\\
\hline
\makebox[.4truein]{\hfill $-$       300} \vrule
\makebox[.4truein]{\hfill \hphantom{$-$}~       300}\\
\hline
\end{tabular}

\begin{tabular}{|c|}
\hline
\makebox[.4truein]{\hfill $-$       304} \vrule
\makebox[.4truein]{\hfill \hphantom{$-$}~       304}\\
\hline
\makebox[.4truein]{\hfill $-$       306} \vrule
\makebox[.4truein]{\hfill \hphantom{$-$}~       306}\\
\hline
\makebox[.4truein]{\hfill $-$       312} \vrule
\makebox[.4truein]{\hfill \hphantom{$-$}~       312}\\
\hline
\makebox[.4truein]{\hfill        $\star $} \vrule
\makebox[.4truein]{\hfill \hphantom{$-$}~       316}\\
\hline
\makebox[.4truein]{\hfill        $\star $} \vrule
\makebox[.4truein]{\hfill \hphantom{$-$}~       318}\\
\hline
\makebox[.4truein]{\hfill $-$       320} \vrule
\makebox[.4truein]{\hfill \hphantom{$-$}~       320}\\
\hline
\makebox[.4truein]{\hfill        $\star $} \vrule
\makebox[.4truein]{\hfill \hphantom{$-$}~       322}\\
\hline
\makebox[.4truein]{\hfill $-$       324} \vrule
\makebox[.4truein]{\hfill \hphantom{$-$}~       324}\\
\hline
\makebox[.4truein]{\hfill $-$       330} \vrule
\makebox[.4truein]{\hfill \hphantom{$-$}~       330}\\
\hline
\makebox[.4truein]{\hfill $-$       336} \vrule
\makebox[.4truein]{\hfill \hphantom{$-$}~       336}\\
\hline
\makebox[.4truein]{\hfill $-$       340} \vrule
\makebox[.4truein]{\hfill \hphantom{$-$}~       340}\\
\hline
\makebox[.4truein]{\hfill $-$       342} \vrule
\makebox[.4truein]{\hfill \hphantom{$-$}~       342}\\
\hline
\makebox[.4truein]{\hfill $-$       344} \vrule
\makebox[.4truein]{\hfill \hphantom{$-$}~       $\star$}\\
\hline
\makebox[.4truein]{\hfill $-$       348} \vrule
\makebox[.4truein]{\hfill \hphantom{$-$}~       348}\\
\hline
\makebox[.4truein]{\hfill       $\star $} \vrule
\makebox[.4truein]{\hfill \hphantom{$-$}~       352}\\
\hline
\makebox[.4truein]{\hfill       $\star $} \vrule
\makebox[.4truein]{\hfill \hphantom{$-$}~       354}\\
\hline
\makebox[.4truein]{\hfill $-$       356} \vrule
\makebox[.4truein]{\hfill \hphantom{$-$}~       356}\\
\hline
\makebox[.4truein]{\hfill $-$       360} \vrule
\makebox[.4truein]{\hfill \hphantom{$-$}~       360}\\
\hline
\makebox[.4truein]{\hfill $-$       364} \vrule
\makebox[.4truein]{\hfill \hphantom{$-$}~       364}\\
\hline
\makebox[.4truein]{\hfill $-$       368} \vrule
\makebox[.4truein]{\hfill \hphantom{$-$}~       368}\\
\hline
\makebox[.4truein]{\hfill $-$       372} \vrule
\makebox[.4truein]{\hfill \hphantom{$-$}~       372}\\
\hline
\makebox[.4truein]{\hfill $-$       376} \vrule
\makebox[.4truein]{\hfill \hphantom{$-$}~       376}\\
\hline
\makebox[.4truein]{\hfill           $\star$ } \vrule
\makebox[.4truein]{\hfill \hphantom{$-$}~       380}\\
\hline
\makebox[.4truein]{\hfill $-$       384} \vrule
\makebox[.4truein]{\hfill \hphantom{$-$}~       384}\\
\hline
\makebox[.4truein]{\hfill $-$       390} \vrule
\makebox[.4truein]{\hfill \hphantom{$-$}~       $\star$}\\
\hline
\makebox[.4truein]{\hfill $-$       396} \vrule
\makebox[.4truein]{\hfill \hphantom{$-$}~       396}\\
\hline
\makebox[.4truein]{\hfill $-$       408} \vrule
\makebox[.4truein]{\hfill \hphantom{$-$}~       408}\\
\hline
\makebox[.4truein]{\hfill $-$       416} \vrule
\makebox[.4truein]{\hfill \hphantom{$-$}~       $\star$}\\
\hline
\makebox[.4truein]{\hfill $-$       420} \vrule
\makebox[.4truein]{\hfill \hphantom{$-$}~       420}\\
\hline
\makebox[.4truein]{\hfill $-$       426} \vrule
\makebox[.4truein]{\hfill \hphantom{$-$}~       426}\\
\hline
\makebox[.4truein]{\hfill $-$       432} \vrule
\makebox[.4truein]{\hfill \hphantom{$-$}~       432}\\
\hline
\makebox[.4truein]{\hfill $-$       444} \vrule
\makebox[.4truein]{\hfill \hphantom{$-$}~       444}\\
\hline
\makebox[.4truein]{\hfill $-$       450} \vrule
\makebox[.4truein]{\hfill \hphantom{$-$}~       450}\\
\hline
\end{tabular}

\begin{tabular}{|c|}
\hline
\makebox[.4truein]{\hfill $-$      456} \vrule
\makebox[.4truein]{\hfill \hphantom{$-$}~       456}\\
\hline
\makebox[.4truein]{\hfill $-$       468} \vrule
\makebox[.4truein]{\hfill \hphantom{$-$}~       468}\\
\hline
\makebox[.4truein]{\hfill      $\star$} \vrule
\makebox[.4truein]{\hfill \hphantom{$-$}~       476}\\
\hline
\makebox[.4truein]{\hfill $-$       480} \vrule
\makebox[.4truein]{\hfill \hphantom{$-$}~       480}\\
\hline
\makebox[.4truein]{\hfill $-$       492} \vrule
\makebox[.4truein]{\hfill \hphantom{$-$}~       492}\\
\hline
\makebox[.4truein]{\hfill $-$       504} \vrule
\makebox[.4truein]{\hfill \hphantom{$-$}~       504}\\
\hline
\makebox[.4truein]{\hfill      $\star$} \vrule
\makebox[.4truein]{\hfill \hphantom{$-$}~       510}\\
\hline
\makebox[.4truein]{\hfill       $\star$} \vrule
\makebox[.4truein]{\hfill \hphantom{$-$}~       512}\\
\hline
\makebox[.4truein]{\hfill $-$       516} \vrule
\makebox[.4truein]{\hfill \hphantom{$-$}~       516}\\
\hline
\makebox[.4truein]{\hfill $-$       528} \vrule
\makebox[.4truein]{\hfill \hphantom{$-$}~       528}\\
\hline
\makebox[.4truein]{\hfill $-$       540} \vrule
\makebox[.4truein]{\hfill \hphantom{$-$}~       540}\\
\hline
\makebox[.4truein]{\hfill $-$       552} \vrule
\makebox[.4truein]{\hfill \hphantom{$-$}~       552}\\
\hline
\makebox[.4truein]{\hfill $-$       564} \vrule
\makebox[.4truein]{\hfill \hphantom{$-$}~       564}\\
\hline
\makebox[.4truein]{\hfill $-$       576} \vrule
\makebox[.4truein]{\hfill \hphantom{$-$}~       576}\\
\hline
\makebox[.4truein]{\hfill $-$       588} \vrule
\makebox[.4truein]{\hfill \hphantom{$-$}~       588}\\
\hline
\makebox[.4truein]{\hfill $-$       612} \vrule
\makebox[.4truein]{\hfill \hphantom{$-$}~       612}\\
\hline
\makebox[.4truein]{\hfill $-$       624} \vrule
\makebox[.4truein]{\hfill \hphantom{$-$}~       624}\\
\hline
\makebox[.4truein]{\hfill $-$       636} \vrule
\makebox[.4truein]{\hfill \hphantom{$-$}~       636}\\
\hline
\makebox[.4truein]{\hfill     $\star $} \vrule
\makebox[.4truein]{\hfill \hphantom{$-$}~       648}\\
\hline
\makebox[.4truein]{\hfill $-$       660} \vrule
\makebox[.4truein]{\hfill \hphantom{$-$}~       660}\\
\hline
\makebox[.4truein]{\hfill $-$       672} \vrule
\makebox[.4truein]{\hfill \hphantom{$-$}~       672}\\
\hline
\makebox[.4truein]{\hfill $-$       720} \vrule
\makebox[.4truein]{\hfill \hphantom{$-$}~       720}\\
\hline
\makebox[.4truein]{\hfill $-$       732} \vrule
\makebox[.4truein]{\hfill \hphantom{$-$}~       732}\\
\hline
\makebox[.4truein]{\hfill $-$       744} \vrule
\makebox[.4truein]{\hfill \hphantom{$-$}~       744}\\
\hline
\makebox[.4truein]{\hfill $-$       804} \vrule
\makebox[.4truein]{\hfill \hphantom{$-$}~       804}\\
\hline
\makebox[.4truein]{\hfill        $\star $} \vrule
\makebox[.4truein]{\hfill \hphantom{$-$}~       840}\\
\hline
\makebox[.4truein]{\hfill $-$       900} \vrule
\makebox[.4truein]{\hfill \hphantom{$-$}~       900}\\
\hline
\makebox[.4truein]{\hfill $-$       960} \vrule
\makebox[.4truein]{\hfill \hphantom{$-$}~       960}\\
\hline
\makebox[.4truein]{\hfill $$       } \vrule
\makebox[.4truein]{\hfill \hphantom{$-$}~       }\\
\hline
\makebox[.4truein]{\hfill $$       } \vrule
\makebox[.4truein]{\hfill \hphantom{$-$}~       }\\
\hline
\makebox[.4truein]{\hfill $$       } \vrule
\makebox[.4truein]{\hfill \hphantom{$-$}~       }\\
\hline
\makebox[.4truein]{\hfill $$       } \vrule
\makebox[.4truein]{\hfill \hphantom{$-$}~       }\\
\hline
\makebox[.4truein]{\hfill $$       } \vrule
\makebox[.4truein]{\hfill \hphantom{$-$}~       }\\
\hline
\end{tabular}
\end{tabular}
\end{small}
\end{center}


\vfill \eject

\begin{center}
\parbox{6.4truein}{\noindent {\bf Table 2}~~
{\it Three generation models that arise in the set of
Landau--Ginzburg vacua}.
\smallskip}
\end{center}
\vskip .2truein

\begin{center}
\begin{scriptsize}
\begin{tabular}{|r|l|l|r|r|r|}
\hline
\multicolumn{1}{|c|}{$N^0$}
&\multicolumn{2}{|l|}{\sf Manifold   \ \ \ \ \ \ \ \ \ \ \ \
\ \ \ \ \ \ \ \ \ \ \ \ \ \ \ \ \ Polynomial }
&\multicolumn{1}{|c|}{$\chi$}
&\multicolumn{1}{|c|}{$h^{(1,1)}$}
&\multicolumn{1}{|c|}{$h^{(1,2)}$} \tabroom \\[.31 mm]
\hline
\hline &&&&&\\[-3.5 mm]

   1& $\IP(     3,     8,    30,    79,   117)[ 237]$ &
 $
 z_{   1}^{  79}
 +
 z_{   2}^{  15}
 z_{   5}
 +
 z_{   3}^{   4}
 z_{   5}
 +
 z_{   4}^{   3}
 +
 z_{   5}^{   2}
 z_{   1}
 +
 z_{   3}^{   3}
 z_{   2}^{  18}
 z_{   1}
 $&
$    6$&$   77$&$   74$ \tabroom \\ [1 mm]
   2& $\IP(     4,     5,    26,    65,    95)[ 195]$ &
 $
 z_{   1}^{  25}
 z_{   5}
 +
 z_{   2}^{  39}
 +
 z_{   3}^{   5}
 z_{   4}
 +
 z_{   4}^{   3}
 +
 z_{   5}^{   2}
 z_{   2}
 $&
$    6$&$   70$&$   67$ \\ [1 mm]
   3& $\IP(     1,    21,    30,    38,    45)[ 135]$ &
 $
 z_{   1}^{ 135}
 +
 z_{   2}^{   5}
 z_{   3}
 +
 z_{   3}^{   3}
 z_{   5}
 +
 z_{   4}^{   3}
 z_{   2}
 +
 z_{   5}^{   3}
 $&
$    6$&$   50$&$   47$ \\ [1 mm]
   4& $\IP(     1,    16,    23,    29,    34)[ 103]$ &
 $
 z_{   1}^{ 103}
 +
 z_{   2}^{   5}
 z_{   3}
 +
 z_{   3}^{   3}
 z_{   5}
 +
 z_{   4}^{   3}
 z_{   2}
 +
 z_{   5}^{   3}
 z_{   1}
 $&
$    6$&$   50$&$   47$ \\ [1 mm]
   5& $\IP(     5,     6,    14,    45,    65)[ 135]$ &
 $
 z_{   1}^{  27}
 +
 z_{   2}^{  15}
 z_{   4}
 +
 z_{   3}^{   5}
 z_{   5}
 +
 z_{   4}^{   3}
 +
 z_{   5}^{   2}
 z_{   1}
 $&
$    6$&$   45$&$   42$ \\ [1 mm]
   6& $\IP(     2,     9,    12,    23,    37)[  83]$ &
 $
 z_{   1}^{  37}
 z_{   2}
 +
 z_{   3}^{   5}
 z_{   4}
 +
 z_{   4}^{   2}
 z_{   5}
 +
 z_{   5}^{   2}
 z_{   2}
 +
 z_{   4}
 z_{   1}^{  30}
 +
 z_{   5}
 z_{   1}^{  23}
 $&
$    6$&$   40$&$   37$ \\ [1 mm]
   7& $\IP(     4,     7,    13,    41,    58)[ 123]$ &
 $
 z_{   1}^{  29}
 z_{   2}
 +
 z_{   3}^{   5}
 z_{   5}
 +
 z_{   4}^{   3}
 +
 z_{   5}^{   2}
 z_{   2}
 +
 z_{   3}^{   3}
 z_{   1}^{  21}
 +
 z_{   5}
 z_{   1}^{  13}
 z_{   3}
 +
 z_{   3}^{   3}
 z_{   2}^{  12}
 $&
$    6$&$   40$&$   37$ \\ [1 mm]
   8& $\IP(     4,     6,    15,    35,    45)[ 105]$ &
 $
 z_{   1}^{  15}
 z_{   5}
 +
 z_{   2}^{  15}
 z_{   3}
 +
 z_{   3}^{   7}
 +
 z_{   4}^{   3}
 +
 z_{   5}^{   2}
 z_{   3}
 +
 z_{   5}
 z_{   2}^{  10}
 $&
$    6$&$   37$&$   34$ \\ [1 mm]
   9& $\IP(     2,     6,     9,    17,    17)[  51]$ &
 $
 z_{   1}^{  21}
 z_{   3}
 +
 z_{   2}^{   7}
 z_{   3}
 +
 z_{   4}^{   3}
 +
 z_{   5}^{   3}
 +
 z_{   1}^{  17}
 z_{   4}
 $&
$    6$&$   34$&$   31$ \\ [1 mm]
  10& $\IP(     2,     9,    19,    24,    27)[  81]$ &
 $
 z_{   1}^{  36}
 z_{   2}
 +
 z_{   2}^{   9}
 +
 z_{   3}^{   3}
 z_{   4}
 +
 z_{   4}^{   3}
 z_{   2}
 +
 z_{   5}^{   3}
 +
 z_{   3}
 z_{   1}^{  31}
 $&
$    6$&$   32$&$   29$ \\ [1 mm]
  11& $\IP(     5,     8,    12,    15,    35)[  75]$ &
 $
 z_{   1}^{  15}
 +
 z_{   2}^{   5}
 z_{   5}
 +
 z_{   3}^{   5}
 z_{   4}
 +
 z_{   4}^{   5}
 +
 z_{   5}^{   2}
 z_{   1}
 $&
$    6$&$   30$&$   27$ \\ [1 mm]
  12& $\IP(     3,     4,     6,    13,    13)[  39]$ &
 $
 z_{   1}^{  13}
 +
 z_{   2}^{   9}
 z_{   1}
 +
 z_{   3}^{   6}
 z_{   1}
 +
 z_{   4}^{   3}
 +
 z_{   5}^{   3}
 +
 z_{   3}
 z_{   2}^{   5}
 z_{   4}
 $&
$    6$&$   29$&$   26$ \\ [1 mm]
  13& $\IP(     4,     5,    10,    11,    19)[  49]$ &
 $
 z_{   1}^{  11}
 z_{   2}
 +
 z_{   3}^{   3}
 z_{   5}
 +
 z_{   4}^{   4}
 z_{   2}
 +
 z_{   5}^{   2}
 z_{   4}
 +
 z_{   4}^{   3}
 z_{   1}^{   4}
 +
 z_{   5}
 z_{   2}^{   6}
 $&
$    6$&$   23$&$   20$ \\ [1 mm]
  14& $\IP(     5,     6,     9,    10,    21)[  51]$ &
 $
 z_{   1}^{   9}
 z_{   2}
 +
 z_{   2}^{   7}
 z_{   3}
 +
 z_{   4}^{   3}
 z_{   5}
 +
 z_{   5}^{   2}
 z_{   3}
 +
 z_{   3}^{   4}
 z_{   1}^{   3}
 +
 z_{   5}
 z_{   1}^{   6}
 +
 z_{   5}
 z_{   2}^{   5}
 $&
$    6$&$   23$&$   20$ \\ [1 mm]
  15& $\IP(    10,    12,    13,    15,    25)[  75]$ &
 $
 z_{   1}^{   6}
 z_{   4}
 +
 z_{   2}^{   5}
 z_{   4}
 +
 z_{   3}^{   5}
 z_{   1}
 +
 z_{   4}^{   5}
 +
 z_{   5}^{   3}
 +
 z_{   2}
 z_{   1}^{   5}
 z_{   3}
 +
 z_{   3}^{   3}
 z_{   2}^{   3}
 $&
$    6$&$   20$&$   17$ \\ [1 mm]
   16& $\IC^{\star}(     3,     4,     6,     6,    7,   7 ,  9 ) [21]$ &
 $
 z_{   1}^{   7}
 +
 z_{   2}^{   3}
 z_{   7}
 +
 z_{   3}^{   3}
 z_{   1}
 +
 z_{   4}^{   3}
 z_{   1}
 +
 z_{   5}^{   3}
 +
 z_{   6}^{   3}
 +
 z_{   7}^{   2}
 z_{   1}
 +
 z_{   3}^{   2}
 z_{   7}
 +
 z_{   7}
 z_{   4}^{   2}
 +
 z_{   4}
 z_{   2}^{   2}
 z_{   5}
 $&
$    6$&$   19$&$   16$ \\ [1 mm]

  17& $\IP(     5,     6,     9,    14,    17)[  51]$ &
 $
 z_{   1}^{   9}
 z_{   2}
 +
 z_{   2}^{   7}
 z_{   3}
 +
 z_{   4}^{   3}
 z_{   3}
 +
 z_{   5}^{   3}
 +
 z_{   3}^{   4}
 z_{   1}^{   3}
 +
 z_{   4}^{   2}
 z_{   2}^{   3}
 z_{   1}
 $&
$    6$&$   18$&$   15$ \\ [1 mm]
  18& $\IP(     5,     8,     9,    11,    12)[  45]$ &
 $
 z_{   1}^{   9}
 +
 z_{   2}^{   5}
 z_{   1}
 +
 z_{   3}^{   5}
 +
 z_{   4}^{   3}
 z_{   5}
 +
 z_{   5}^{   3}
 z_{   3}
 $&
$    6$&$   16$&$   13$ \\ [1 mm]
  19& $\IP(     3,    10,    18,    59,    87)[ 177]$ &
 $
 z_{   1}^{  59}
 +
 z_{   2}^{   9}
 z_{   5}
 +
 z_{   3}^{   5}
 z_{   5}
 +
 z_{   4}^{   3}
 +
 z_{   5}^{   2}
 z_{   1}
 +
 z_{   3}^{   3}
 z_{   2}^{  12}
 z_{   1}
 $&
$   -6$&$   57$&$   60$ \\ [1 mm]
  20& $\IP(     2,     8,    29,    49,    59)[ 147]$ &
 $
 z_{   1}^{  59}
 z_{   3}
 +
 z_{   2}^{  11}
 z_{   5}
 +
 z_{   4}^{   3}
 +
 z_{   5}^{   2}
 z_{   3}
 +
 z_{   5}
 z_{   1}^{  44}
 $&
$   -6$&$   48$&$   51$ \\ [1 mm]
  21& $\IP(     1,    18,    32,    39,    45)[ 135]$ &
 $
 z_{   1}^{ 135}
 +
 z_{   2}^{   5}
 z_{   5}
 +
 z_{   3}^{   3}
 z_{   4}
 +
 z_{   4}^{   3}
 z_{   2}
 +
 z_{   5}^{   3}
 $&
$   -6$&$   47$&$   50$ \\ [1 mm]
  22& $\IP(     1,    13,    23,    28,    32)[  97]$ &
 $
 z_{   1}^{  97}
 +
 z_{   2}^{   5}
 z_{   5}
 +
 z_{   3}^{   3}
 z_{   4}
 +
 z_{   4}^{   3}
 z_{   2}
 +
 z_{   5}^{   3}
 z_{   1}
 $&
$   -6$&$   47$&$   50$ \\ [1 mm]
  23& $\IP(     3,     5,     8,    24,    35)[  75]$ &
 $
 z_{   1}^{  25}
 +
 z_{   2}^{  15}
 +
 z_{   3}^{   9}
 z_{   1}
 +
 z_{   4}^{   3}
 z_{   1}
 +
 z_{   5}^{   2}
 z_{   2}
 +
 z_{   3}^{   5}
 z_{   5}
 $&
$   -6$&$   40$&$   43$ \\ [1 mm]
  24& $\IP(     2,     6,    15,    23,    31)[  77]$ &
 $
 z_{   1}^{  31}
 z_{   3}
 +
 z_{   2}^{   9}
 z_{   4}
 +
 z_{   4}^{   2}
 z_{   5}
 +
 z_{   5}^{   2}
 z_{   3}
 +
 z_{   4}
 z_{   1}^{  27}
 +
 z_{   5}
 z_{   1}^{  23}
 $&
$   -6$&$   37$&$   40$ \\ [1 mm]
  25& $\IP(     3,     6,     8,    23,    37)[  77]$ &
 $
 z_{   1}^{  23}
 z_{   3}
 +
 z_{   2}^{   9}
 z_{   4}
 +
 z_{   3}^{   5}
 z_{   5}
 +
 z_{   4}^{   3}
 z_{   3}
 +
 z_{   5}^{   2}
 z_{   1}
 +
 z_{   4}
 z_{   1}^{  18}
 $&
$   -6$&$   37$&$   40$ \\ [1 mm]
  26& $\IP(     3,     9,    17,    22,    24)[  75]$ &
 $
 z_{   1}^{  25}
 +
 z_{   2}^{   8}
 z_{   1}
 +
 z_{   3}^{   3}
 z_{   5}
 +
 z_{   4}^{   3}
 z_{   2}
 +
 z_{   5}^{   3}
 z_{   1}
 +
 z_{   5}^{   2}
 z_{   2}^{   3}
 $&
$   -6$&$   35$&$   38$ \\ [1 mm]
  27& $\IP(     3,     4,    14,    21,    21)[  63]$ &
 $
 z_{   1}^{  21}
 +
 z_{   2}^{  15}
 z_{   1}
 +
 z_{   3}^{   3}
 z_{   4}
 +
 z_{   4}^{   3}
 +
 z_{   5}^{   3}
 $&
$   -6$&$   32$&$   35$ \\ [1 mm]
  28& $\IP(     3,     7,    18,    26,    27)[  81]$ &
 $
 z_{   1}^{  27}
 +
 z_{   2}^{   9}
 z_{   3}
 +
 z_{   3}^{   3}
 z_{   5}
 +
 z_{   4}^{   3}
 z_{   1}
 +
 z_{   5}^{   3}
 $&
$   -6$&$   29$&$   32$ \\ [1 mm]
  29& $\IP(     3,     8,    21,    30,    31)[  93]$ &
 $
 z_{   1}^{  31}
 +
 z_{   2}^{   9}
 z_{   3}
 +
 z_{   3}^{   3}
 z_{   4}
 +
 z_{   4}^{   3}
 z_{   1}
 +
 z_{   5}^{   3}
 $&
$   -6$&$   29$&$   32$ \\ [1 mm]
  30& $\IP(     3,     8,    13,    15,    36)[  75]$ &
 $
 z_{   1}^{  25}
 +
 z_{   2}^{   9}
 z_{   1}
 +
 z_{   3}^{   3}
 z_{   5}
 +
 z_{   4}^{   5}
 +
 z_{   5}^{   2}
 z_{   1}
 +
 z_{   5}
 z_{   2}^{   3}
 z_{   4}
 $&
$   -6$&$   29$&$   32$ \\ [1 mm]
  31& $\IP(     3,    12,    15,    16,    17)[  63]$ &
 $
 z_{   1}^{  21}
 +
 z_{   2}^{   5}
 z_{   1}
 +
 z_{   3}^{   4}
 z_{   1}
 +
 z_{   4}^{   3}
 z_{   3}
 +
 z_{   5}^{   3}
 z_{   2}
 +
 z_{   3}
 z_{   2}^{   4}
 $&
$   -6$&$   29$&$   32$ \\ [1 mm]
  32& $\IP(     3,     4,    12,    17,    19)[  55]$ &
 $
 z_{   1}^{  17}
 z_{   2}
 +
 z_{   3}^{   3}
 z_{   5}
 +
 z_{   4}^{   3}
 z_{   2}
 +
 z_{   5}^{   2}
 z_{   4}
 +
 z_{   4}^{   2}
 z_{   1}^{   7}
 +
 z_{   5}
 z_{   1}^{  12}
 +
 z_{   5}
 z_{   2}^{   9}
 $&
$   -6$&$   26$&$   29$ \\ [1 mm]
  33& $\IP(     3,     5,     8,     9,    20)[  45]$ &
 $
 z_{   1}^{  15}
 +
 z_{   2}^{   9}
 +
 z_{   3}^{   5}
 z_{   2}
 +
 z_{   4}^{   5}
 +
 z_{   5}^{   2}
 z_{   2}
 +
 z_{   5}
 z_{   3}^{   2}
 z_{   4}
 $&
$   -6$&$   23$&$   26$ \\ [1 mm]
  34& $\IP(     5,    12,    13,    15,    20)[  65]$ &
 $
 z_{   1}^{  13}
 +
 z_{   2}^{   5}
 z_{   1}
 +
 z_{   3}^{   5}
 +
 z_{   4}^{   4}
 z_{   1}
 +
 z_{   5}^{   3}
 z_{   1}
 +
 z_{   4}^{   3}
 z_{   5}
 +
 z_{   5}^{   2}
 z_{   2}
 z_{   3}
 $&
$   -6$&$   23$&$   26$ \\ [1 mm]
  35& $\IP(     4,     4,    11,    17,    19)[  55]$ &
 $
 z_{   1}^{  11}
 z_{   3}
 +
 z_{   2}^{  11}
 z_{   3}
 +
 z_{   3}^{   5}
 +
 z_{   4}^{   3}
 z_{   1}
 +
 z_{   5}^{   2}
 z_{   4}
 +
 z_{   1}^{   9}
 z_{   5}
 +
 z_{   4}^{   3}
 z_{   2}
 +
 z_{   5}
 z_{   2}^{   9}
 $&
$   -6$&$   21$&$   24$ \\ [1 mm]
  36& $\IP(     4,     4,     7,    13,    15)[  43]$ &
 $
 z_{   1}^{   9}
 z_{   3}
 +
 z_{   2}^{   9}
 z_{   3}
 +
 z_{   3}^{   4}
 z_{   5}
 +
 z_{   4}^{   3}
 z_{   1}
 +
 z_{   5}^{   2}
 z_{   4}
 +
 z_{   1}^{   7}
 z_{   5}
 +
 z_{   4}^{   3}
 z_{   2}
 +
 z_{   5}
 z_{   2}^{   7}
 $&
$   -6$&$   20$&$   23$ \\ [1 mm]
  37& $\IP(     3,     5,     8,    14,    15)[  45]$ &
 $
 z_{   1}^{  15}
 +
 z_{   2}^{   9}
 +
 z_{   3}^{   5}
 z_{   2}
 +
 z_{   4}^{   3}
 z_{   1}
 +
 z_{   5}^{   3}
 $&
$   -6$&$   20$&$   23$ \\ [1 mm]
  38& $\IP(     4,     5,     7,     8,    19)[  43]$ &
 $
 z_{   1}^{   9}
 z_{   3}
 +
 z_{   2}^{   7}
 z_{   4}
 +
 z_{   3}^{   5}
 z_{   4}
 +
 z_{   4}^{   3}
 z_{   5}
 +
 z_{   5}^{   2}
 z_{   2}
 +
 z_{   2}^{   3}
 z_{   1}^{   7}
 +
 z_{   5}
 z_{   1}^{   6}
 +
 z_{   3}^{   4}
 z_{   2}^{   3}
 $&
$   -6$&$   20$&$   23$ \\ [1 mm]
  39& $\IP(     4,     4,     5,     5,     7)[  25]$ &
 $
 z_{   1}^{   5}
 z_{   3}
 +
 z_{   2}^{   5}
 z_{   3}
 +
 z_{   3}^{   5}
 +
 z_{   4}^{   5}
 +
 z_{   5}^{   3}
 z_{   1}
 +
 z_{   1}^{   5}
 z_{   4}
 +
 z_{   5}^{   3}
 z_{   2}
 $&
$   -6$&$   17$&$   20$ \\ [1 mm]
  40& $\IP(     4,     7,     9,    10,    15)[  45]$ &
 $
 z_{   1}^{   9}
 z_{   3}
 +
 z_{   2}^{   5}
 z_{   4}
 +
 z_{   3}^{   5}
 +
 z_{   4}^{   3}
 z_{   5}
 +
 z_{   5}^{   3}
 +
 z_{   2}^{   3}
 z_{   1}^{   6}
 $&
$   -6$&$   13$&$   16$ \\ [1 mm]
\hline
\end{tabular}
\end{scriptsize}
\end{center}


\vfill \eject

\noindent
\section*{Acknowledgements}

\noindent
It is a pleasure to thank D.Dais, T.H\"ubsch and S.S.Roan for
discussions. A. Klemm likes to thank the MPI f\"ur Mathematik in
Bonn for hospitality. We also like to thank
M.Kreuzer and H.Skarke for informing
us about their investigation which is parallel to
ours to some extent.

 \end{document}